\documentclass[preprint,12pt]{elsarticle}

\PassOptionsToPackage{margin=0.81in}{geometry}

\usepackage{amssymb, bm}
\usepackage{amsmath}
\usepackage{subfig}
\usepackage{caption}
\usepackage{subfig}
\usepackage{graphicx}
\usepackage{epstopdf}
\usepackage{fancyhdr}
\usepackage{geometry}
\usepackage{setspace}
\usepackage{booktabs}
\usepackage{nomencl}
\usepackage{tikz}
\usepackage{multirow}
\makenomenclature
\RequirePackage{ifthen}
\renewcommand{\nomgroup}[1]{\ifthenelse{\equal{#1}{R}}{\item[\textbf{Roman letters}]}{\ifthenelse{\equal{#1}{S}}{\item[\textbf{Subscripts}]}{\ifthenelse{\equal{#1}{G}}{\item[\textbf{Greek letters}]}{\ifthenelse{\equal{#1}{A}}{\item[\textbf{Acronyms}]}}}}}

\begin{document}

\begin{frontmatter}



\title{An accurate methodology for surface tension modeling in $\texttt{OpenFOAM}^{\textregistered}$}


\author[a]{A.E. Saufi}
\author[b]{O. Desjardins}
\author[a]{A. Cuoci}

\address[a]{Department of Chemistry, Materials, and Chemical Engineering “G. Natta”, P.zza Leonardo da Vinci 32, Milano, Italy}
\address[b]{Sibley School of Mechanical and Aerospace Engineering, Cornell University, Ithaca, NY, United States}

\begin{abstract}
In this paper a numerical methodology for surface tension modeling is presented, with an emphasis on the implementation in the $\texttt{OpenFOAM}^{\textregistered}$ framework. The methodology relies on a combination of (i) a well-balanced approach based on the Ghost Fluid Method (GFM), including the jump of density and pressure directly in the numerical discretization of the pressure equation, and (ii)  Height Functions to evaluate the interface curvature, implemented, to the authors' knowledge, for the first time in $\texttt{OpenFOAM}^{\textregistered}$. The method is able to significantly reduce spurious currents (almost to machine accuracy) for a stationary droplet, showing second order convergence both for the curvature and the interface shape. Accurate results are also obtained for additional test cases such as translating droplets, capillary oscillations and rising bubbles, for which numerical results are comparable to what obtained by other numerical codes in the same conditions. Finally, the Height Functions method is extended to include the treatment of contact angles, both for  sessile droplets and droplets suspended under the effect of gravity, showing a very good agreement  with the theoretical prediction.\\ The code works in parallel mode and  details on the actual implementation in $\texttt{OpenFOAM}^{\textregistered}$ are included to facilitate the reproducibility of the results.

\end{abstract}

\nomenclature[G]{$\rho$ }{density $\left[\frac{kg}{m^3}\right]$}
\nomenclature[R]{$\textbf{g}$ }{gravitational acceleration $\left[\frac{m}{s^2}\right]$}
\nomenclature[S]{$L$}{liquid}
\nomenclature[S]{$G$}{gas}
\nomenclature[R]{$\textbf{v}$ }{velocity $\left[\frac{m}{s}\right]$}
\nomenclature[R]{$p$ }{pressure $\left[Pa\right]$}
\nomenclature[R]{$p_{d}$ }{dynamic pressure $\left[Pa\right]$}
\nomenclature[G]{$\alpha$ }{VOF marker function $[-]$}
\nomenclature[G]{$\beta$ }{reciprocal of density  $\left[\frac{m^3}{kg}\right]$}
\nomenclature[R]{$t$}{time $\left[s\right]$}
\nomenclature[G]{$\phi$}{velocity face flux $\left[m^3/s\right]$}
\nomenclature[R]{$\textbf{x}$}{position vector $\left[m\right]$}
\nomenclature[G]{$\kappa$}{curvature $\left[\frac{1}{m}\right]$}
\nomenclature[G]{$\delta_s$}{Dirac delta $\left[\frac{1}{m}\right]$}
\nomenclature[G]{$\mu$}{dynamic viscosity $\left[\frac{kg}{ms}\right]$}
\nomenclature[R]{$D$}{diameter $\left[m\right]$}
\nomenclature[R]{$R$}{radius $\left[m\right]$}
\nomenclature[R]{$h$}{height $\left[m\right]$}
\nomenclature[R]{$F$}{integral function}
\nomenclature[R]{$T$}{time scale $[s]$}
\nomenclature[R]{$v_{\sigma}$}{capillary wave velocity $[m/s]$}
\nomenclature[R]{$L$}{error norm}
\nomenclature[R]{$Ca$}{capillary number $[-]$}
\nomenclature[R]{$La$}{Laplace number $[-]$}
\nomenclature[R]{$We$}{Weber number $[-]$}
\nomenclature[R]{$Eo$}{Eotvos number $[-]$}
\nomenclature[R]{$p$}{perimeter $[m]$}
\nomenclature[R]{$c$}{circularity $[-]$}
\nomenclature[G]{$\nu$}{kinematic viscosity $[m/s^2]$}
\nomenclature[R]{$\textbf{d}$}{distance vector $\left[m\right]$}
\nomenclature[R]{$H$}{total pressure jump $\left[Pa\right]$}
\nomenclature[R]{$\textbf{S}$}{surface vector $\left[m^2\right]$}
\nomenclature[R]{$\textbf{w}$}{vertex position vector $\left[m\right]$}
\nomenclature[R]{$\textbf{n}$}{interface normal $\left[-\right]$}
\nomenclature[S]{$f$}{face-centered value}
\nomenclature[G]{$\lambda$}{relative position of the interface $\left[-\right]$}
\nomenclature[G]{$\Theta_s$}{Heaviside function $\left[-\right]$}
\nomenclature[G]{$\psi$}{distance $\left[m\right]$}
\nomenclature[G]{$\epsilon$}{deformation factor $\left[-\right]$}
\nomenclature[G]{$\theta$}{angle [degree, rad]}
\nomenclature[G]{$\omega$}{frequency $\left[s^{-1}\right]$}
\nomenclature[S]{$P$}{cell-centered value}
\nomenclature[S]{$O$}{owner}
\nomenclature[S]{$N$}{neighbour}
\nomenclature[S]{$w$}{wet}
\nomenclature[S]{$ex$}{exact}
\nomenclature[S]{$d$}{dry}
\nomenclature[S]{$s$}{smoothed}
\nomenclature[G]{$\Delta x,y$}{cell size $[m]$}
\nomenclature[S]{$0$}{initial, reference}
\nomenclature[G]{$\sigma$}{surface tension $\left[\frac{N}{m}\right]$}
\nomenclature[A]{VOF}{Volume Of Fluid}
\nomenclature[A]{PLIC}{Piecewise Linear Interface Calculation}
\nomenclature[A]{GFM}{Ghost Fluid Method}
\nomenclature[A]{RMS}{Root Mean Square}
\nomenclature[A]{CSF}{Continuum Surface Force}


\end{frontmatter}

	
\section{Introduction}
The research on multiphase flows is of practical and fundamental interest for the scientific community, especially regarding energy and propulsion science. Liquid sprays and atomization processes are ubiquitous and the fundamental understanding of the physical phenomena involved paves the way to a more rational and focused design of many engineering devices (e.g for combustion, irrigation, coating).  The transformation of a macro-scale liquid structure into small droplets involves several physical steps, including break-up, fragmentation and droplet coalescence, in addition to the phase-change process \cite{abramzon1989droplet, sirignano1999fluid}. Bubbly flows are also present in numerous applications, including boiling heat transfer, gas absorption and stirring in chemical engineering processes \cite{juric1998computations, delale2005direct}. The correct design of these systems requires a knowledge on the collective rising velocity of the bubbles, their distribution and the interaction with the liquid phase. The numerical analysis of these phenomena is very complex and requires a multiphase code  able to efficiently handle the interface transport (based on VOF, level-set or front-tracking methods \cite{hirt1981volume, sussman1998improved, tryggvason2001front}), large density ratios and a detailed description of the surface tension force. In particular, this latter will be the main focus of this paper. \\ The accurate representation of surface tension is extremely delicate, especially at small scales, due to the singular nature of this force. This makes its implementation within a continuum mechanics context very challenging, in particular:

\begin{itemize}
	\item The surface tension force only exists at the liquid-gas interface, posing a great problem concerning its discretization. It is well known that incorrect, approximate or trivial discretizations of the surface tension force lead to the formation of unphysical velocities around the interface (called spurious currents) due to the local numerical imbalance between the pressure gradient and the surface tension force \cite{francois2006balanced,  popinet2009accurate, popinet2018numerical}. Conversely, well-balanced numeric schemes are  able to recover equilibrium solutions of specific cases. The simplest case is given by a motionless spherical droplet at zero gravity, described by the following (continuous) equation:
	
	\begin{equation}
	\frac{\partial\left(\rho\textbf{v}\right)}{\partial t} = -\nabla p + \sigma\kappa\delta_s\textbf{n} = \textbf{0}
	\label{equationStatic}
	\end{equation} 
	
	where $\delta_s$ is a Dirac delta of the interface position,  $\kappa$ is the curvature and $\textbf{n}$ is the normal to the interface. The equilibrium solution for a two-dimensional  droplet of diameter $D$ is a discontinuous pressure field, with a pressure jump $[p]$:
	
	\begin{equation}
	[p] = \frac{2\sigma}{D}
	\label{pressurejumpanalytical}
	\end{equation} 
	
	If the numerical discretizations of the pressure gradient $\nabla p$ and surface tension force $\sigma\kappa\delta_s\textbf{n}$ do not cancel out:
	
	\begin{equation}
	\frac{\partial\left(\rho\textbf{v}\right)}{\partial t} =-\nabla p + \sigma\kappa\delta_s\textbf{n} \neq \textbf{0}
	\end{equation}
	
	a non-physical flow field (spurious current) will be generated as a consequence of this numerical imbalance. This flow generally grows in time, eventually destroying the droplet. The Continuum Surface Force (CSF) method \cite{brackbill1992continuum} suffers particularly from this issue, due to the volumetric representation of the surface tension force via delta functions. Alternative methods directly approach the sharp discontinuity, such as the Ghost Fluid Method (GFM). The main strategy is an explicit treatment of the discontinuity, including the jump conditions directly in the numerical discretization. Introduced by Fedwik et al. \cite{fedkiw1999non} to handle discontinuities in compressible solvers, it has been extended to study detonations \cite{fedkiw1999ghost}, to efficiently solve variable coefficients Poisson equations \cite{liu2000boundary} and finally to treat the large density ratio typical of multiphase flows \cite{olsson2005conservative, desjardins2008accurate, bo2014volume, vukvcevic2017implementation}, as well as the pressure jump due to surface tension (Equation \ref{pressurejumpanalytical}). An extensive review by Lalanne et al. \cite{lalanne2015computation} analyzes the treatment of the viscosity jump with the GFM approach, based on the works of Kang \cite{kang2000boundary} and Sussmann \cite{sussman2007sharp};

	\item The surface tension force is directly proportional to the interface curvature (Equation \ref{equationStatic}). Even if the surface tension force is discretized with a perfectly balanced method, this is generally not enough to eliminate spurious currents. The curvature gradients along the interface generate a flow, which may be physical (e.g. when a deformed droplet tends towards a spherical shape) or unphysical, when these variations are actually due to numerical errors in the curvature evaluation \cite{brackbill1992continuum, cummins2005estimating}. This unphysical flow (spurious current) deforms the droplet, further complicating the curvature computation. This issue is particularly evident within a VOF method, because it requires the differentiation of a discontinuous function (i.e. $\alpha$):
	
	\begin{equation}
	\kappa=\nabla\cdot\textbf{n}=\nabla\cdot\left(\frac{\nabla\alpha}{|\nabla \alpha|}\right)
	\label{curvatureDefinition}
	\end{equation}
	
	On the other hand, level-set methods adopt an iso-contour of a smooth  function to track the interface, allowing an easier evaluation of the curvature through Equation \ref{curvatureDefinition}. \\ The first solutions relied on the filtering of $\alpha$, in order to handle smoother functions. However, many studies showed the inaccuracy and non-consistency of these methods \cite{williams1998accuracy, tryggvason2011direct}. Widely used are Height Functions \cite{cummins2005estimating, bornia2011properties}, in which local heights are computed (from volume fractions) and differentiated in order to obtain the curvature. The main conditions is to have well defined heights, usually available when the droplet is well resolved. When this is not possible, least-squares methods can be used, based on discrete surfaces differentiation \cite{chiodi2017reformulation, marchandise2007stabilized,  desjardins2008accurate}. The interface is locally approximated by a quadratic form, from which the curvature can be analytically computed;
	
	\item Finally, the numerical treatment of gas-liquid-solid contact points is particularly important. The study of the contact angle between the solid and the interface is fundamental in many applications such as sessile and suspended droplets. In particular, in this latter the contact angle is provided by the equilibrium between surface tension force and the droplet weight. In case of transient systems, it also depends on the liquid and gas velocity fields (which ultimately act as additional forces on the droplet). It is necessary to use a stable methodology to predict the interface contact angle, depending on the local operating conditions.  While it is relatively easy to deal with static contact angles \cite{afkhami2008height}, the description of a moving  contact line implies a paradox: Navier Stokes equations with no-slip conditions produce an infinite viscous dissipation \cite{huh1971hydrodynamic}. This has been addressed in several works \cite{afkhami2009mesh, wang20183d, renardy2001numerical} regarding contact angle hysteresis.

\end{itemize}

\begin{figure}
	\centering
	{\includegraphics[width=.9\textwidth,height=0.22\textheight]{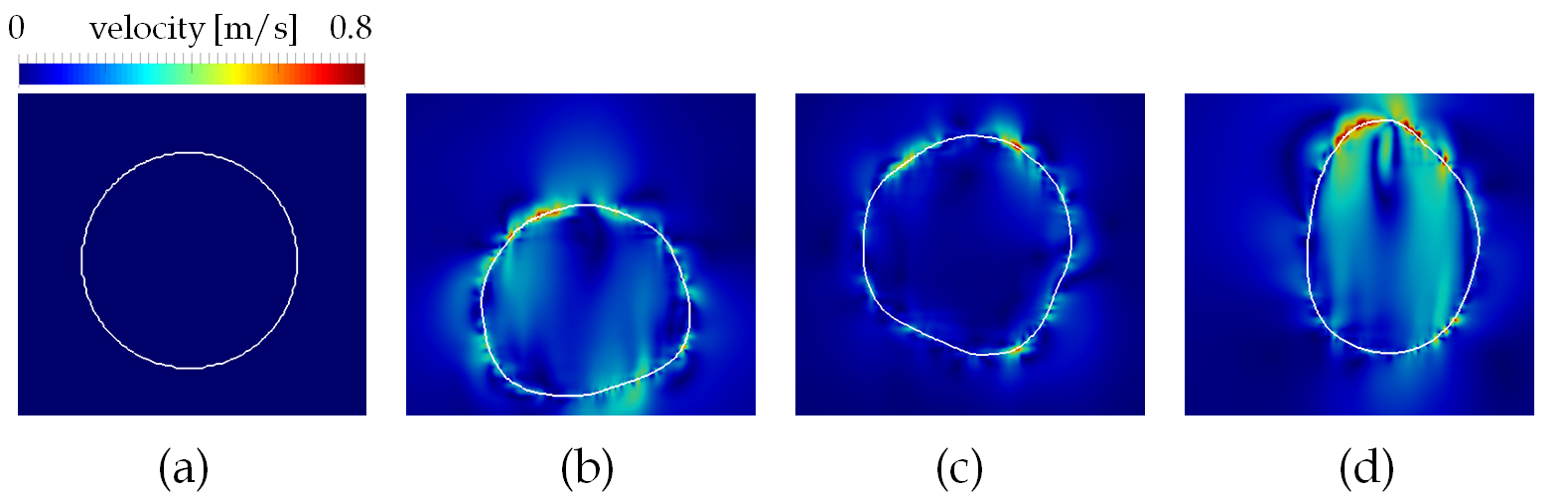}}
	\caption{Spurious currents on a droplet of water  ($D=1$ mm, white contour) using the \texttt{interFoam} multiphase solver available in $\texttt{OpenFOAM}^{\textregistered}$ \cite{greenshields2015openfoam}. Times t = 0 (a), 0.04 (b), 0.1 (c) and 0.2 (d) s.}
	\label{exampleSpuriousCurrents}
\end{figure}

 These aspects have been extensively treated in many works and recent numerical implementations efficiently include all of them (e.g. Gerris \cite{popinet2009accurate}, NGA \cite{desjardins2008accurate}). In this work we want to focus on  the $\texttt{OpenFOAM}^{\textregistered}$ framework, which lacks for a general comprehensive solver for surface tension driven flows. Most of the recent attempts are mainly based on smoothing-filtering techniques \cite{aboukhedr2018simulation, raeini2012modelling} or simplified coupled VOF-Level Set methods \cite{ferrari2017flexible, albadawi2013influence}, which can be hardly generalized (uncertainty on the number-type of filters, choice of the interface thickness, re-initialization method etc.). Comprehensive reviews on the capabilities of the main multiphase solver in $\texttt{OpenFOAM}^{\textregistered}$ (\texttt{interFOAM}) for microfluidic applications \cite{deshpande2012evaluating, jamshidi2019suitability} clearly show large errors in the pressure jump prediction, a significant presence of spurious currents (Figure \ref{exampleSpuriousCurrents}) and no mesh convergence (even first order) on the interface curvature.  The objective of this paper is to fill this gap, providing a stable and accurate numerical methodology for $\texttt{OpenFOAM}^{\textregistered}$  able to overcome the aforementioned issues, allowing to correctly represent the surface tension effects in common systems. This is implemented as an extension of the $\texttt{DropletSMOKE++}$ solver \cite{saufi2019dropletsmoke++}, a CFD multiphase code specifically conceived for the analysis of fuel droplets vaporization. Therefore:
 
 \begin{itemize}
 	\item The Ghost Fluid Method (GFM) has been implemented to handle the density and the pressure gradients at the interface, providing a  balanced and sharp discretization of these fields. This is based on the work of Vukcevic and Jasak \cite{vukvcevic2017implementation}, which has been further extended in this work to account for the pressure jump due to surface tension;
 	\item The Height Functions method has been implemented to evaluate the interface curvature, to the authors' knowledge for the first time in $\texttt{OpenFOAM}^{\textregistered}$. This  method has been chosen due to its simplicity, good convergence properties and accurate results for sufficiently resolved interfaces;
 	\item The Height Functions method is extended to treat contact lines, in order to predict contact angles for static (sessile droplets) and dynamic (suspended droplets) systems. This is partially based on the work of Dupont et al. \cite{dupont2010numerical} and Afkhami et al. \cite{afkhami2008height}.

 \end{itemize}

The paper organization includes: (i) the main mathematical model, (ii) a detailed description of the Ghost Fluid Method and (iii) the Height Functions method. Details about the specific implementation in  $\texttt{OpenFOAM}^{\textregistered}$ are also provided. The methodology is validated with classical test cases, including: static droplet equilibrium, translating droplets, capillary oscillations and bubbles rising in a dense fluid. Many of these tests have been already carried out by other authors and the comparison of their results with ours is provided as well.  Finally, the contact angle treatment is presented, with applications to sessile droplets and suspended droplets. The manuscript finishes with the conclusions.

\section{Main mathematical model}
\subsection{Interface advection}
The VOF methodology is adopted to transport the interface. The two phases are treated as a single fluid whose properties vary abruptly at the phase boundary. A scalar marker function $\alpha$ represents the liquid volumetric fraction, varying from value 0 in the gas-phase to value 1 in the liquid phase. The $\alpha$  advection equation is: 

\begin{equation}
\frac{\partial \alpha}{\partial t} + \nabla\cdot\left(\textbf{v}\alpha\right)= 0
\end{equation}

The interface transport is solved using the \texttt{isoAdvector} library developed by Roenby and Jasak \cite{roenby2016computational}.
\texttt{isoAdvector} performs a geometric advection of the interface, whose quality is superior to the MULES (Multidimensional Universal Limiter with Explicit Solution) compressive scheme by Weller \cite{damian2013extended} usually used in the  $\texttt{OpenFOAM}^{\textregistered}$ multiphase solvers.

\subsection{Navier-Stokes equations}
A single Navier-Stokes equation is solved for both phases (one-fluid approach):

\begin{equation}
\frac{\partial\left(\rho\textbf{v}\right)}{\partial t} + \nabla\cdot\left(\rho\textbf{v}\otimes\textbf{v}\right)= \nabla\cdot\mu\left(\nabla\textbf{v}+\nabla\textbf{v}^T\right)-\nabla p + \rho\textbf{g}+\sigma\kappa\delta_s\textbf{n}
\label{navierstokes}
\end{equation}

Taking out the continuity equation $\frac{\partial \rho}{\partial t}+\nabla\cdot\left(\rho\textbf{v}\right)=0$ :

\begin{equation}
\frac{\partial\textbf{v}}{\partial t} + \textbf{v}\cdot\nabla\textbf{v}= \beta\nabla\cdot\mu\left(\nabla\textbf{v}+\nabla\textbf{v}^T\right)-\beta\nabla p + \textbf{g}+\beta\sigma\kappa\delta_s\textbf{n}
\label{navierstokesII}
\end{equation}

where $\beta=\frac{1}{\rho}$. Following the work of Vukcevic and Jasak \cite{vukvcevic2017implementation}, we can write Equation \ref{navierstokesII} as:

\begin{equation}
\frac{\partial\textbf{v}}{\partial t} + \textbf{v}\cdot\nabla\textbf{v}= \beta\nabla\cdot\mu\left(\nabla\textbf{v}+\nabla\textbf{v}^T\right)-\beta\nabla p_d
\label{finalnavierstokesequation}
\end{equation}

where $p_d = p - \frac{1}{\beta}\textbf{g}\cdot\textbf{x}-\sigma\kappa \Theta_s$ is the dynamic pressure ($\Theta_s$ is the Heaviside function of the interface). In this way gravity and surface tension are considered as contributions to the dynamic pressure jump. 
The main novelty of this work with respect to that of Vukcevic et al.  \cite{vukvcevic2017implementation} is the introduction of the surface tension term $\sigma\kappa \Theta_s$, which was not considered in their work.

\subsection{Pressure equation}
The PIMPLE algorithm \cite{greenshields2015openfoam}, a combination between SIMPLE  (Semi-Implicit Method for Pressure-Linked Equations) and PISO  (Pressure Implicit Splitting of Operators) is used to manage the pressure-velocity coupling in $\texttt{OpenFOAM}^{\textregistered}$, computing a velocity field which satisfies both momentum and continuity equation through an iterative procedure. Equation \ref{finalnavierstokesequation} is written in a semi-discretized form \cite{jasak1996error}:

\begin{equation}
	a_P\textbf{v}_P=\textbf{H}\left(\textbf{v}_N\right)-\beta\nabla p_d
\end{equation}

where $\textbf{H}\left(\textbf{v}_N\right)$ matrix includes neighbor transport coefficients and source terms, while $a_p$ are the diagonal coefficients of the matrix. The pressure gradient is not discretized at this point. We can derive the velocities at the cell centers:

\begin{equation}
\textbf{v}_{P}=\frac{\textbf{H}\left(\textbf{v}_N\right)}{a_{P}}-\frac{1}{a_{P}}\beta\nabla p_d
\label{cellVelocities}
\end{equation}

and the face fluxes $\phi_f$:

\begin{equation}
\phi_f=\frac{\textbf{H}\left(\textbf{v}_N\right)_f}{a_{P,f}}\cdot\textbf{S}_f -\left(\frac{1}{a_{P}}\right)_f\beta_f\left(\nabla p_d\right)_f\cdot\textbf{S}_f
\label{faceFluxes}
\end{equation}

The continuity equation for incompressible flows is: 
\begin{equation}
       \nabla\cdot\textbf{v}=0
       \label{continuityEquation}
\end{equation}

The final form of the pressure equation is (substituting Equation \ref{cellVelocities} in Equation \ref{continuityEquation}):

\begin{equation}
 \nabla\cdot\left(\frac{\textbf{H}\left(\textbf{v}_N\right)}{a_P}\right) =\nabla\cdot\left(\frac{1}{a_{P}}\beta\nabla p_d\right)
\label{finalpressureequationWithDivergence}
\end{equation}

The finite volume discretization of this equation is \cite{jasak1996error}:
 
\begin{equation}
 \sum_{f}^{}\frac{\textbf{H}\left(\textbf{v}_N\right)_f}{a_{P,f}}\cdot\textbf{S}_f =\sum_{f}^{}\left(\frac{1}{a_{P}}\right)_f\beta_f\left(\nabla p_d\right)_f\cdot\textbf{S}_f
\label{finalpressureequation}
\end{equation}

The solution of Equation \ref{finalpressureequation} gives a  pressure field $p_d$ that, inserted in \ref{finalnavierstokesequation}, provides a conservative velocity field.
The face fluxes $\phi_f$ are then reconstructed based on the new pressure gradient via Equation \ref{faceFluxes}. \\
The pressure equation (Equation \ref{finalpressureequation}) involves three discontinuous fields at the free surface:  (i) $\beta_f=\left(\frac{1}{\rho}\right)_f$, (ii) the dynamic pressure $p_d$ and (iii) the viscosity $\mu$ included in the matrix $\textbf{H}\left(\textbf{v}_N\right)$. Their jump $[~]$ at the interface $f$ is:

\begin{equation}
	\left[\frac{1}{\beta_f}\right]=\rho_L-\rho_G
	\label{densityjump}
\end{equation}

\begin{equation}
 H_f=\left[p_d\right] = -\left[\frac{1}{\beta_f}\right]\textbf{g}\cdot\textbf{x}_f-\sigma\kappa_f=-\left(\rho_L-\rho_G\right)\textbf{g}\cdot\textbf{x}_f-\sigma\kappa_f
\label{pressurejump}
\end{equation}

\begin{equation}
\left[\mu\right]=\mu_L-\mu_G
\label{viscosityjump}
\end{equation}

The density and pressure discontinuities will be treated with the GFM. Due to the inherent complexity of treating the viscosity jump within a GFM approach \cite{lalanne2015computation}, a Continuum Surface Force method (CSF) is adopted for this term. The jump is smoothed across a few interfacial cells, making the viscosity field continuous:

\begin{equation}
	\mu=\alpha\mu_L+\left(1-\alpha\right)\mu_G
\end{equation}

\section{The Ghost Fluid Method (GFM)}

\begin{figure}
	\centering
	\subfloat[]
	{\includegraphics[width=.4\textwidth,height=0.2\textheight]{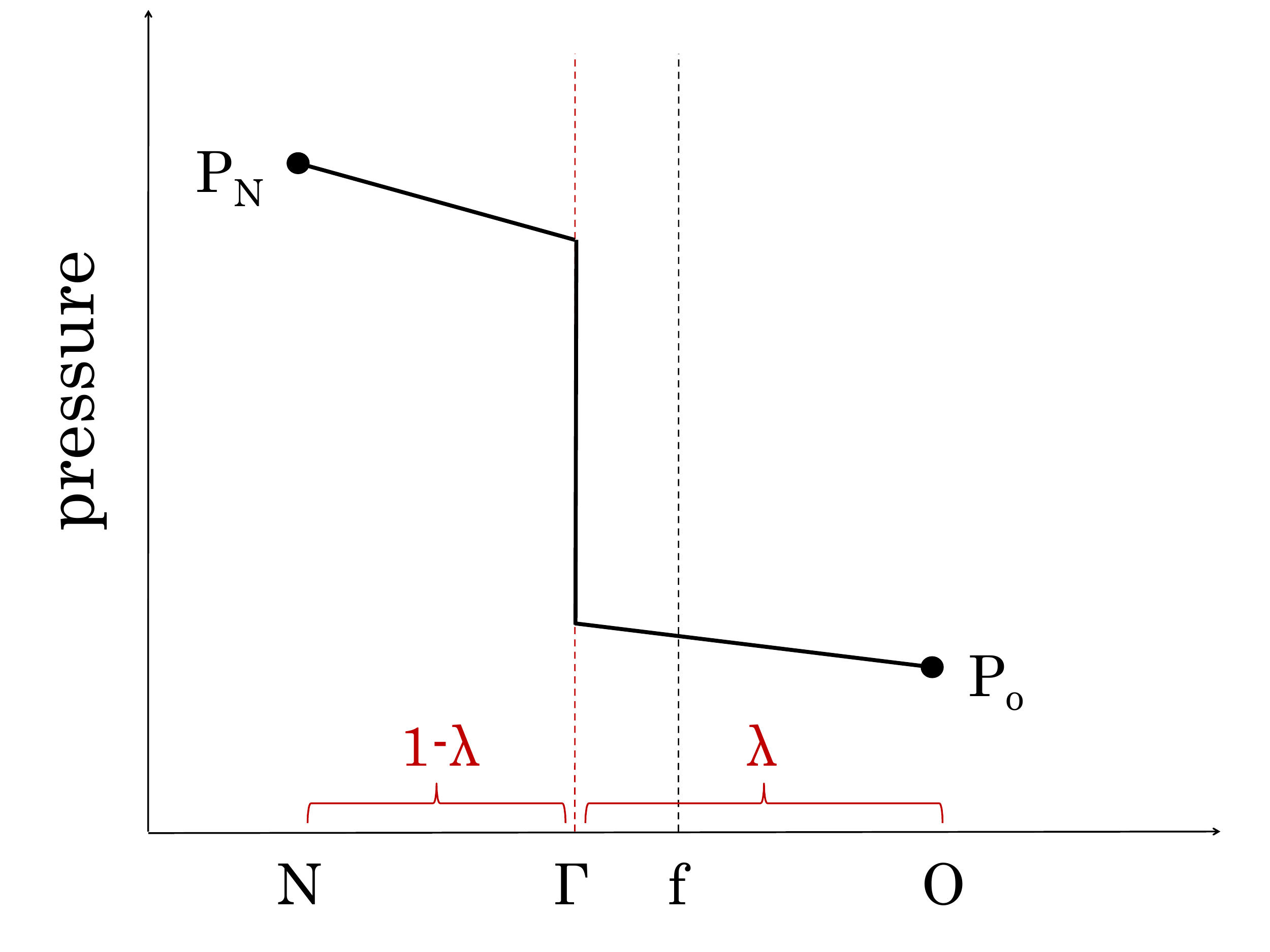}}
	\caption{Pressure jump at the interface $\Gamma$, defined on the cell face $f$. $\lambda$ is the relative interface position.}
	\label{lambda}
\end{figure}

 Figure \ref{lambda} reports a qualitative plot of the pressure jump $[p]$ at the free surface $\Gamma$.  In $\texttt{OpenFOAM}^{\textregistered}$ every face $f$ is shared between two adjacent cells called owner (O) and a neighbour (N). Following Vukcevic et al. \cite{vukvcevic2017implementation}, an interfacial face $f$ is defined when two adjacent cells $N$ (neighbour) and $O$ (owner)  are respectively "wet" ($\alpha> 0.5$) and "dry" ($\alpha< 0.5$), satisfying the following equation:

\begin{equation}
	\left(\alpha_N-0.5\right)	\left(\alpha_O-0.5\right)< 0
	\label{criterionInterfacialFace}
\end{equation}

The interface $\Gamma$ will be between the cells $O$ and $N$, but we do not know where.  The dimensionless relative position of the interface (with respect to cell  $O$) is $\lambda$ and it is defined as:

\begin{equation}
	\lambda=\frac{\alpha_O-0.5}{\alpha_O-\alpha_N}
	\label{lambdaEquation}
\end{equation}

In Equation \ref{finalpressureequation} the term $\beta_f\left(\nabla p_d\right)_f$ is the only one which requires special attention. An ordinary discretization of the pressure gradient at the cell face $\left(\nabla p_d\right)_f$ would be (for orthogonal meshes):

\begin{equation}
	\left(\nabla p_d\right)_f = \frac{p_O-p_N}{|\textbf{d}|}
\end{equation}

\begin{figure}
	\centering
	\subfloat[]
	{\includegraphics[width=.4\textwidth,height=0.22\textheight]{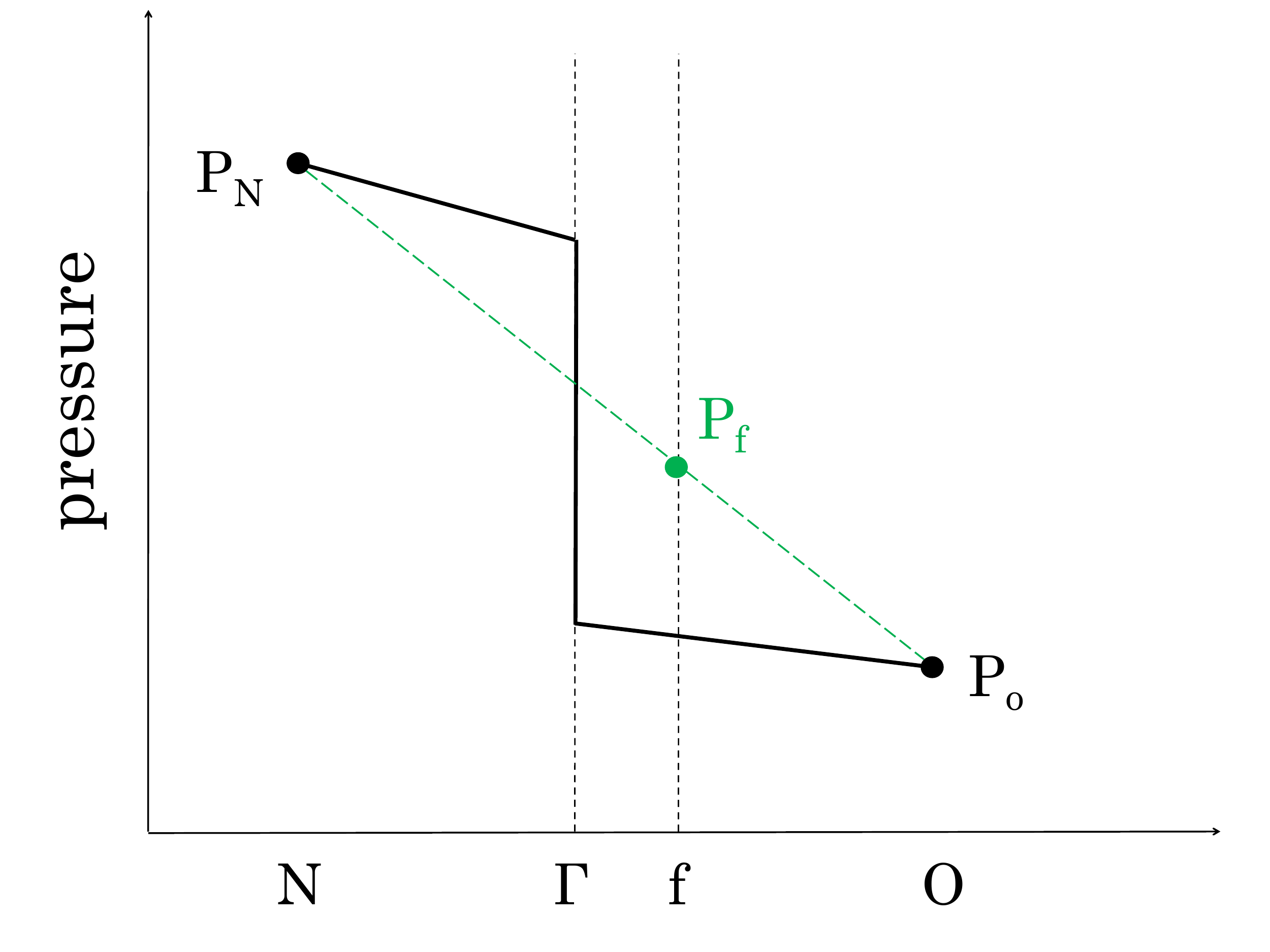}}~
	\subfloat[]
	{\includegraphics[width=.4\textwidth,height=0.22\textheight]{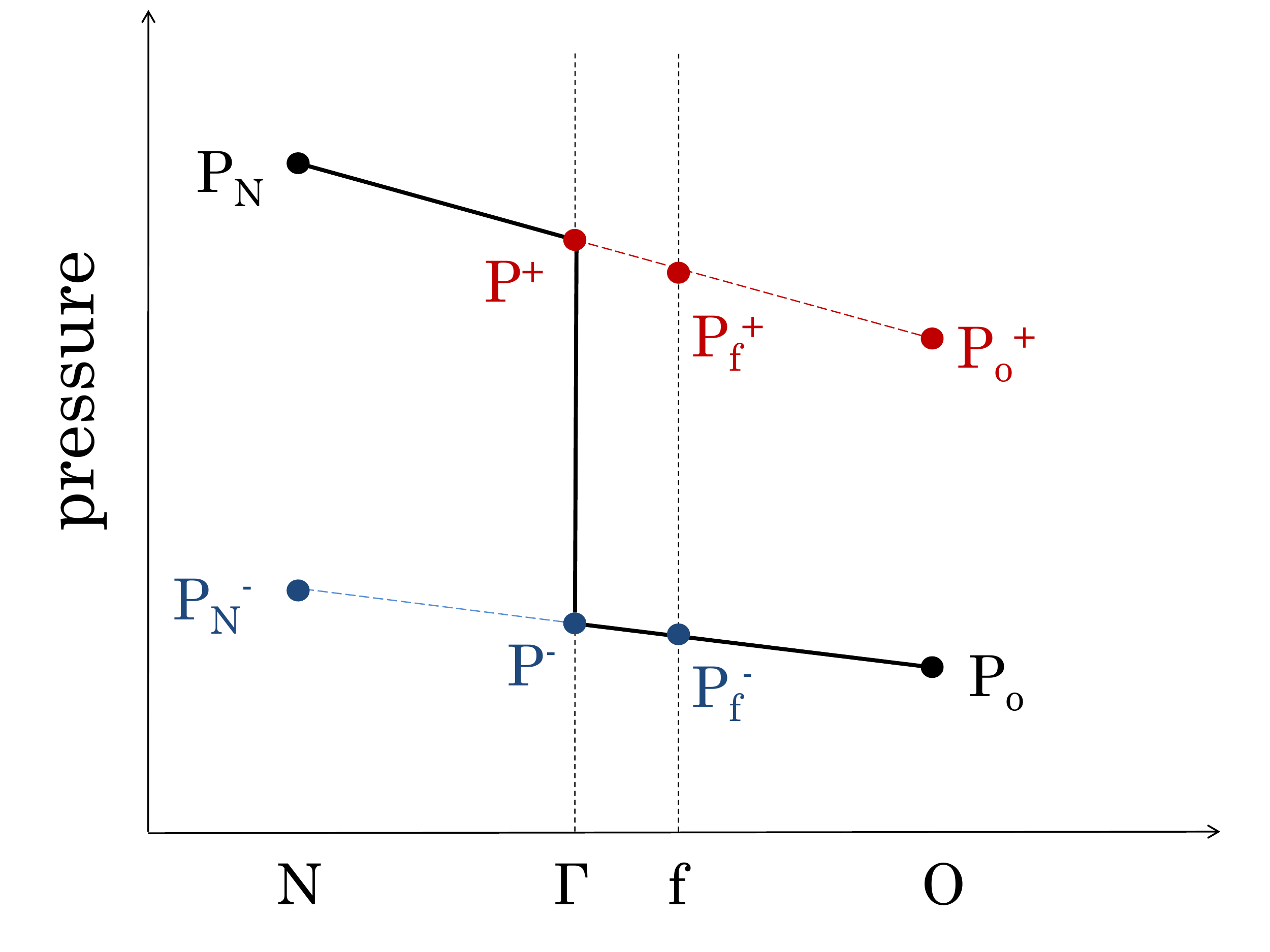}}
	\caption{Ordinary interpolation in presence of a jump at the free surface, leading to incorrect gradients (Figure a). Correct gradient evaluation through one-sided extrapolation of interface values (Figure b).}
	\label{explanationGFM}
\end{figure}

where $\textbf{d}$ is the distance vector (from $O$ to $N$). As can be seen in Figure \ref{explanationGFM} (a), this implementation leads to incorrect values of the face gradient. On the other hand, the GFM exploits the knowledge of the pressure jump at the interface $\Gamma$ (Equation \ref{pressurejump}), allowing to derive one-sided extrapolated values of the pressure field which can be used to discretize the pressure gradient. Let us refer to Figure \ref{explanationGFM} (b), considering cell $N$ a liquid cell. Knowing the values of $p_N$  and $p^+$  it is easy to derive the value $p_o^+$ at cell $O$ by linear extrapolation. This latter is called "ghost value", since it is defined on the other side of the interface $\Gamma$ (on the gas phase in this case). The same is valid for the gas side (cell $O$) and its liquid side extrapolation $p_N^-$ at cell $N$. The pressure gradient has now two different formulations, depending on the position with respect to the interface $\Gamma$:

\begin{equation}
\left. \left(\nabla p_d\right)_f \right|_N = \frac{p_O^+-p_N}{|\textbf{d}|}
\end{equation}

\begin{equation}
\left. \left(\nabla p_d\right)_f \right|_O = \frac{p_N^- - p_O}{|\textbf{d}|}
\end{equation}

where the extrapolated values $p_O^+$ and $p_N^-$ are linked by  interfacial pressure jump $H_f$ (Equation \ref{pressurejump}). The jump $H_f$ requires the  position vector $\textbf{x}_f$ of the interface $\Gamma$ (easily defined based on $\lambda$):

\begin{equation}
    \textbf{x}_f = \textbf{x}_O+\lambda|\textbf{d}|
    \label{interfacePosition}
\end{equation}

and the face interpolated values of the curvature $\kappa_f$ (accessible knowing the cell-centered curvature values in cells $N$ and $O$). The interfacial jump $H_f$ is therefore directly included in the discretization of the pressure equation. 
The detailed derivation of the extrapolated values  $p_O^+$ and $p_N^-$ can be found in the reference work of Vukcevic et al. \cite{vukvcevic2017implementation}. The final GFM discretization of the pressure Laplacian (for orthogonal meshes) in Equation \ref{finalpressureequationWithDivergence} is the following:

\begin{itemize}
	\item If the owner cell $O$ is "wet" $\left(\alpha>0.5\right)$:
	
	\begin{equation}
	\left. \nabla\cdot\left(\frac{1}{a_{P}}\beta\nabla p_d\right)\right|_O = \sum_{f}^{}\left(\frac{1}{a_{P}}\right)_f\frac{|\textbf{S}_f|}{|\textbf{d}|}\frac{\beta^L\beta^G}{\beta_w}\left(p_N-p_O-H_f\right)
	\label{pressureGFM_wet_own}
	\end{equation}
	
	\begin{equation}
	\left. \nabla\cdot\left(\frac{1}{a_{P}}\beta\nabla p_d\right)\right|_N = \sum_{f}^{}\left(\frac{1}{a_{P}}\right)_f\frac{|\textbf{S}_f|}{|\textbf{d}|}\frac{\beta^L\beta^G}{\beta_w}\left(p_O-p_N+H_f\right)
	\label{pressureGFM_wet_neigh}
	\end{equation}
	
	where $\beta_w=\lambda\beta^G+\left(1-\lambda\right)\beta^L$.
	\item  If  the owner cell $O$ is "dry" $\left(\alpha<0.5\right)$:
	
	\begin{equation}
	\left. \nabla\cdot\left(\frac{1}{a_{P}}\beta\nabla p_d\right)\right|_O = \sum_{f}^{}\left(\frac{1}{a_{P}}\right)_f\frac{|\textbf{S}_f|}{|\textbf{d}|}\frac{\beta^L\beta^G}{\beta_d}\left(p_N-p_O-H_f\right)
	\label{pressureGFM_dry_own}
	\end{equation}
	
	\begin{equation}
	\left. \nabla\cdot\left(\frac{1}{a_{P}}\beta\nabla p_d\right)\right|_N = \sum_{f}^{}\left(\frac{1}{a_{P}}\right)_f\frac{|\textbf{S}_f|}{|\textbf{d}|}\frac{\beta^L\beta^G}{\beta_d}\left(p_O-p_N+H_f\right)
	\label{pressureGFM_dry_neigh}
	\end{equation}
	
	where $\beta_d=\lambda\beta^L+\left(1-\lambda\right)\beta^G$.
\end{itemize}

 The face flux $\phi_{f}$ is reconstructed based on the new pressure gradient (discretized explicitly with GFM) according to Equation \ref{faceFluxes}:

\begin{itemize}
	\item If the owner $O$ is "wet":
	
	\begin{equation}
	\phi_{f}=\frac{\textbf{H}\left(\textbf{v}_N\right)_f}{a_{P,f}}\cdot\textbf{S}_f-\left(\frac{1}{a_{P}}\right)_f\frac{|\textbf{S}_f|}{|\textbf{d}|}\frac{\beta^L\beta^G}{\beta_w}\left(p_N-p_O-H_f\right)
	\label{reconstructedVelocityGFM_wet}
	\end{equation}
	
	\item If the owner $O$ is "dry":
	
	\begin{equation}
	\phi_{f}=\frac{\textbf{H}\left(\textbf{v}_N\right)_f}{a_{P,f}}\cdot\textbf{S}_f-\left(\frac{1}{a_{P}}\right)_f\frac{|\textbf{S}_f|}{|\textbf{d}|}\frac{\beta^L\beta^G}{\beta_d}\left(p_N-p_O-H_f\right)
	\label{reconstructedVelocityGFM_dry}
	\end{equation}	
	
\end{itemize}

Finally, the cell-centered velocity $\textbf{v}$ can be reconstructed from the face fluxes $\phi_{f}$. Equations \ref{pressureGFM_wet_own} to \ref{reconstructedVelocityGFM_dry} have to be used for interfacial faces only (defined by Equation \ref{criterionInterfacialFace}). For the other faces, ordinary discretization applies.

\subsection{Details on the $\texttt{OpenFOAM}^{\textregistered}$ implementation}

The main algorithm for the implementation of the Ghost Fluid Method in $\texttt{OpenFOAM}^{\textregistered}$ is here reported, including the main code lines:

\begin{enumerate}
    \item Localize the interfacial faces: create a surface scalar field \texttt{interfacialFace}, which assumes value 1 if Equation \ref{criterionInterfacialFace} is satisfied and 0 otherwise;
    \item Construct the surface scalar field $\lambda$, the relative position of the interface (Equation \ref{lambdaEquation});
    \item Construct the surface vector field \texttt{interfacePosition}, the interface position  $\textbf{x}_f$  (Equation \ref{interfacePosition});
    \item Compute the interpolated value of the curvature $\kappa_f=\kappa_O\lambda+\kappa_N\left(1-\lambda\right)$. The cell values of $\kappa$ are computed with the Height Functions method, as explained in the dedicated section. Construct the surface scalar field \texttt{surfaceTensionJump} as $\sigma \kappa_f$;
    \item Construct the surface scalar field $\texttt{Jump}$, the jump $H_f$ across the interfacial face (Equation \ref{pressurejump}). Consider that the jump on the face has opposite sign depending if it is "seen" from the owner side or from the neighbour side. In $\texttt{OpenFOAM}^{\textregistered}$ we have:
    
    \begin{verbatim}
      if (alpha1.ref()[owner[facei]] > 0.5)
      Jump.ref()[facei] = 
      (rhoL.ref()[owner[facei]]-rhoG.ref()[neighbour[facei]])
      *(g.value() & interfacePosition.ref()[facei]) 
      + surfaceTensionJump.ref()[facei];
      
      else
      Jump.ref()[facei] =
      -(rhoL.ref()[neighbour[facei]]-rhoG.ref()[owner[facei]])
      *(g.value() & interfacePosition.ref()[facei])
      - surfaceTensionJump.ref()[facei];
    \end{verbatim}   
    
    \item Compute the values  $\beta_w=\lambda\beta^G+\left(1-\lambda\right)\beta^L$ and $\beta_d=\lambda\beta^L+\left(1-\lambda\right)\beta^G$;
    \item Construct the Laplacian finite volume matrix  \texttt{fvmLaplacian} (Equations \ref{pressureGFM_wet_own} to  \ref{pressureGFM_dry_neigh}):
    
\begin{enumerate}
	\item   The resulting discretization matrix is symmetric, it is sufficient to specify the upper matrix coefficient $a_{upp}$ of the Laplacian (describing the effect of the neighbour cell on the owner). If not defined, the lower coefficient will be automatically set the same. For every face:
	\begin{itemize}	
		
		\item 	If the owner is "wet":
		
		\begin{equation}
		a_{upp}=\left(\frac{1}{a_{P}}\right)_f\frac{|\textbf{S}_f|}{|\textbf{d}|}\frac{\beta^L\beta^G}{\beta_w}
		\end{equation}
		
		in $\texttt{OpenFOAM}^{\textregistered}$ the term  $\left(\frac{1}{a_{P}}\right)_f|\textbf{S}_f|$ is called \texttt{raUfMagSf}. So:
		
		\begin{verbatim}
		   fvmLaplacian.upper()[facei]
		   = deltaCoeff.ref()[facei]*raUfMagSf.ref()[facei]*
		   betaL.ref()[facei]*betaG.ref()[facei]/betaW.ref()[facei];
		\end{verbatim}
		
		\item 	If the owner is "dry":
		\begin{equation}
		a_{upp}=\left(\frac{1}{a_{P}}\right)_f\frac{|\textbf{S}_f|}{|\textbf{d}|}\frac{\beta^L\beta^G}{\beta_d}
		\end{equation}	

		in $\texttt{OpenFOAM}^{\textregistered}$:
		\begin{verbatim}
		   fvmLaplacian.upper()[facei]
	       = deltaCoeff.ref()[facei]*raUfMagSf.ref()[facei]*
	   	   betaL.ref()[facei]*betaG.ref()[facei]/betaD.ref()[facei];
		\end{verbatim}
		
		\item If the face is not interfacial, standard  interpolation applies to $\beta$:
		
		\begin{equation}
		a_{upp}=\left(\frac{1}{a_{P}}\right)_f\frac{|\textbf{S}_f|}{|\textbf{d}|}{\beta}_f
		\end{equation}	
		
		in $\texttt{OpenFOAM}^{\textregistered}$:
		\begin{verbatim}
		fvmLaplacian.upper()[facei]
		= deltaCoeff.ref()[facei]*raUfMagSf.ref()[facei]*	beta.ref()[facei];
		\end{verbatim}					
				
	\end{itemize}
      
	\item The diagonal term $d$ is:
	
	\begin{equation}
	d=-\sum_{f}^{}a_{upp}
	\end{equation}
	
	It is very easy in $\texttt{OpenFOAM}^{\textregistered}$ to do this operation:
		\begin{verbatim}
         fvmLaplacian.negSumDiag();
		\end{verbatim}
		
	\item 
	The source terms are antisymmetric. This means that for each interfacial face the additional flux (due to the jump $H_f$) from the owner $O$ and the neighbour $N$ cancel out, providing a perfectly balanced scheme. For every interfacial face:
	\begin{itemize}
	\item If the owner cell $O$ is wet (Equations \ref{pressureGFM_wet_own} and \ref{pressureGFM_wet_neigh}):
	\begin{itemize}
		\item Owner contribution:
		\begin{equation}
			\left(\frac{1}{a_{P}}\right)_f\frac{|\textbf{S}_f|}{|\textbf{d}|}\frac{\beta^L\beta^G}{\beta_w}H_f
		\end{equation}
		\item Neighbour contribution: 
			\begin{equation}
			-\left(\frac{1}{a_{P}}\right)_f\frac{|\textbf{S}_f|}{|\textbf{d}|}\frac{\beta^L\beta^G}{\beta_w}H_f
			\end{equation}
	\end{itemize}
	\item If the owner cell $O$ is dry (Equations \ref{pressureGFM_dry_own} and \ref{pressureGFM_dry_neigh}):
	
	\begin{itemize}
		\item Owner contribution:
		\begin{equation}
		\left(\frac{1}{a_{P}}\right)_f\frac{|\textbf{S}_f|}{|\textbf{d}|}\frac{\beta^L\beta^G}{\beta_d}H_f
		\end{equation}
		\item Neighbour contribution: 
		\begin{equation}
		-\left(\frac{1}{a_{P}}\right)_f\frac{|\textbf{S}_f|}{|\textbf{d}|}\frac{\beta^L\beta^G}{\beta_d}H_f
		\end{equation}
	\end{itemize}
			
	\end{itemize}

The face contributions must be summed to provide the cell centered source term. In 	$\texttt{OpenFOAM}^{\textregistered}$ we created a volume scalar field \texttt{sourceLaplacian} in which:

	\begin{verbatim}
    forAll(owner, facei) 
    {
      sourceLaplacian.ref()[owner[facei]] 
      +=  fvmLaplacian.upper()[facei]*Jump.ref()[facei];
      sourceLaplacian.ref()[neighbour[facei]] 
      -=  fvmLaplacian.upper()[facei]*Jump.ref()[facei];
    }
	\end{verbatim}
		
There is no need to distinguish between wet and dry cells, it has been already done constructing $\texttt{fvmLaplacian}$. Finally:

\begin{verbatim}
fvmLaplacian.source() = sourceLaplacian;
\end{verbatim}
	
\end{enumerate}

 \item Solve the pressure equation (Equation \ref{finalpressureequation}) using $\texttt{fvmLaplacian}$;   
 
 \item Reconstruct the face fluxes $\texttt{phi}$ on the faces (Equations \ref{reconstructedVelocityGFM_wet} and \ref{reconstructedVelocityGFM_dry}). In 	$\texttt{OpenFOAM}^{\textregistered}$:

\begin{verbatim}
phi.ref()[facei] 
= phiHbyA.ref()[facei] - fvmLaplacian.upper()[facei]*
 (
   p.ref()[neighbour[facei]]-p.ref()[owner[facei]]-Jump.ref()[facei]
 );
\end{verbatim} 

again, no need to distinguish between wet and dry cells, the difference is included in $\texttt{fvmLaplacian.upper()}$;
  
\item Reconstruct the cell-centered velocity values $\textbf{v}$. In $\texttt{OpenFOAM}^{\textregistered}$:

\begin{verbatim}
 U = fvc::reconstruct(phi);
\end{verbatim} 
    
\end{enumerate}

It is important to follow these steps not only for the internal faces, but also for the boundary ones, in order to be able to perform parallel computations.

\begin{figure}
	\centering
	{\includegraphics[width=.85\textwidth,height=0.25\textheight]{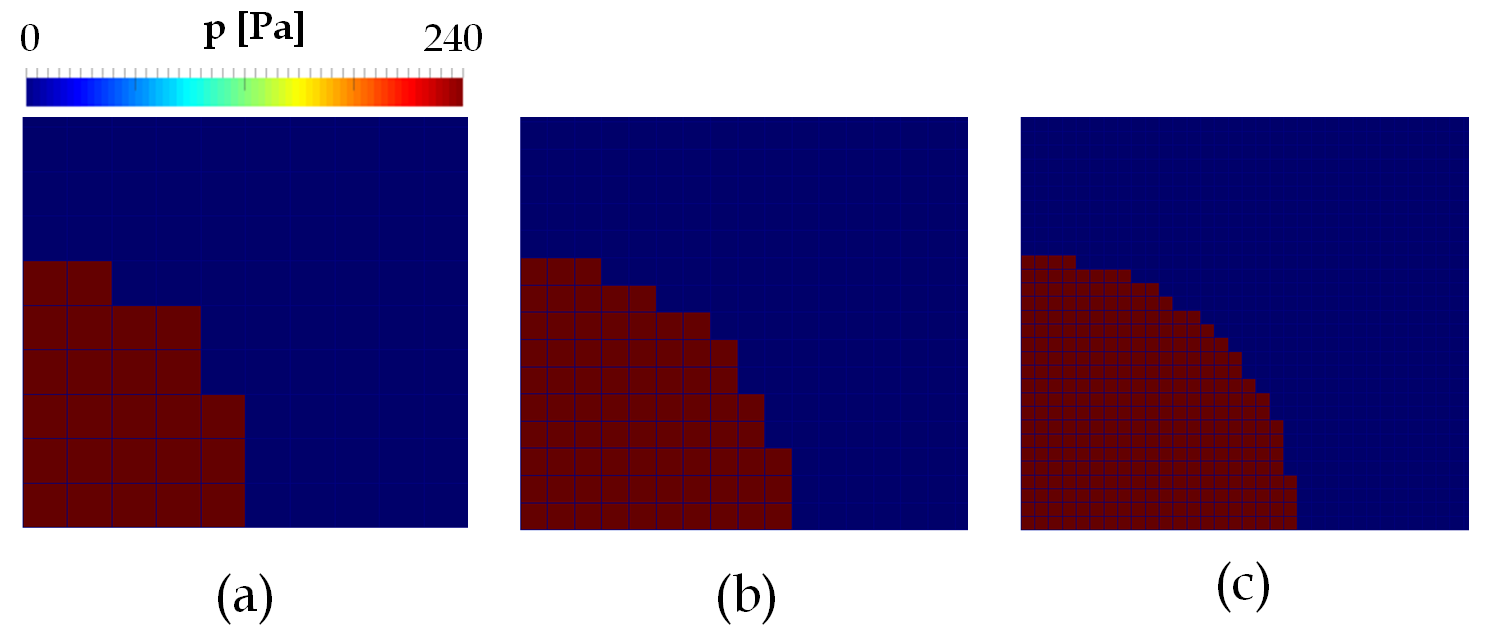}}
	\caption{Numerical equilibrium solution (pressure $p$) for a 2D droplet at zero gravity using the Ghost Fluid Method (GFM) for pressure discretization. The curvature is prescribed $\kappa_f=2000$ $m^{-1}$, providing a pressure jump equal to  $\left[p\right]=240$ Pa. Three droplet resolutions $\frac{D}{\Delta x}=10$ (a), 20 (b), 40 (c) at time $t=0.01$ s.}
	\label{GFMStaticDroplet}
\end{figure}

\subsection{Static test case}

As reported by Francois et al. \cite{francois2006balanced}, a consistent discretization method should be able to recover an exact balanced solution for a static droplet (Equation \ref{equationStatic}) when the interface curvature $\kappa_f$ is prescribed exactly. 
 A 2D droplet ($D=1$ mm) is considered, with equal density and viscosity for both gas and liquid $\rho=1$ $kg/m^3$, $\mu=10^{-4}$ $Pa\cdot s$. The curvature $\kappa_f$ is imposed equal to the analytical value $\frac{1}{R}=2000$ $m^{-1}$. The surface tension is $\sigma=0.012$ N/m. The relative Laplace number $La$:

\begin{equation}
La=\frac{\rho D \sigma}{\mu^2}
\end{equation}

is equal to 1200. The pressure jump due to surface tension equals the total pressure jump (since gravity is zero):

\begin{equation}
	H_f=\sigma\kappa_f = 240~Pa
\end{equation}

Four droplet resolutions have been adopted $\frac{D}{\Delta x}=10, 20, 40, 100$. The results are presented in Figure \ref{GFMStaticDroplet} and in Table \ref{spuriousCurrentsGFM} in terms of maximum velocity $|\textbf{v}|_{max}$ and capillary number $Ca$:

\begin{equation}
Ca=\frac{\mu|\textbf{v}|_{max}}{\sigma}
\end{equation}

  The equilibrium solution is numerically well recovered, providing a perfectly sharp pressure field. Spurious currents are reduced to machine precision (round off), because of the well-balanced discretization adopted. This holds for all the grid resolutions.

\begin{table}
	\centering
	
	\begin{tabular}{llll}
		\toprule
		
		La \quad\quad \quad\quad& $D/\Delta x$ \quad\quad\quad\quad& $|\textbf{v}|_{max}$ [m/s] \quad\quad\quad\quad & Ca \\
		\midrule
		\multirow{3}{*}{1200}& 10 & $2.60\cdot10^{-13}$ & $2.16\cdot10^{-15}$  \\
		& 20 & $1.60\cdot10^{-12}$ & $1.33\cdot10^{-14}$   \\
		& 40 & $2.72\cdot10^{-12}$ & $2.27\cdot10^{-14}$ \\
		& 100 & $1.44\cdot10^{-12}$ & $1.20\cdot10^{-14}$ \\ 	                    		                    	
		\bottomrule
		
	\end{tabular}	
	
	\caption{Spurious currents analysis for a Laplace number $La=1200$  at four different droplet resolutions $D/\Delta x=10, 20, 40, 100$, by means of $|\textbf{v}|_{max}$ and capillary number $Ca$. Time $t=0.01$ s.}
	\label{spuriousCurrentsGFM}
\end{table}

\subsection{Dynamic test case}
\begin{figure}
	\centering
	\subfloat[]
	{\includegraphics[width=.24\textwidth,height=0.22\textheight]{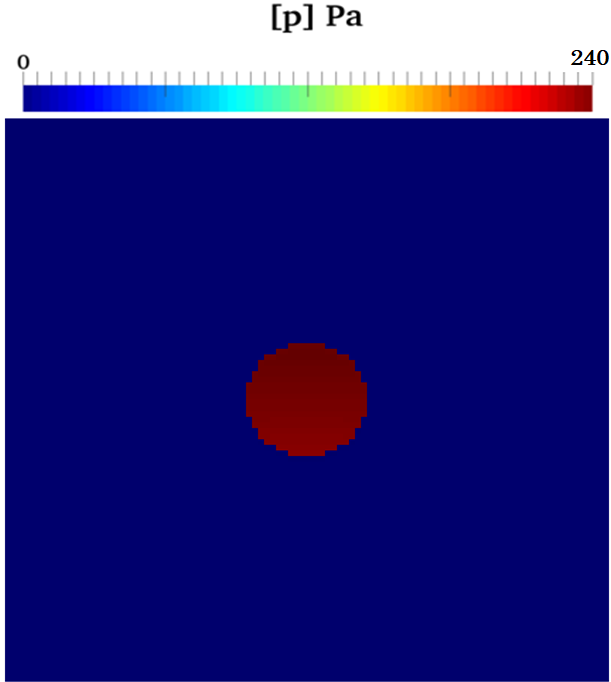}}~
	\subfloat[]
	{\includegraphics[width=.24\textwidth,height=0.18\textheight]{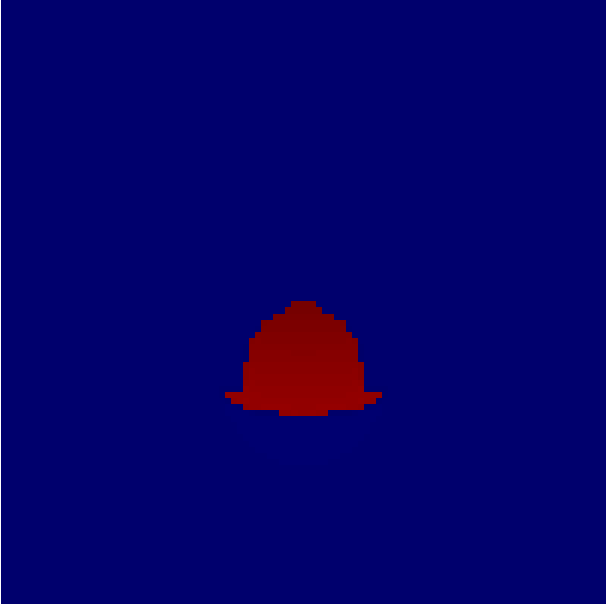}}~
	\subfloat[]
	{\includegraphics[width=.24\textwidth,height=0.18\textheight]{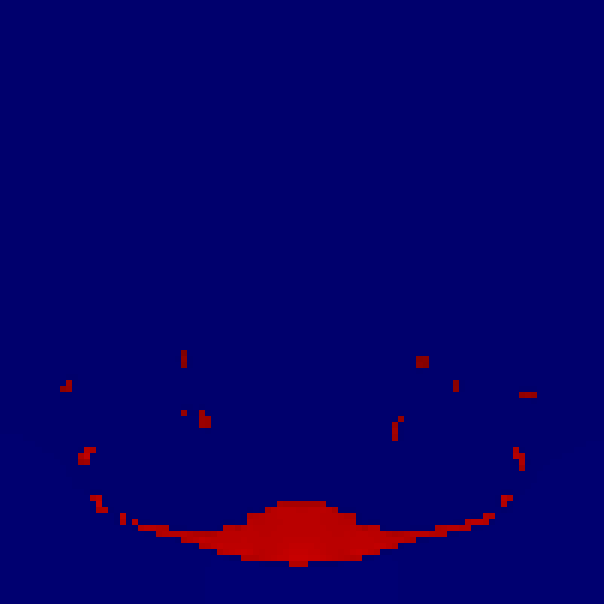}}~
	\subfloat[]
	{\includegraphics[width=.24\textwidth,height=0.18\textheight]{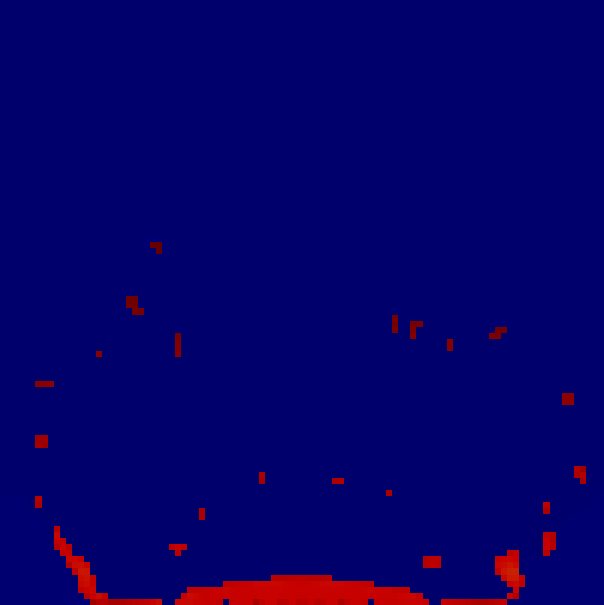}}\\
	\subfloat[]
	{\includegraphics[width=.24\textwidth,height=0.18\textheight]{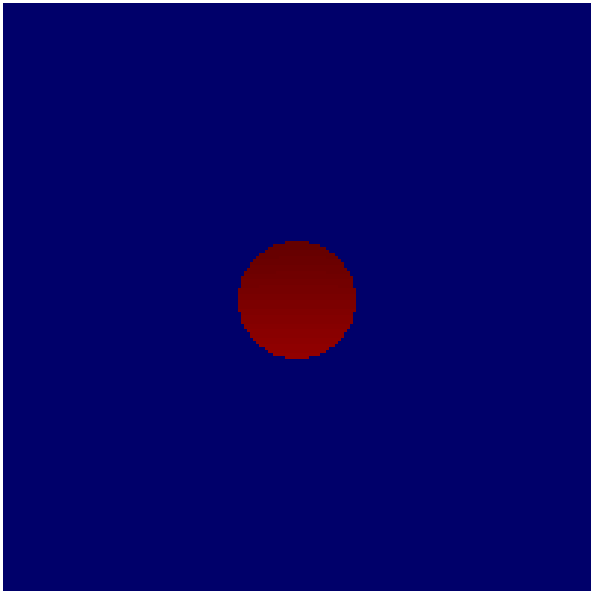}}~
	\subfloat[]
	{\includegraphics[width=.24\textwidth,height=0.18\textheight]{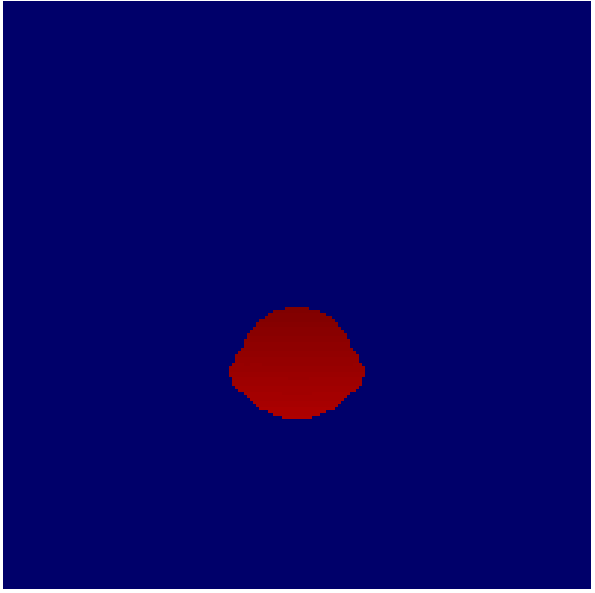}}~
	\subfloat[]
	{\includegraphics[width=.24\textwidth,height=0.18\textheight]{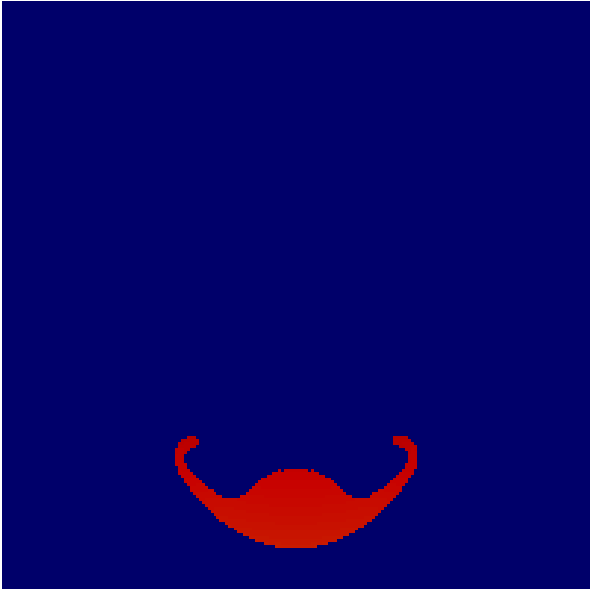}}~
	\subfloat[]
	{\includegraphics[width=.24\textwidth,height=0.18\textheight]{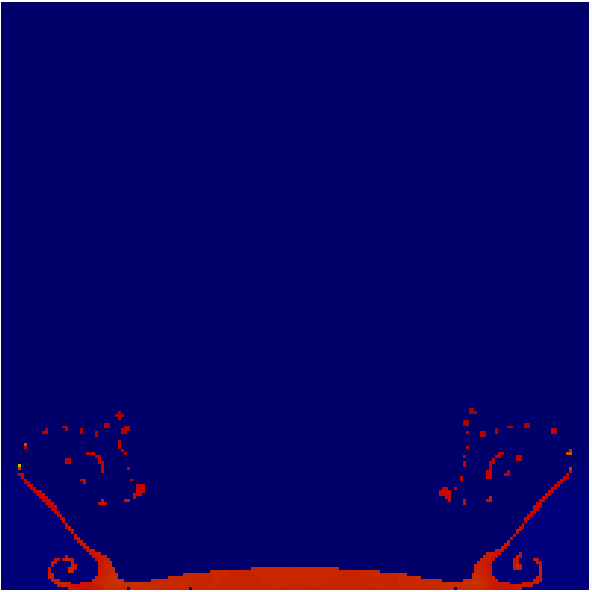}}\\
	\subfloat[]
	{\includegraphics[width=.24\textwidth,height=0.18\textheight]{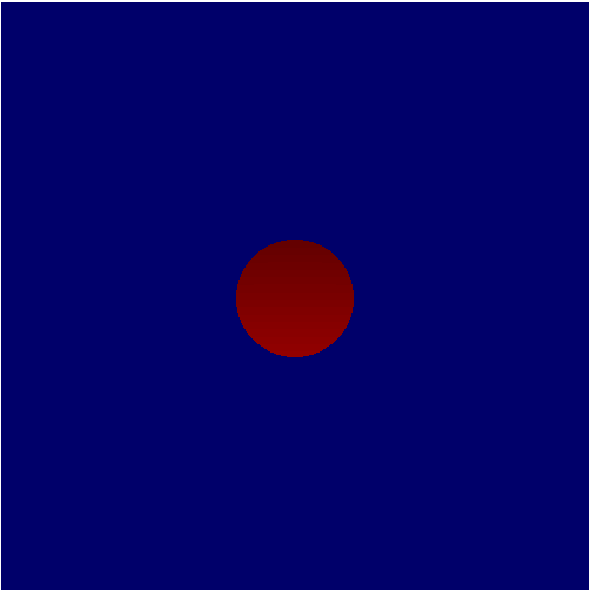}}~
	\subfloat[]
	{\includegraphics[width=.24\textwidth,height=0.18\textheight]{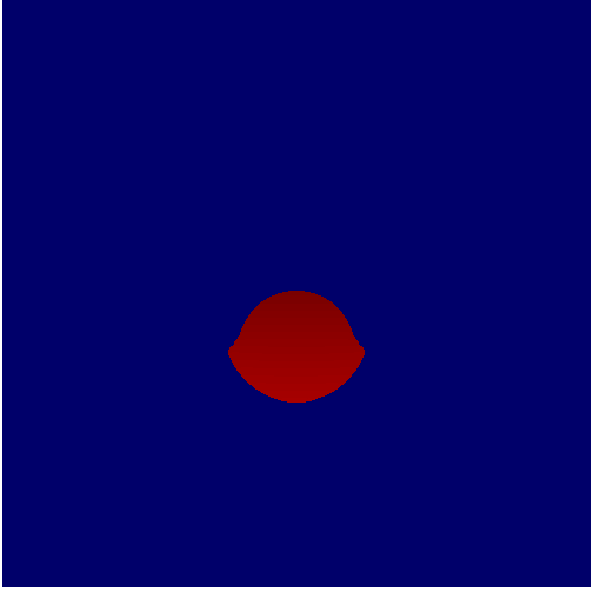}}~
	\subfloat[]
	{\includegraphics[width=.24\textwidth,height=0.18\textheight]{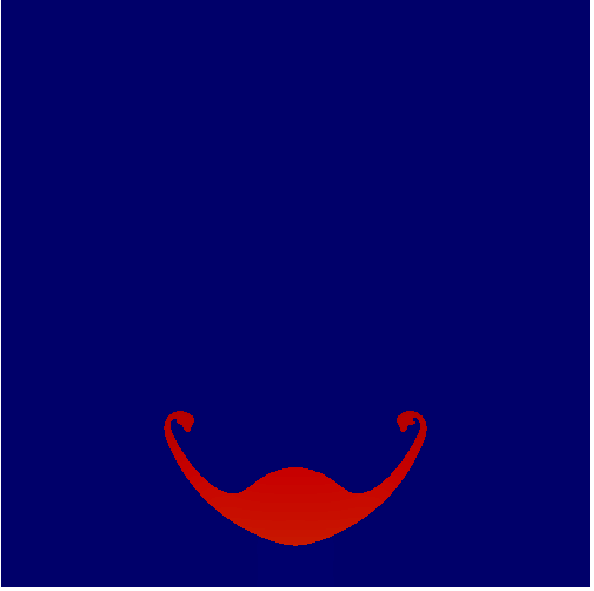}}~
	\subfloat[]
	{\includegraphics[width=.24\textwidth,height=0.18\textheight]{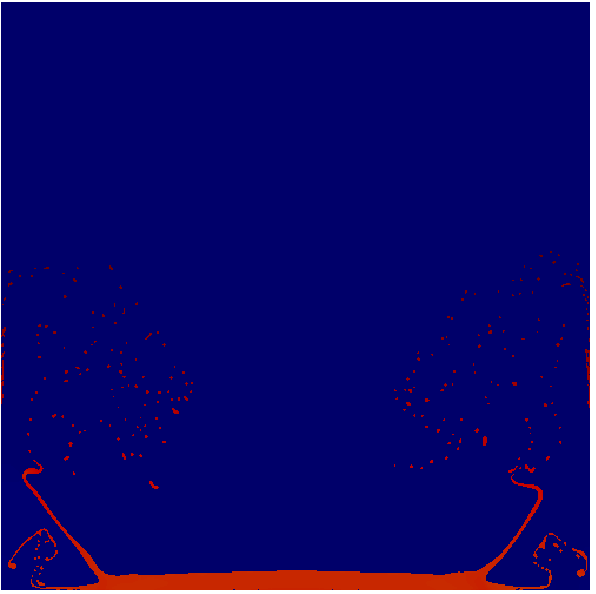}}
	\caption{Numerical simulation of a falling droplet at three different droplet resolutions $\frac{D}{\Delta x}=20$ (a, b, c, d), $\frac{D}{\Delta x}=40$ (e, f, g, h)  and $\frac{D}{\Delta x}=100$ (i, j, k, l). The surface tension pressure jump is imposed equal $\left[p\right]=240$ Pa. Times $t= 0.002$ s (a, e, i), $t= 0.012$ s (b, f, j), $t= 0.024$ s (c, g, k), $t= 0.036$ s (d, h, l).}
	\label{GFMDynamic}
\end{figure}

In order to test (only qualitatively) a dynamic case, gravity has been activated maintaining the same initial conditions (Figure \ref{GFMStaticDroplet}). The surface tension pressure jump across the interface is still $\sigma\kappa_f=240$ $Pa$.  Figure \ref{GFMDynamic} reports the pressure field of the falling droplet at four different times, for three droplet resolutions $\frac{D}{\Delta x}=20, 40, 100$. Differently from the static case, the total pressure jump $H_f$ is not constant because it includes the gravity contribution (Equation \ref{pressurejump}). The GFM  maintains a perfectly sharp representation of the pressure field. The droplet falls as if there was no surface tension acting on the interface, because we are disregarding the variation of curvature along the interface. This will be the focus of the next section.

\section{Curvature evaluation: Height Functions}

\begin{figure}
	\centering
	{\includegraphics[width=.6\textwidth,height=0.3\textheight]{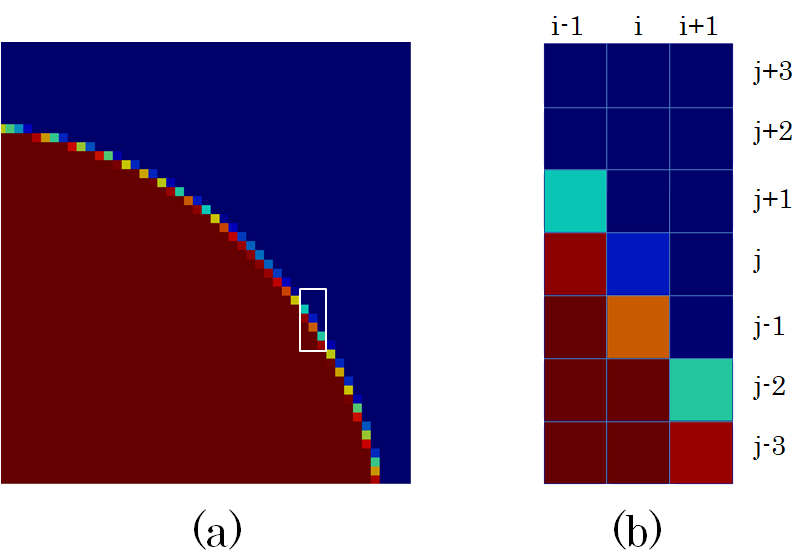}}~
	\caption{A 3x7 vertical stencil constructed around the interfacial cell (i,j). }
	\label{heightFunction}
\end{figure}

Every interfacial face $f$, defined in Equation \ref{criterionInterfacialFace}, is shared between two adjacent cells: owner $O$ and neighbour $N$. As stated in Equation \ref{pressurejump},  the jump at the interface $H_f$ requires the value of the face interpolated curvature $\kappa_f$, therefore the cell-centered values of the curvature $\kappa$ are necessary in cells $O$ and $N$ for every interfacial face $f$. We will call these cells "interfacial cells". In this work, we implemented the Height Functions method for the cell-centered curvature calculation.\\
In the Height Functions method the curvature is obtained by a direct differentiation of local heights, defined for every interfacial cell. Referring to Figure \ref{heightFunction}, a 3x7 vertical stencil is constructed for every interfacial cell $(i,j)$. The liquid volume fractions (VOF function) are summed up for every cell (multiplied by the vertical cell size $\Delta y$) for each of the $i$ columns, obtaining three local heights $h_i$:

\begin{equation}
	h_i=\sum_{j-3}^{j+3}\alpha_{i,j}\Delta y
	\label{heightscalculation}
\end{equation}

In order to avoid differentiating multi-valued functions, we can also choose an horizontal stencil: the choice is based on the largest component of the interface normal \textbf{n} in the $(i,j)$ cell:

\begin{itemize}
	\item  if $|n_y|>|n_x|$ we choose a 3x7 vertical stencil;
	\item  if $|n_x|>|n_y|$ we choose a 3x7 horizontal stencil.
\end{itemize}

The heights represent the position of the interface with respect to the stencil base. Adopting a second order finite difference scheme, we can calculate the first derivative $h'_{i,j}$:

\begin{equation}
	h'_{i,j}=\frac{h_{i+1}-h_{i-1}}{2\Delta x}
\end{equation}

the second derivative $h''_{i,j}$:

\begin{equation}
h''_{i,j}=\frac{h_{i+1}-2h_i+h_{i-1}}{\Delta x^2}
\end{equation}

and finally the cell-centered curvature $\kappa_{i,j}$ (in 2D):

\begin{equation}
	\kappa_{i,j}=\frac{h''_{i,j}}{\left(1+h'^2_{i,j}\right)^{3/2}}
	\label{curvaturecalculation}
\end{equation}

The curvature is then interpolated at the interface $\Gamma$, based on the relative interface position $\lambda$ (Equation \ref{lambdaEquation}):

\begin{equation}
	\kappa_f=\lambda\kappa_N + \left(1-\lambda\right)\kappa_O
	\label{curvatureInterpolation}
\end{equation}

Finally, this face-centered curvature value can be used to construct the pressure jump in Equation \ref{pressurejump} for the GFM pressure discretization.\\
This is a standard Height Functions method, because it adopts a fixed stencil for every interfacial cell $(i,j)$. Alternative methods include adaptive stencils \cite{popinet2009accurate, sussman2007improvements}, for which the size changes depending on the local interface topology (using from 3x3 up to 3x9 stencils) and rotated stencil which are aligned to the interface normal \cite{owkes2015mesh}.

\subsection{Details on the $\texttt{OpenFOAM}^{\textregistered}$ implementation}
The concept behind the use of Height Functions is simple and  relatively  easy to implement (if compared  to discrete surface fitting  or spline interpolation \cite{renardy2002prost, ginzburg2001two}). As described before, we use fixed 3x7 stencils (vertical or horizontal, depending on the interface normal $\textbf{n}$ direction) for all the interfacial cells. To our knowledge there was no attempt in literature to develop a Height Functions method for the $\texttt{OpenFOAM}^{\textregistered}$ framework. While in structured finite-difference CFD codes it is immediate to construct a stencil around a cell (using the  $\left(i,j\right)$ indices), the $\texttt{polyMesh}$ mesh description of  $\texttt{OpenFOAM}^{\textregistered}$ (based around faces) makes it more difficult. Hence, the need of a specific description.

\begin{figure}
	\centering
	{\includegraphics[width=.85\textwidth,height=0.32\textheight]{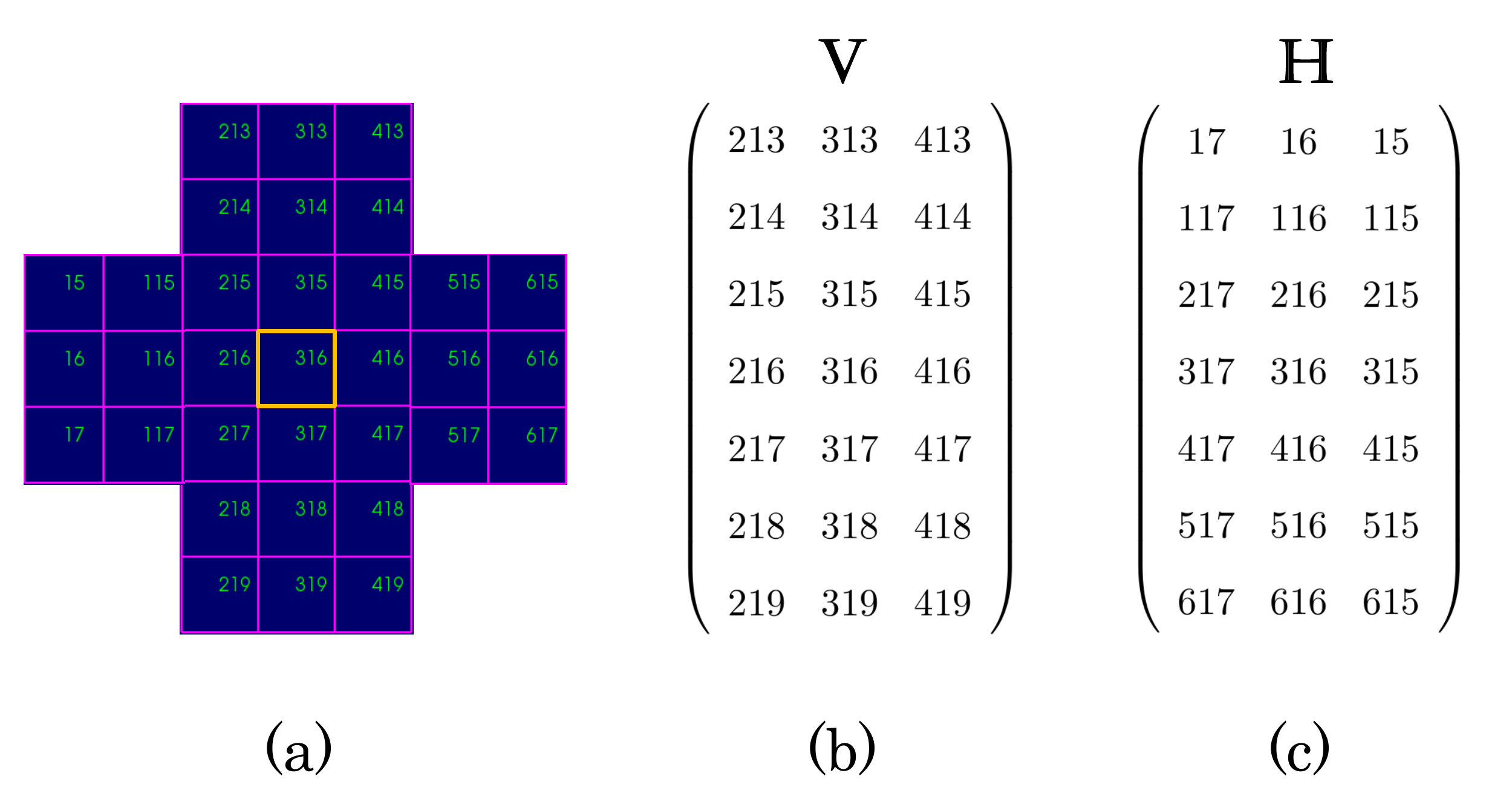}}
	\caption{Example of a vertical and horizontal stencils for the cell 316 (a). Corresponding $\textbf{V}$ (b) matrix, for the vertical stencil and $\textbf{H}$ (c) matrix for the horizontal stencil.}
	\label{Matrices}
\end{figure}

\subsubsection{Stencils construction}

 In this work we build all the necessary stencils \textit{in pre-processing}, to be more efficient during the simulation. The main algorithm for the stencils construction is coded in a pre-processing function, which only requires the mesh:

\begin{enumerate}
	\item For every cell in the entire domain construct a 3x3 square stencil around it. This is easily done creating an object of the class \texttt{CPCCellToCellStencil}, for which only the mesh information is needed:
	
	\begin{verbatim}
	 CPCCellToCellStencil initialStencil(mesh);
	 labelList squareStencil = initialStencil[celli];
	\end{verbatim}
	
	The $\texttt{squareStencil}$ list contains the \textit{global} indices of the initial 3x3 stencil in $\texttt{celli}$;
	
	\item Extend the stencil in the vertical direction, adding two rows at the top and at the bottom. For each cell  $\texttt{cellj}$ in the 3x3 stencil ($\texttt{squareStencil}$):
	\begin{enumerate}
		\item Find the 4 adjacent cells:
	\begin{verbatim}
	labelList adjacentCells = mesh.cellCells()[ squareStencil[cellj] ];
	\end{verbatim}
	
	\item Choose the cell  $\texttt{cellk}$ in the vertical direction, checking that the vector:
	\begin{verbatim}
	vector normal = mesh.C()[adjacentCells[cellk]] - mesh.C()[squareStencil[cellj]];
	\end{verbatim}	
	
    points upwards (or downwards) and that $\texttt{cellk}$ is not already included in the stencil. We have now a 3x5 stencil;
    
    \item Repeat the extension (a, b) using the 3x5 stencil, to obtain a 3x7 stencil;
	 
	\end{enumerate}
	
	\item Store the 21 cells indices of the vertical stencil in a 3x7 matrix $\textbf{V}$.  The matrix elements are sorted as presented in Figure \ref{Matrices} (b);
	\item Do the same operation on the $\texttt{squareStencil}$ (point 2) in the horizontal direction (adding two columns to the left and two columns to the right);
	\item Store the 21 cells indices of the horizontal stencil in a 3x7 matrix $\textbf{H}$. The matrix elements are sorted as presented in Figure \ref{Matrices} (c);
	\item If a cell in the domain does not allow to build the complete stencil (e.g. close to the boundary), ignore the cell. We will focus on these cells when dealing with contact angles.
\end{enumerate}

Adopting this pre-processing methodology, every cell in the domain is already linked to its two stencils ($\textbf{V}$ and $\textbf{H}$).  Therefore, we do not need to construct a stencil at every iteration during run-time, since everything is already tabulated.  The great advantage arises when parallel computations are handled. A single processor only has access to its internal cells and to the boundary cells of adjacent processors, making the stencil construction difficult and expensive. In our case this is not necessary, since the stencils are already pre-computed for every cell before the domain partitioning and remain available to the processor throughout the simulation.

\subsubsection{Heights calculation}
At this point we have two matrices for every cell $\texttt{celli}$ of the domain:
\begin{itemize}
	\item a  3x7 matrix $\textbf{V}$ containing the 21 global indices of the vertical stencil (Figure \ref{Matrices} b);
	\item a 3x7 matrix $\textbf{H}$ containing the 21 global indices of the horizontal stencil (Figure \ref{Matrices} c);
\end{itemize}

During the simulation, the stencil cells (and their $\alpha$ values) must be called. For each interfacial face (Equation \ref{criterionInterfacialFace}):

\begin{enumerate}
	\item Select the owner and neighbour cell of the face. For each one:
	\item Calculate the interface normal as $\textbf{n}=\frac{\nabla \alpha_s}{|\nabla \alpha_s|}$. $\alpha_s$ is a smoothed $\alpha$ field, which helps to calculate the normals with better accuracy. A recursive interpolation smoothing is used in this work:

	\begin{verbatim}
	alphaSmooth = fvc::average(fvc::interpolate(alpha));
	\end{verbatim}
	
	which is repeated a few times (3-4 times is enough);
	
	\item Choose the correct stencil:
	\begin{enumerate}
		\item If $n_y$ is the largest component of the interface normal $\textbf{n}$, choose the vertical stencil matrix $\textbf{V}$;
		\item If $n_x$ is the largest component of the interface normal $\textbf{n}$, choose the horizontal stencil matrix $\textbf{H}$;
	\end{enumerate}
	 
	\item Calculate the heights (Equation \ref{heightscalculation}),  and the cell-centered curvature (Equation \ref{curvaturecalculation});
	\item Interpolate the curvature to the face $f$ (Equation \ref{curvatureInterpolation}) to obtain $\kappa_f$;
	\item Proceed to the GFM discretization of the pressure equation.   
\end{enumerate}

It is important to point out that sometimes 3x3 or 3x5 stencils are sufficient to compute discrete heights and obtain a the same curvature value provided by the 3x7 stencil (if the upper/lower rows only add full/empty cells). However, as stated by Popinet \cite{popinet2009accurate} 3x7 stencils may not be enough in some cases and 3x9 stencils can be necessary. In the cases presented in this work strong deformations of the interface are  not present and 3x7 stencils are safe to compute build consistent heights for curvature calculation. Moreover, in order to have  defined heights for all interfacial cells, the droplet must be well resolved. If the curvature radius approaches the mesh size, switching to curve fitting methods would be the best option.

\section{Equilibrium of a circular droplet}
In a circular droplet at zero gravity the pressure jump at the interface should perfectly balance the surface tension force. The test case from Popinet \cite{popinet2009accurate} is adopted and used for comparison: a circular interface is initialized at the center of the domain and it is given enough time to relax to its numerical equilibrium shape. Only a quarter of the circle is considered. In order to reach the numerical equilibrium (Equation \ref{pressurejumpanalytical}) three conditions are required:
 \begin{itemize}
 	\item A perfectly balanced discretization method;
 	\item An accurate methodology for curvature calculation;
 	\item An \textit{exact} initialization of the circular interface.
 \end{itemize}

The Ghost Fluid Method and the Height Functions are used for pressure  discretization and curvature calculation. The last point requires a brief discussion.

\begin{figure}
	\centering
	\subfloat[]
	{\includegraphics[width=.2\textwidth,height=0.14\textheight]{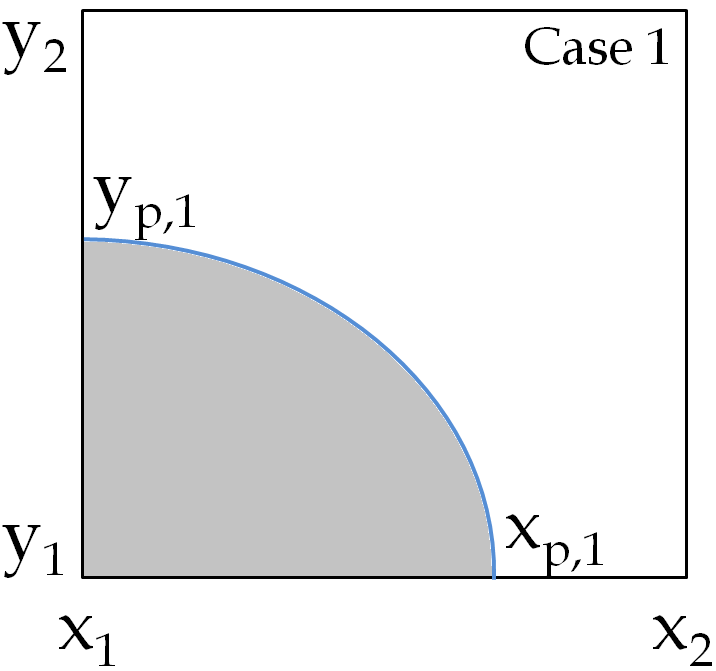}}\quad
	\subfloat[]
	{\includegraphics[width=.2\textwidth,height=0.14\textheight]{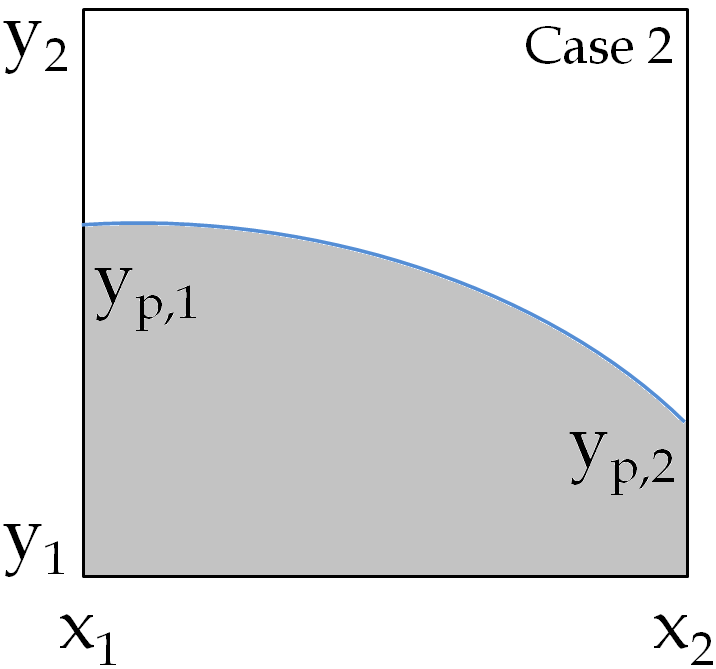}}\quad
	\subfloat[]
	{\includegraphics[width=.2\textwidth,height=0.14\textheight]{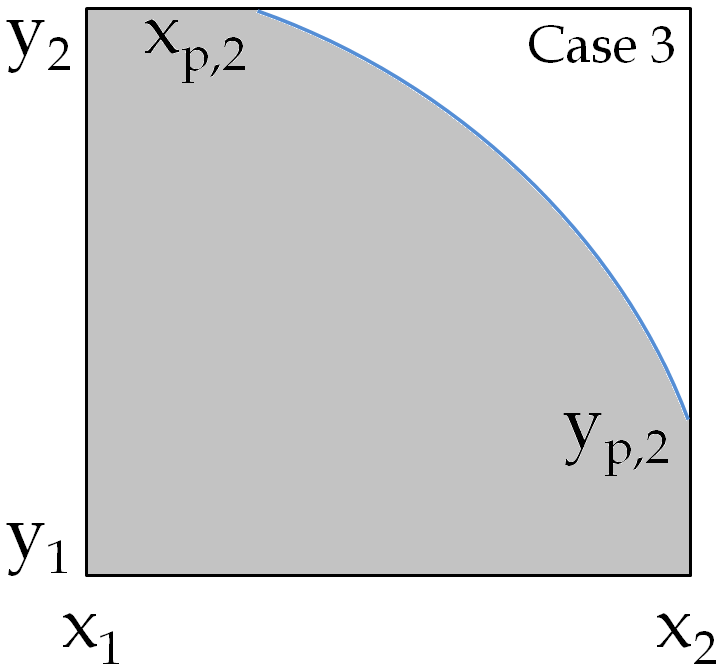}}\quad
	\subfloat[]
	{\includegraphics[width=.2\textwidth,height=0.14\textheight]{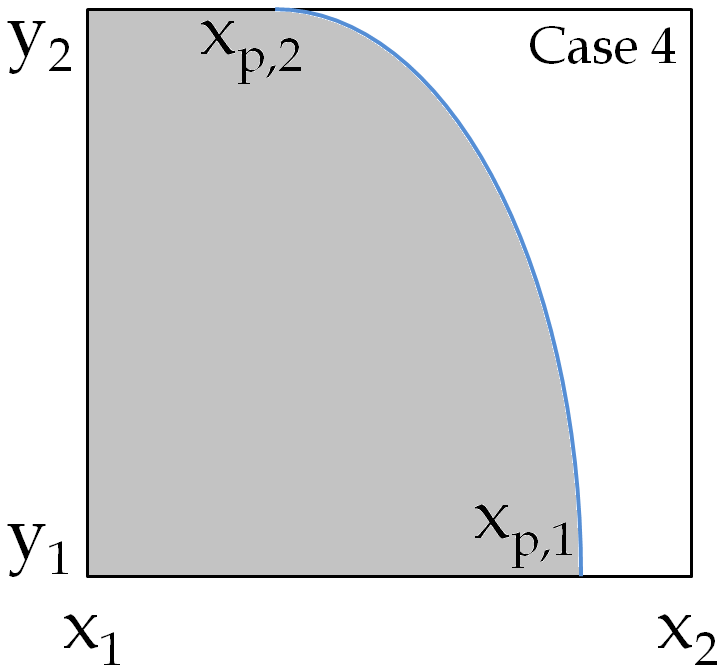}}\quad
	\caption{Four possible configurations for the intersection between a quarter of a circular interface and the cell.}
	\label{intersections}
\end{figure}

\begin{table}
	\centering
	
	\begin{tabular}{ll}
		\toprule
		Case ~~ & Area  \\
		\midrule
		1     & $F\left(x_{p,1}\right)-F\left(x_1\right)-\left(x_{p,1}-x_1\right)y_1$    \\
		2     & $F\left(x_2\right)-F\left(x_1\right)-\left(x_2-x_1\right)y_1$       \\
		3     &    $F\left(x_2\right)-F\left(x_{p,2}\right)+\left(x_{p,2}-x_1\right)y_2-\left(x_2-x_1\right)y_1$    \\  	
		4     &  $F\left(x_{p,1}\right)-F\left(x_{p,2}\right)+\left(x_{p,2}-x_1\right)y_2-\left(x_{p,1}-x_1\right)y_1$    \\				
		\bottomrule
		
	\end{tabular}	
	
	\caption{Computation of the grey area for the four cases in Figure \ref{intersections}.}
	\label{tableForIntersections}
\end{table}

\subsection{Interface initialization}
A circular interface at rest has a constant curvature. An equilibrium solution (\textbf{v}=\textbf{0}) can be reached if the curvature deviations are minimized along the interface, providing a uniform $\kappa$ value. However, this does not imply that the actual curvature value will be accurate (i.e converging). In fact,  mesh convergence towards the exact value is usually not achievable, unless the droplet is \textit{exactly} initialized \cite{coquerelle2016fourth, lopez2009improved}. This is not a trivial task and many recent works aimed at giving an accurate methodology for general surfaces initialization on polyhedral meshes \cite{bna2015numerical, bna2016vofi}. In this work we propose a relatively simple methodology, based on the direct integration of the analytical expression $f\left(x\right)$ of a quarter of a circle:
\begin{equation}
	f\left(x\right)=\sqrt{R^2-x^2}dx
	\label{circularInterface}
\end{equation}

 where $R$ is the radius. The following algorithm is implemented for the circle initialization.  For each cell in the domain:

\begin{itemize}
	\item The four vertex coordinates $\textbf{w}_1(x_1,y_1)$, $\textbf{w}_2(x_2,y_1)$, $\textbf{w}_3(x_2,y_2)$,  $\textbf{w}_4(x_1,y_2)$ of the cell are extracted;
	\item For each vertex $\textbf{w}_i$ the distance function $\psi_i$ is computed:
	
		\begin{equation}
		\psi_i = |\textbf{w}_i|-R
		\end{equation}
		
	\item If all the vertices are inside the droplet ($\psi<0$), set $\alpha=1$;

	\item If all the vertices are outside the droplet ($\psi>0$), set $\alpha=0$;
	\item Otherwise, the cell contains the interface. In this case:
	\begin{itemize}
		\item Calculate the intersections $x_{p,1}, x_{p,2}, y_{p,1}, y_{p,2}$ of the circle with the cell faces $x_1, x_2, y_1, y_2$;
		\item Depending on the position of the intersections (Figure \ref{intersections}), compute the area delimited by the curve and the cell faces. The integration of \ref{circularInterface} is analytical:
		\begin{equation}
			F(x)=\int_{0}^{x}f(x)=\frac{R^2}{2}\left[\frac{x}{R}\sqrt{1-\left(\frac{x}{R}\right)^2}+asin\left(\frac{x}{R}\right)\right]
		\end{equation}
		
		The area formulae are reported in Table \ref{tableForIntersections}.
		
	\end{itemize}
\end{itemize}

This methodology can be rapidly implemented, it is very accurate and does not depend on the mesh adopted. The obvious drawback is the non-generality of the method, only valid for this specific case.

\subsection{Test case setup}
The droplet diameter is $D=1$ mm. Density and viscosity are equal for gas and liquid: $\rho=1$ $kg/m^3$, $\mu=10^{-4}$ $Pa\cdot s$. Three surface tension values are used ($\sigma=0.0012, 0.012, 0.12$ N/m) to obtain three cases at different Laplace numbers $La = 120, 1200, 12000$. As reported by Popinet \cite{popinet2009accurate} the velocity scale of the problem regards the capillary waves originated from the droplet:

\begin{equation}
	v_{\sigma}=\sqrt{\frac{\sigma}{\rho D}}
\end{equation}

while the dissipation of this velocity occurs in a time scale $T_{\nu}$:

\begin{equation}
	T_{\nu}=\frac{D^2}{\nu}
\end{equation}

where $\nu$ is the kinematic viscosity. In order to test the accuracy of the surface tension implementation it is fundamental to run the test case for a time $t \sim T_{\nu}$, in order to properly dissipate the initial interface perturbation. The surface tension implementation is explicit and the CFL condition for surface tension driven flows requires a minimum time step size \cite{brackbill1992continuum} for stability:

\begin{equation}
	\Delta t_{min}=\sqrt{\frac{\rho\Delta x^3}{\pi\sigma}}
\end{equation}

The numerical result are presented in terms the following error norms, for velocity $\textbf{v}$:

\begin{equation}
L_{\infty}\left(\textbf{v}\right)=max\left(|\textbf{v}|\right)
\end{equation}

\begin{table}
	\centering
	
	\begin{tabular}{llll}
		\toprule
		
		La \quad\quad \quad\quad& $D/\Delta x$ \quad\quad\quad\quad& $|\textbf{v}|_{max}$ [m/s] \quad\quad\quad\quad & Ca \\
		\midrule
		\multirow{3}{*}{120}& 10 & $6.81\cdot10^{-9}$ & $5.67\cdot10^{-10}$  \\
		& 20 & $7.95\cdot10^{-9}$ & $6.62\cdot10^{-10}$   \\
		& 40 & $5.70\cdot10^{-11}$ & $4.75\cdot10^{-12}$  \\
		& 100 & $9.95\cdot10^{-11}$ & $8.29\cdot10^{-12}$  \\		                    
		\midrule
		\multirow{3}{*}{1200}& 10 & $1.32\cdot10^{-9}$ & $1.10\cdot10^{-11}$  \\
		& 20 & $8.68\cdot10^{-9}$ & $7.23\cdot10^{-11}$    \\
		& 40 & $2.94\cdot10^{-8}$ & $2.45\cdot10^{-10}$  \\
		& 100 & $1.18\cdot10^{-8}$ & $9.83\cdot10^{-11}$  \\		                    
		\midrule
		\multirow{3}{*}{12000}& 10 & $6.70\cdot10^{-9}$ & $5.58\cdot10^{-12}$  \\
		& 20 & $3.14\cdot10^{-7}$ &  $2.62\cdot10^{-10}$   \\
		& 40 & $1.48\cdot10^{-7}$ &  $1.23\cdot10^{-10}$ \\
		& 100 & $6.36\cdot10^{-7}$ & $5.30\cdot10^{-10}$  \\		                    	
		\bottomrule
		
	\end{tabular}	
	
	\caption{Spurious currents analysis for the 2D static droplet. Results at $La=120, 1200, 12000$ and four different droplet resolutions $D/\Delta x=10, 20, 40, 100$, by means of $L_{\infty}\left(\textbf{v}\right)$ error norm $|\textbf{v}|_{max}$ and capillary number Ca. Time $t=T_{\nu}=0.01$ s.}
	\label{spuriousCurrentsTable}
\end{table}

curvature $\kappa$:

\begin{equation}
L_{\infty}\left(\kappa\right)=\frac{1}{\kappa_{ex}}max\left(|\kappa_i-\kappa_{ex}|\right)
\end{equation}

and interface shape:

\begin{equation}
L_2\left(shape\right)=\sqrt{\frac{\sum_{i}^{}\left(\alpha_i-\alpha_{ex}\right)^2}{\sum_{i}^{}}}
\end{equation}

\begin{equation}
L_{\infty}\left(shape\right)=max\left(|\alpha_i-\alpha_{ex}|\right)
\end{equation}

where $\alpha_{ex}$ is the exact $\alpha$ field from the initialization. 
The three cases ($La=120, 1200, 12000$) are run at four different droplet resolutions $\frac{D}{\Delta x} = 10, 20, 40, 100$ to investigate the mesh convergence.

\begin{figure}
	\centering
	{\includegraphics[width=.45\textwidth,height=0.25\textheight]{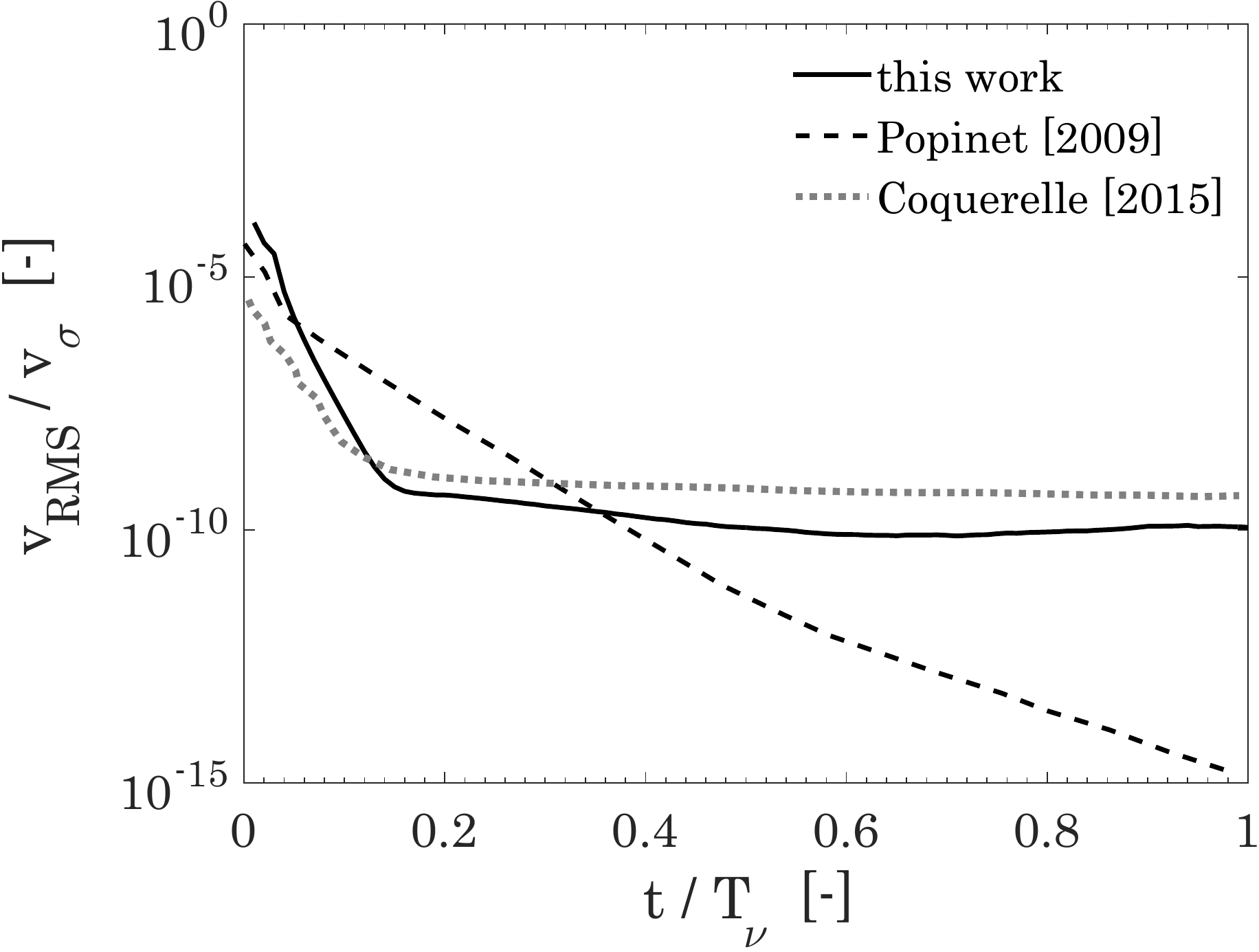}} 
	\caption{Dimensionless root mean square (RMS) velocity profile over time, for $La=1200$. Comparison with the results from  Popinet \cite{popinet2009accurate} and Coquerelle  \cite{coquerelle2016fourth}.}
	\label{velocityComparison}
\end{figure}

\subsection{Spurious currents}

Table \ref{spuriousCurrentsTable} reports the $L_{\infty}$ error norm for the velocity field $|\textbf{v}|_{max}$ and the relative capillary number $Ca$ for the three Laplace numbers (120, 1200, 12000) at four different resolution of the droplet. The residual parasitic currents are extremely low, even reaching the machine precision in some cases. As expected, the $|\textbf{v}|_{max}$ values are located along the diagonal axis (at $\sim 45^\circ$, where $n_y\sim n_x$) due to the lower accuracy of Height Functions in this region \cite{popinet2009accurate}. The capillary number remains well below $10^{-9}$ and the result do not significantly change further reducing the time step size.

\begin{figure}
	\centering
	\subfloat[]
	{\includegraphics[width=.38\textwidth,height=0.2\textheight]{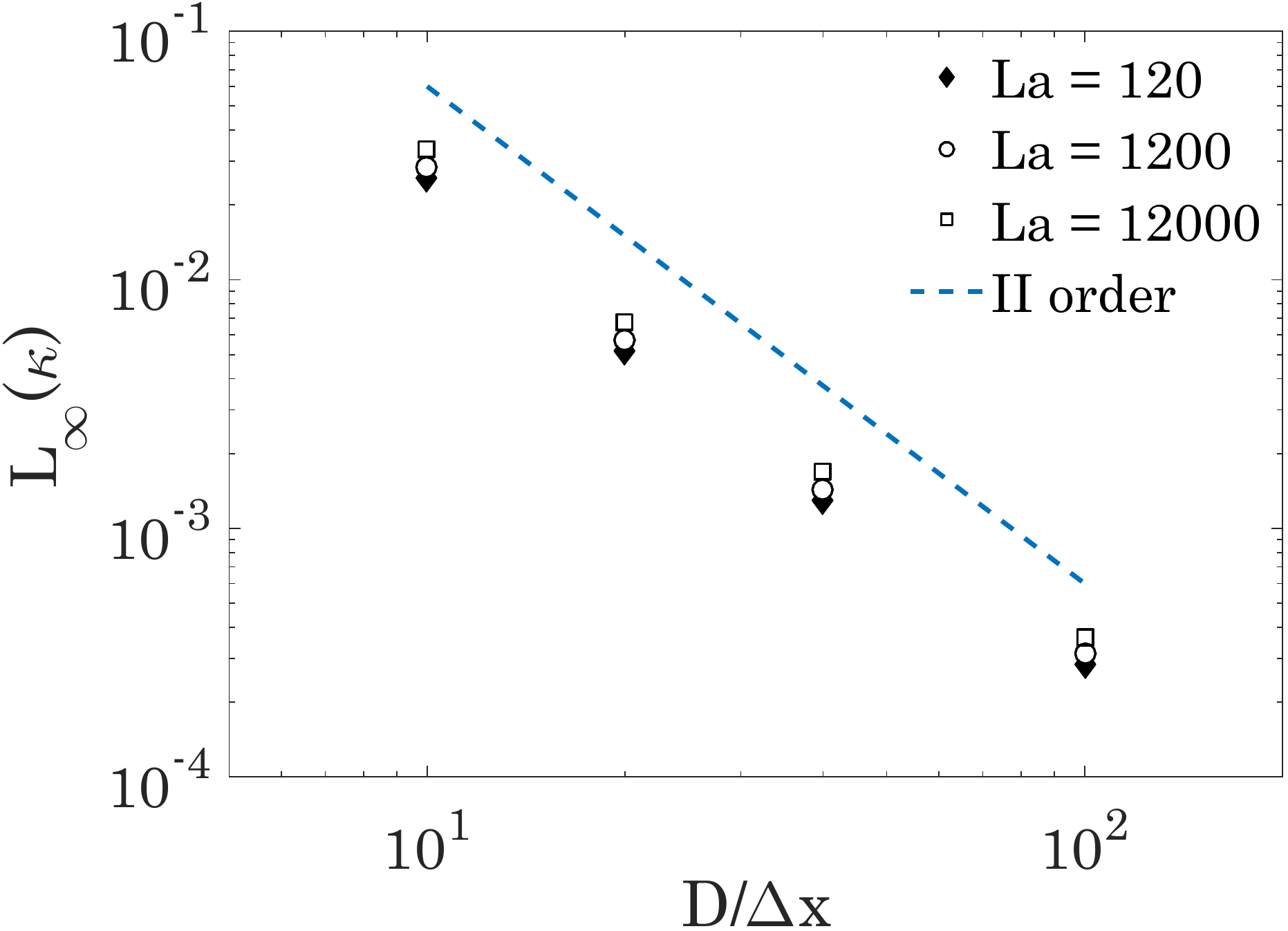}}\quad\quad
	\subfloat[]
	{\includegraphics[width=.38\textwidth,height=0.2\textheight]{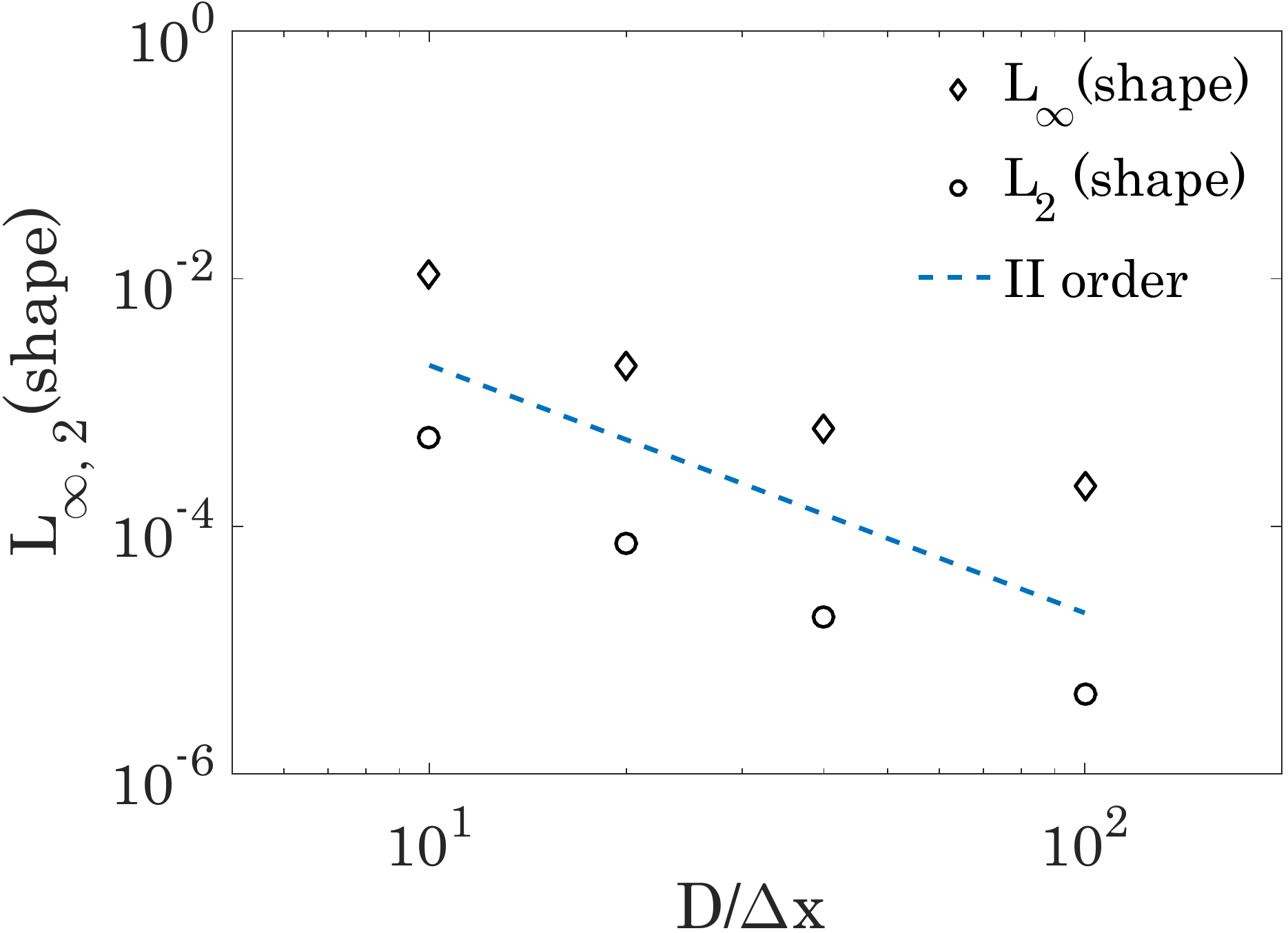}}
	\caption{Mesh convergence analysis for the $L_{\infty}$ error norm for curvature $\kappa$ (a) at different Laplace numbers. $L_{\infty}$ and $L_2$ error norms for shape (b) at $La=12000$.}
	\label{meshConvergence}
\end{figure}

Parasitic currents are also analyzed over time in Figure \ref{velocityComparison} for $La=1200$. The RMS velocity is  made dimensionless using $v_{\sigma}$, while the dimensionless time $\tau$ is referred to $T_{\nu}$. The RMS velocity initially shows a strong exponential decrease (due to viscous dissipation) followed by a slower one, around the value $\sim 10^{-10}$. The same  occurs for all $La$ numbers. The results are between the ones obtained by Popinet \cite{popinet2009accurate} and Coquerelle \cite{coquerelle2016fourth} for the same case. Even though parasitic currents do not completely vanish within $T_{\nu}$, they remain very low and we did not notice any interference with the accuracy of the curvature calculation. 

\subsection{Mesh convergence}
The accuracy of the simulations is assessed by a mesh convergence analysis on the curvature $\kappa$ and the numerical droplet shape. Results are presented in  Figure \ref{meshConvergence}. We observe close to second-order accuracy both for curvature and droplet shape, with a very weak dependence on the Laplace number. The $L_2(\kappa)$ norm (not shown) coincides with $L_{\infty}(\kappa)$, indicating a very  uniform value of the curvature along the interface. Even for coarse resolutions the error values remain very small and comparable to what obtained by other studies \cite{popinet2009accurate}. 

\begin{figure}
	\centering
	\quad\quad\quad\quad\quad	{\includegraphics[width=.48\textwidth,height=0.25\textheight]{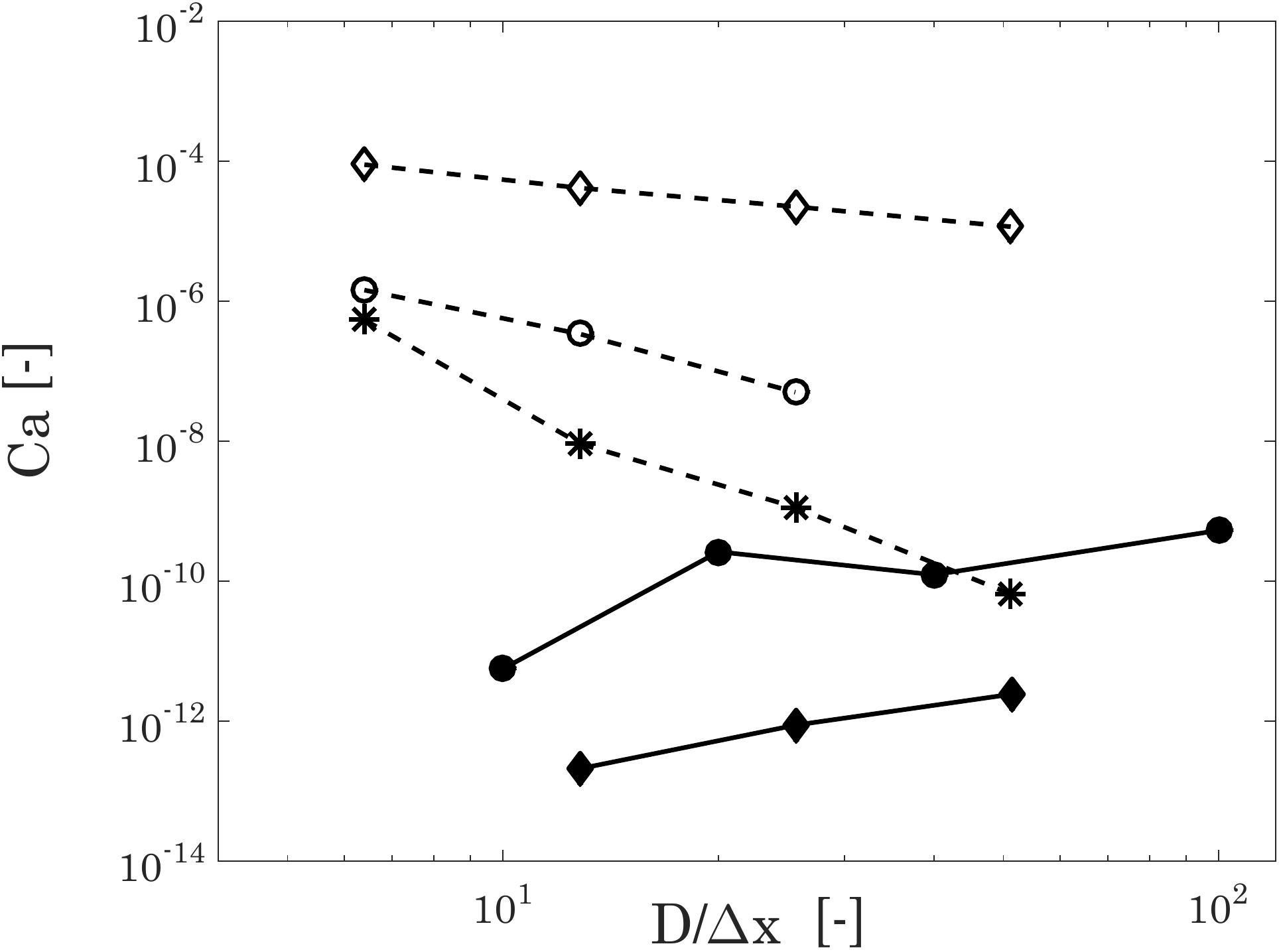}}\quad
	{\includegraphics[width=.25\textwidth,height=0.17\textheight]{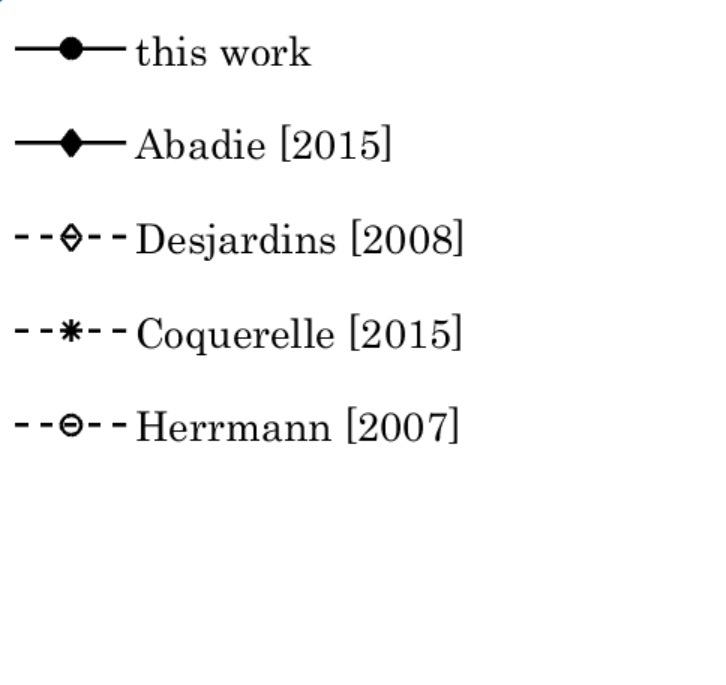}}	
	\caption{Capillary number convergence with mesh refinement, $La=12000$. Comparison with the works of Abadie \cite{abadie2015combined}, Desjardins \cite{desjardins2008accurate}, Coquerelle \cite{coquerelle2016fourth} and Herrmann \cite{herrmann2008balanced}. Dashed lines for level-set based methods, solid lines for VOF methods. }
	\label{spuriousCurrentsMeshConvergence}
\end{figure}

Concerning the velocity field, in Table \ref{spuriousCurrentsTable} spurious currents become close to round off errors refining the mesh for $La=120$. However, we could not recover a similar mesh convergence for  $La=1200$ and $La=12000$, in which spurious currents slightly increase with mesh refinement. This behavior is not new in literature \cite{harvie2006analysis} and apparently tends to occur at high Laplace numbers. To better investigate this aspect we compared our results at $La=12000$ with other available studies in Figure \ref{spuriousCurrentsMeshConvergence} in terms of capillary number $Ca$. The works of Desjardins \cite{desjardins2008accurate},  Herrmann \cite{herrmann2008balanced} and Coquerelle \cite{coquerelle2016fourth} are based on a Level-Set approach and show mesh convergence at different orders (from $\sim$ 1 to $\sim$ 3).  Abadie  \cite{abadie2015combined} instead performed a large comparative study on the effect of the interface advection on surface tension modeling, showing that a combination of a geometric interface reconstruction for VOF (i.e. PLIC) and Height Functions for curvature can actually lead to a non-decreasing velocity error for higher resolutions (Figure 5 in \cite{abadie2015combined}, also reported here in Figure \ref{spuriousCurrentsMeshConvergence} for comparison). Similar results have been reached by Popinet and Zaleski comparing front-tracking methods and PLIC/VOF (Figure 6 in \cite{popinet1999front}). Owkes and Desjardins also show  non-converging capillary numbers for a standard mesh-aligned height functions method (Table 1 in \cite{owkes2015mesh}). Interestingly, they could avoid this adopting normal-aligned stencil in order to reduce curvature errors (especially when $n_x\sim n_y$), even though the issue seems to persist at high $La$ numbers ($>10^6$).\\ As reported by Abadie (Table 1 in \cite{abadie2015combined}), residual spurious currents for VOF methods coupled with Height Functions are most likely due to numerical errors in the advection step rather than inaccurate curvature calculation. Indeed, accurate methods such Height Functions are able to capture small fluctuations of the interface position due to advection errors. In our case   \texttt{isoAdvector} by Roenby and Jasak \cite{roenby2016computational} is used to reconstruct and advect the interface. The authors report slightly worse results in the advection for  low time steps (Co $<$ 0.5) and this was also previously observed by Ubbink and Issa \cite{ubbink1999method}. In our case the time step is controlled by surface tension and it is orders of magnitude lower than the one dictated by the Courant number. Refining the mesh further enhances this effect (since $\Delta t \sim \Delta x^{3/2}$) and we believe it probably introduces additional errors in the advection step. The effect of the interface advection on spurious currents is further analyzed in the next section.

\section{Translating droplet}
In order to investigate the combined accuracy of the advection scheme, the curvature calculation and surface tension discretization, a droplet subjected to a uniform constant velocity field $\textbf{v}_0$ is considered, as firstly proposed in  \cite{popinet2009accurate}. The exact solution in the moving reference frame is the same as for the static case. This test case is much more problematic, since the advection errors directly impact the curvature estimation and therefore the velocity field. The new time scale of the problem is:

\begin{equation}
	T_U=\frac{D}{|\textbf{v}_0|}
\end{equation}

As for the static droplet test, three Laplace numbers $La=120, 1200, 12000$ are considered. Gas and liquid properties are unchanged. The velocity is prescribed in the horizontal direction and it is chosen in order to maintain a constant Weber number:

\begin{table}
	\centering
	
	\begin{tabular}{llll}
		\toprule
		
		La \quad\quad \quad\quad& $D/\Delta x$ \quad\quad\quad\quad& $|\textbf{v}-\textbf{v}_0|_{max}$ [m/s] \quad\quad\quad\quad & Ca \\
		\midrule
		\multirow{3}{*}{120}& 10 & $7.76\cdot10^{-3}$ & $6.46\cdot10^{-4}$  \\
		& 20 & $4.69\cdot10^{-3}$ & $3.39\cdot10^{-4}$   \\
		& 40 & $5.25\cdot10^{-3}$ & $4.37\cdot10^{-4}$ \\	
		& 100 & $1.52\cdot10^{-2}$ & $1.26\cdot10^{-3}$ \\		                    
		\midrule
		\multirow{3}{*}{1200}& 10 & $2.55\cdot10^{-2}$ & $2.10\cdot10^{-4}$  \\
		& 20 & $1.88\cdot10^{-2}$ & $1.56\cdot10^{-4}$    \\
		& 40 & $2.31\cdot10^{-2}$ & $1.92\cdot10^{-4}$   \\	
		& 100 & $8.80\cdot10^{-2}$ & $7.33\cdot10^{-4}$ \\		                    
		\midrule
		\multirow{3}{*}{12000}& 10 & $8.21\cdot10^{-2}$ & $6.84\cdot10^{-5}$  \\
		& 20 & $6.15\cdot10^{-2}$ &  $5.12\cdot10^{-5}$   \\
		& 40 & $1.03\cdot10^{-1}$ &  $8.58\cdot10^{-5}$  \\		       
		& 100 & $4.04\cdot10^{-1}$ & $3.36\cdot10^{-4}$ \\	             	
		\bottomrule
		
	\end{tabular}	
	
	\caption{Spurious currents analysis for the 2D translating droplet. Results at $La=120, 1200, 12000$ and four different droplet resolutions $D/\Delta x=10, 20, 40, 100$, by means of relative velocity error norm $|\textbf{v}-\textbf{v}_0|_{max}$ and capillary number $Ca$. Time $t=T_{U}$.}
	\label{spuriousCurrentsTranslatingTable}
\end{table}

\begin{equation}
	We = \frac{\rho |\textbf{v}_0|^2 D}{\sigma}=0.4
\end{equation}

meaning that for higher surface tensions (i.e. higher $La$ numbers) a stronger velocity field is imposed. We have:

\begin{itemize}
	\item $|\textbf{v}_0| = 0.69$ m/s for $La=120$
	\item $|\textbf{v}_0| = 2.19$ m/s for $La=1200$
	\item $|\textbf{v}_0| = 6.93$ m/s for $La=12000$
\end{itemize}

The simulations are run for a total time $t=T_U$ at four different droplet resolutions $D/\Delta x =10, 20, 40, 100$. The capillary number $Ca$ is based on the moving reference frame:

\begin{equation}
	Ca=\frac{\mu|\textbf{v}-\textbf{v}_0|_{max}}{\sigma}
\end{equation}

\begin{figure}
	\centering
	\subfloat[]
	{\includegraphics[width=.45\textwidth,height=0.24\textheight]{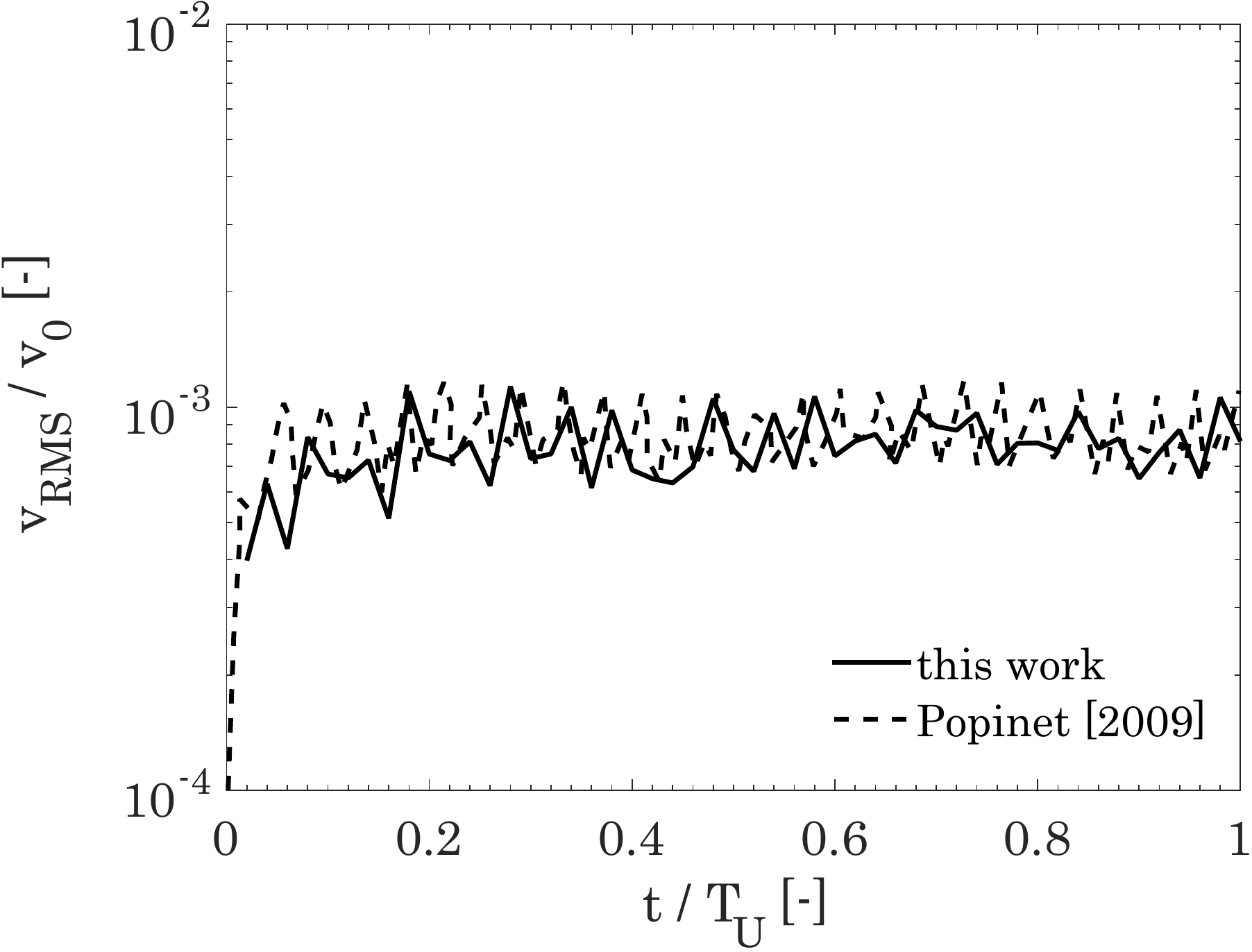}} \quad \quad
	\subfloat[]
	{\includegraphics[width=.43\textwidth,height=0.24\textheight]{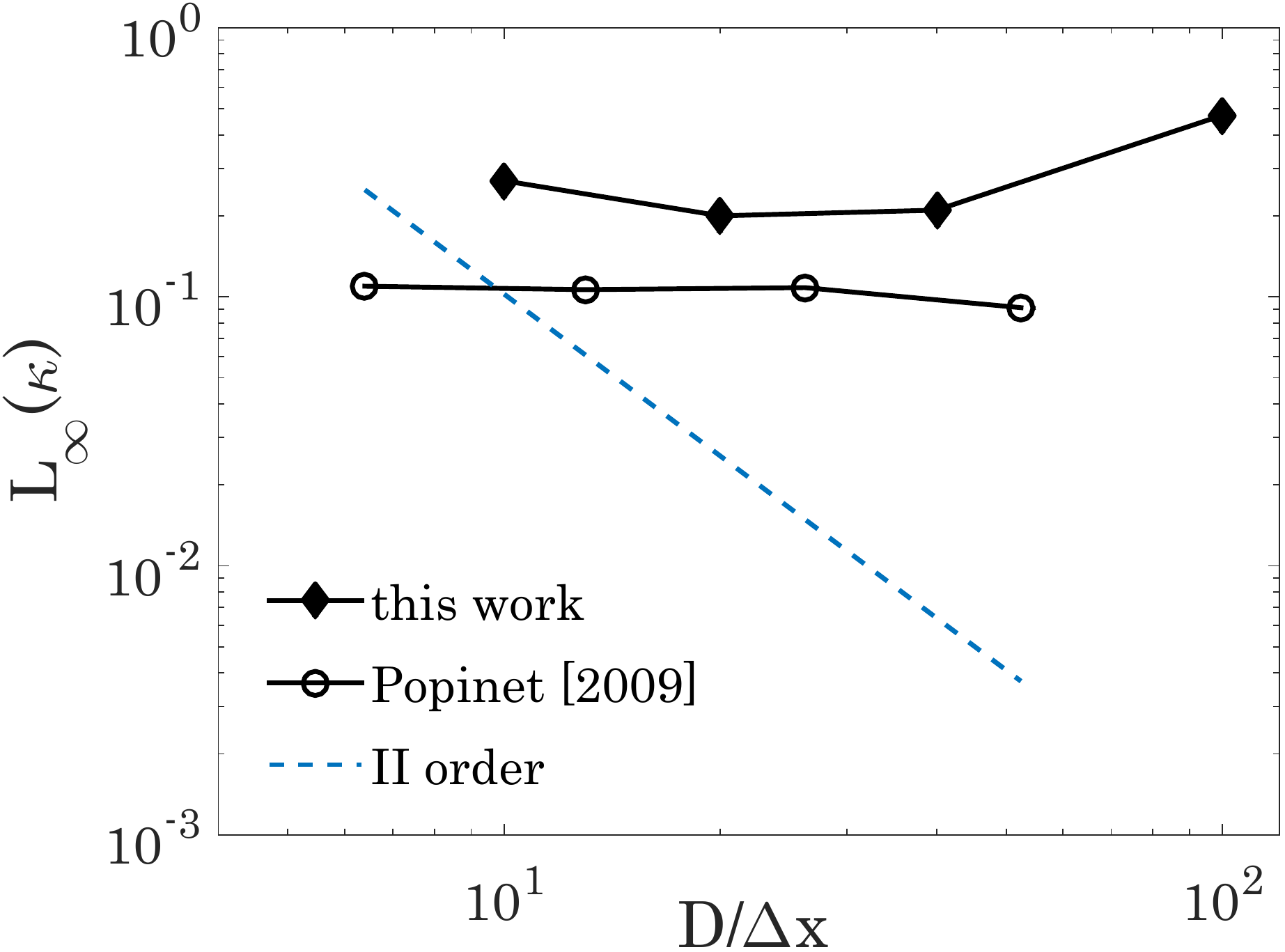}} 
	\caption{RMS velocity profile over the time (a) and $L_\infty$ error norm on curvature $\kappa$ (b) for the translating droplet test. Comparison with the results by Popinet \cite{popinet2009accurate}.}
	\label{velocityComparisonTranslating}
\end{figure}

The results are summarized in Table \ref{spuriousCurrentsTranslatingTable}. The capillary numbers in this case are 7-8 orders of magnitudes higher than for the static test, clearly indicating that the main spurious currents magnifier is the  advection scheme. This is in agreement to what obtained by Abadie (Figure 18 in \cite{abadie2015combined}): in particular for VOF/PLIC/Height Functions methods he shows that the ratio between the capillary number of translating and static droplet can reach up to $10^9$, while it is close to unity for level-set based approaches. We also compared the dimensionless RMS velocity profile to Popinet results \cite{popinet2009accurate} in Figure \ref{velocityComparisonTranslating} (a), showing very similar results: oscillations of frequency $\sim\frac{|\textbf{v}_0|}{\Delta x}$ and no convergence towards the exact solution. On overall, the maximum error on the relative velocity is below $5\%$ for all the cases, which is usually acceptable for common applications. In Figure \ref{velocityComparisonTranslating} (b) the $L_\infty$ norm on the curvature is shown for $La=12000$. Mesh convergence cannot be reached, both for this work and the results from Popinet \cite{popinet2009accurate}.  These results underline the need of a tighter coupling between surface tension scheme and interface advection, in order to reach more accurate solutions in more representative cases.

\section{Oscillating droplet}
A slightly elliptical shape is imposed as initial condition for a 2D liquid droplet, which tends to recover a spherical geometry due to surface tension-induced oscillations. The perturbed surface of the droplet is given in polar coordinates:

\begin{equation}
R\left(\theta\right)=R_0\left[1+\epsilon cos\left(2\theta\right)\right]
\end{equation}

The droplet ($D=1$ mm) is centered at (0,0) and it is deformed by a factor $\epsilon=0.04$. The surface tension is $\sigma=0.07$ N/m, while the density ratio is 100. The viscosity of both fluids is set to zero. The inviscid condition makes the test case more stringent, since spurious currents cannot be limited by a physical viscosity. In fact, oscillations decay is present anyway due to numerical dissipation. The analytical solution for small perturbations has been derived by Lamb \cite{lamb1945hydrodynamics}. The oscillation frequency $\omega$ for a 2D droplet is:

\begin{equation}
\omega=\sqrt{\frac{6\sigma}{\left(\rho_L+\rho_G\right) R^3}}
\end{equation}

\begin{figure}
	\centering
	\subfloat[]
	{\includegraphics[width=.45\textwidth,height=0.23 \textheight]{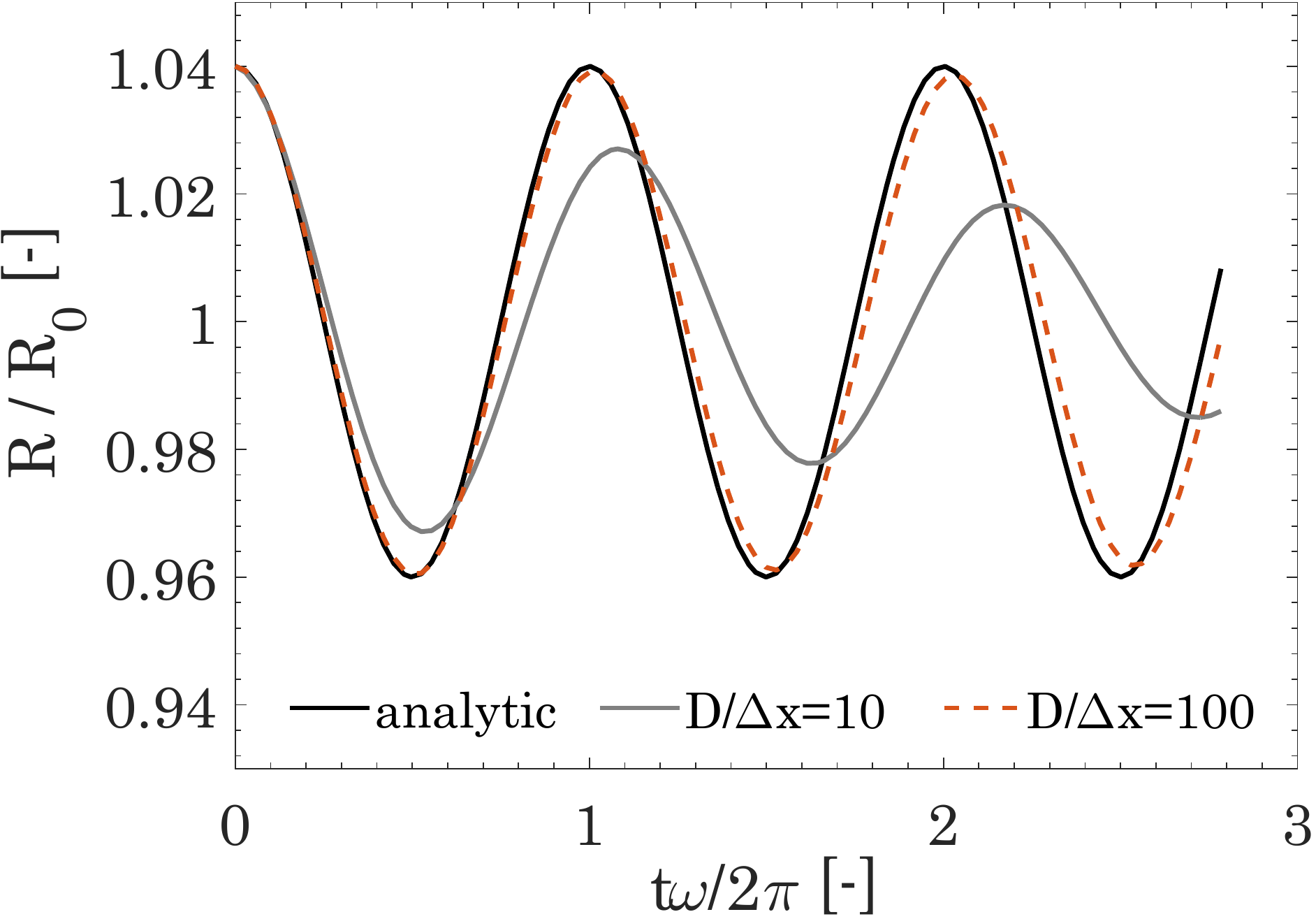}}\quad\quad
	\subfloat[]
	{\includegraphics[width=.4\textwidth,height=0.23\textheight]{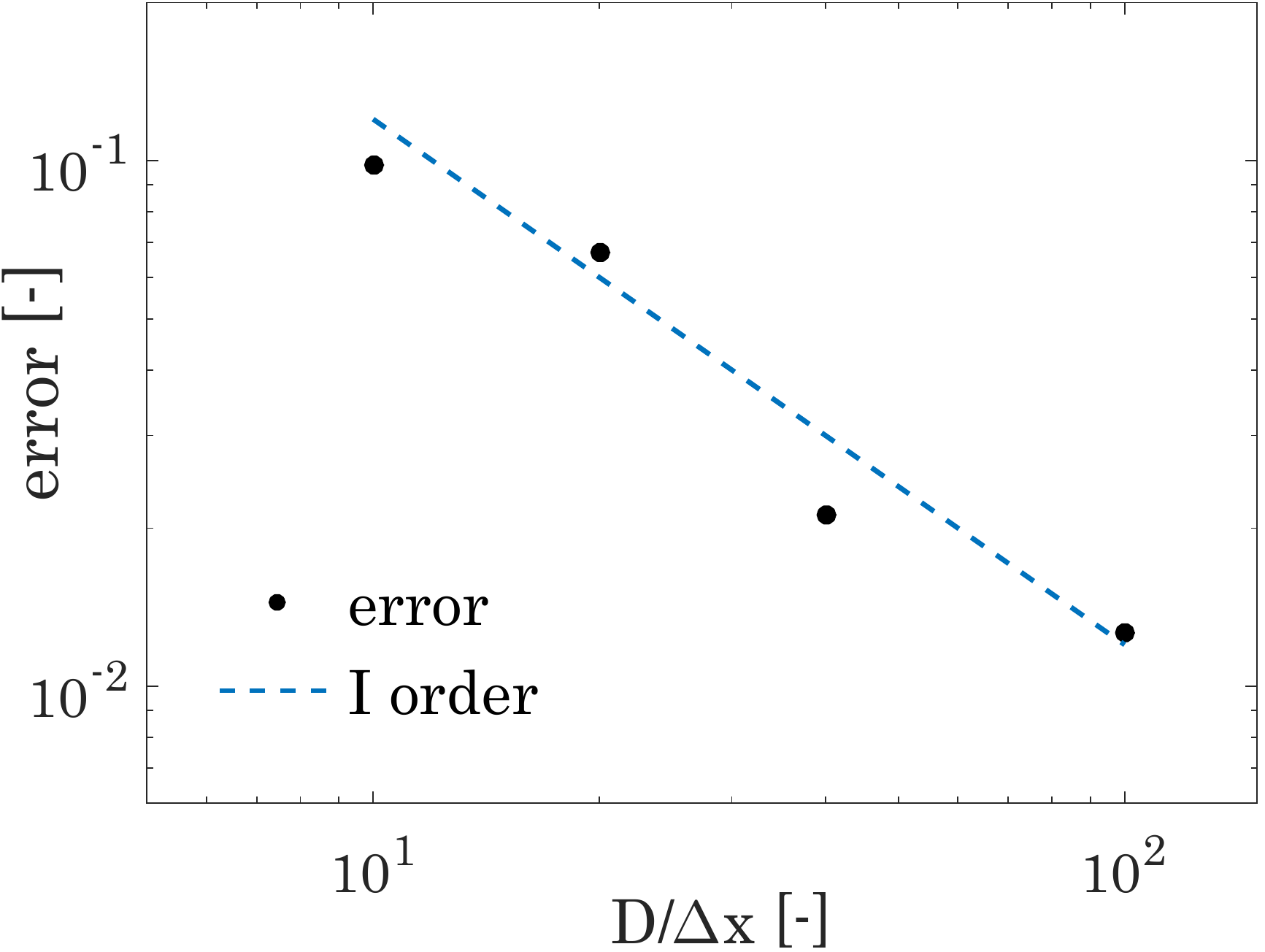}}
	\caption{Oscillations of a 2D inviscid liquid droplet (a): comparison between analytical solution (black line) and numerical results on the coarsest (grey line) and finest (dashed line) mesh. Mesh convergence in (b).}
	\label{dropletoscillations}
\end{figure}

Four droplet resolutions have been adopted $\frac{D}{\Delta x}=10, 20, 40, 100$. The results are reported in Figure \ref{dropletoscillations} in terms of maximum relative position on the \textit{x} axis (a) and convergence of the relative error on the oscillation frequency:

\begin{equation}
error=\frac{|\omega-\omega_{ex}|}{\omega_{ex}}
\end{equation}

where the numerical frequency is averaged over the first three periods.

The oscillation of the interface position is in very good agreement with the analytical solution, reaching an error of $\sim 1\%$ at the finest level. The mesh convergence analysis shows first order convergence on the frequency. The coarse solution suffers from significant numerical diffusion, which is almost totally recovered refining the mesh.

\section{Rising bubble}
A more complex test case is represented by a gas bubble rising in a dense liquid under the effect of gravity. Analytical solutions are not available for this case and a simple comparison of the bubble shapes with experimental or other numerical works is not sufficient to verify the accuracy of the results \cite{popinet2018numerical}. Hysing et al. \cite{hysing2009quantitative} provided an accurate reference solution for this test, adopting different numerical codes under different conditions, representing a valid numerical benchmark for comparison. The domain is [1 x 2] m and the bubble ($D=0.5$ m) is initially positioned at (0.5, 0.5). The physical parameters are defined in Table \ref{bubbleRisingTable}. The Eotvos number (based on the liquid properties) is defined as:

\begin{equation}
Eo=\frac{\rho_L g D^2}{\sigma}
\end{equation}

\begin{table}
	\centering
	
	\begin{tabular}{llllllll}
		\toprule
		
	Test case & $\rho_L$ & $\rho_G$ & $\mu_L$ & $\mu_G$ & $|\textbf{g}|$ & $\sigma$ & Eo \\
		\midrule
		 1 & 1000 & 100 & 10 & 1 & 0.98 & 24.5 & 10 \\	     	
		\bottomrule
		
	\end{tabular}	
	
	\caption{Physical parameters for the bubble rising test case.}
	\label{bubbleRisingTable}
\end{table}

The quantitative results are given in terms of center of mass $y_c$ of the bubble:

\begin{equation}
	y_c=\frac{\int_{S}^{}y dS}{\int_{S}^{} dS}
\end{equation}

rising velocity of the bubble $v_c$:

\begin{equation}
	v_c=\frac{\int_{S}^{}v_y dS}{\int_{S}^{} dS}
\end{equation}

and bubble circularity $c$:

\begin{figure}
	\centering
	\subfloat[]
	{\includegraphics[width=.42\textwidth,height=0.23\textheight]{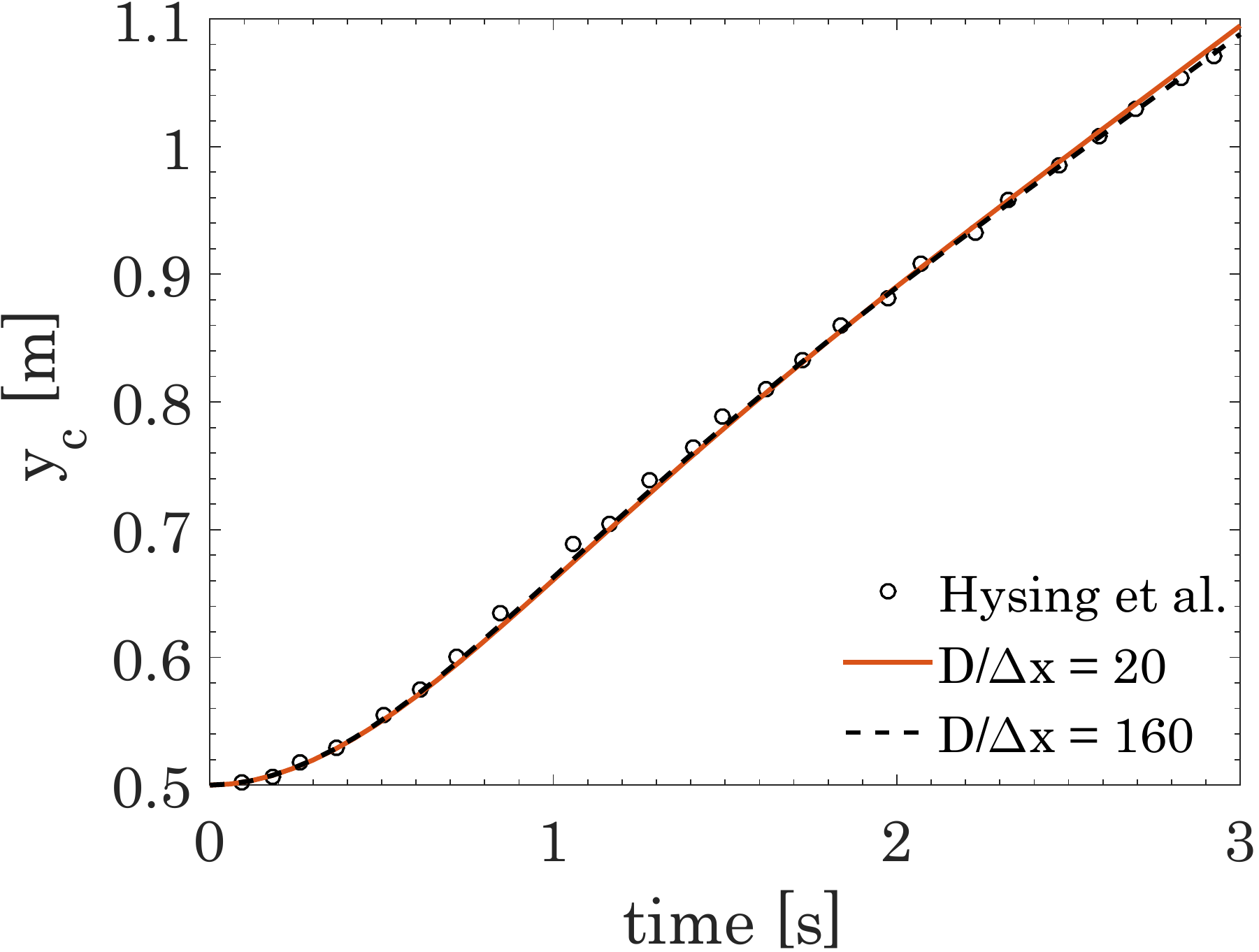}}\quad\quad
	\subfloat[]
	{\includegraphics[width=.42\textwidth,height=0.23\textheight]{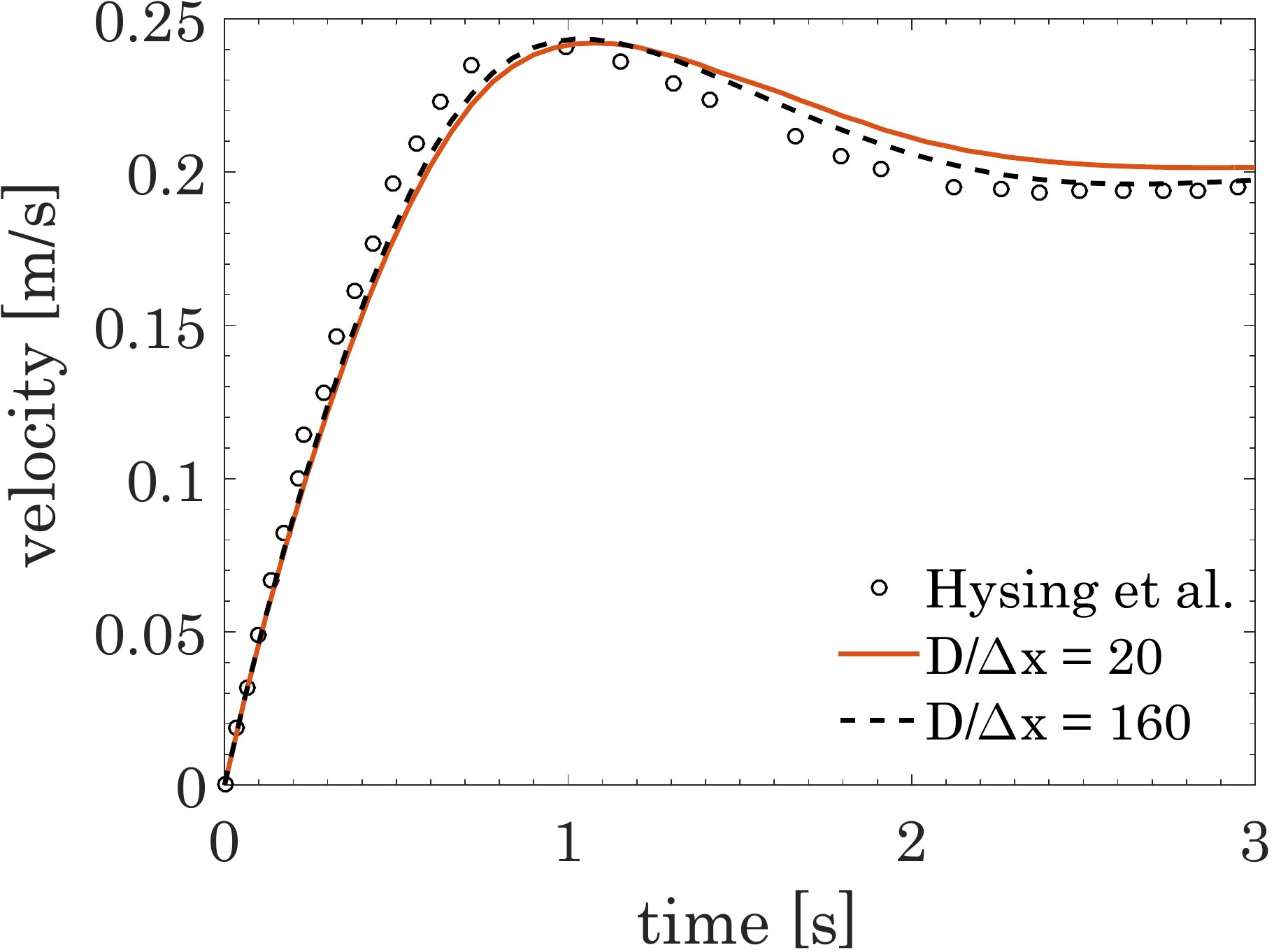}}\\
	\subfloat[]
	{\includegraphics[width=.42\textwidth,height=0.23\textheight]{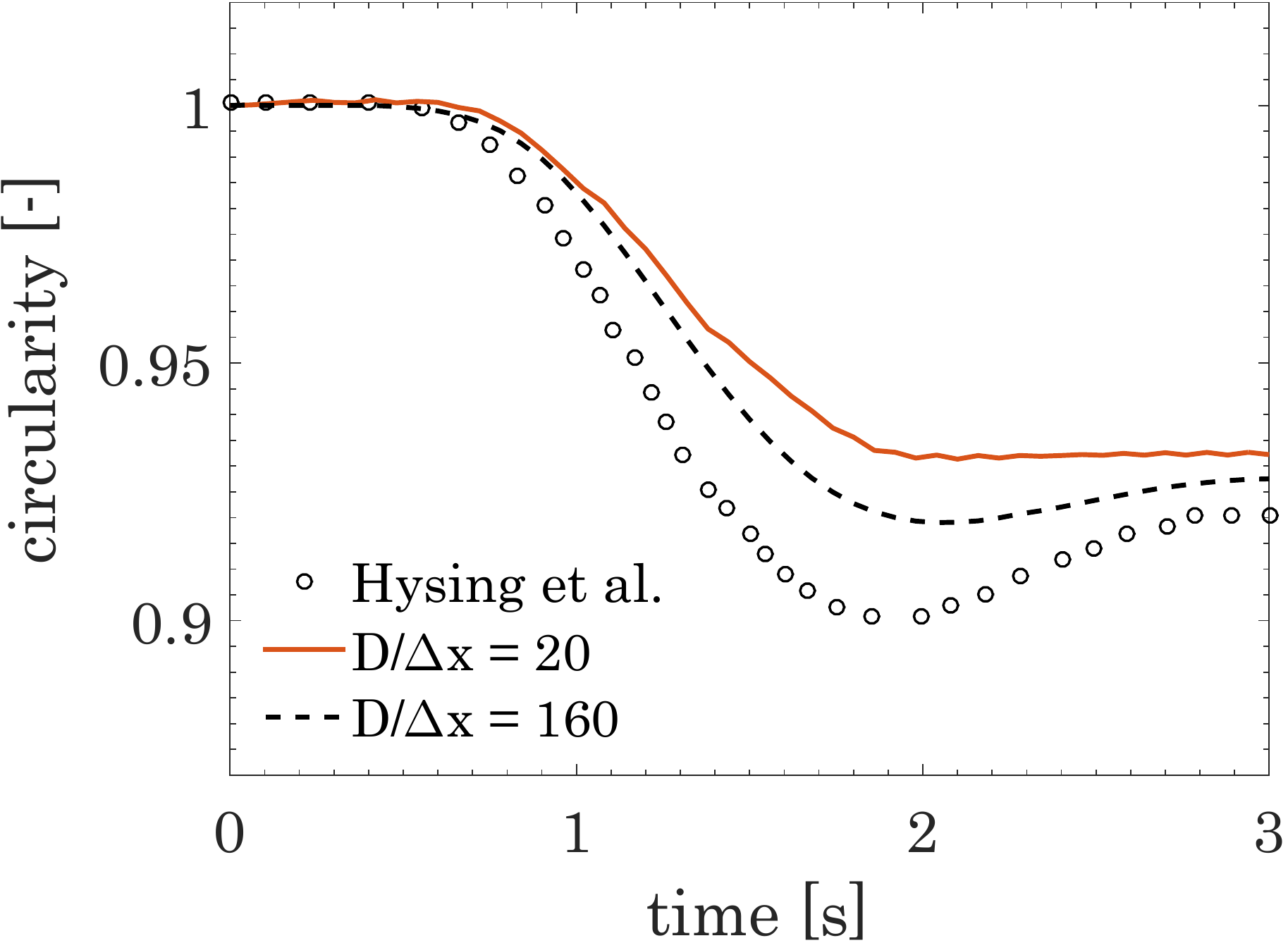}}\quad\quad
	\subfloat[]
	{\includegraphics[width=.42\textwidth,height=0.24\textheight]{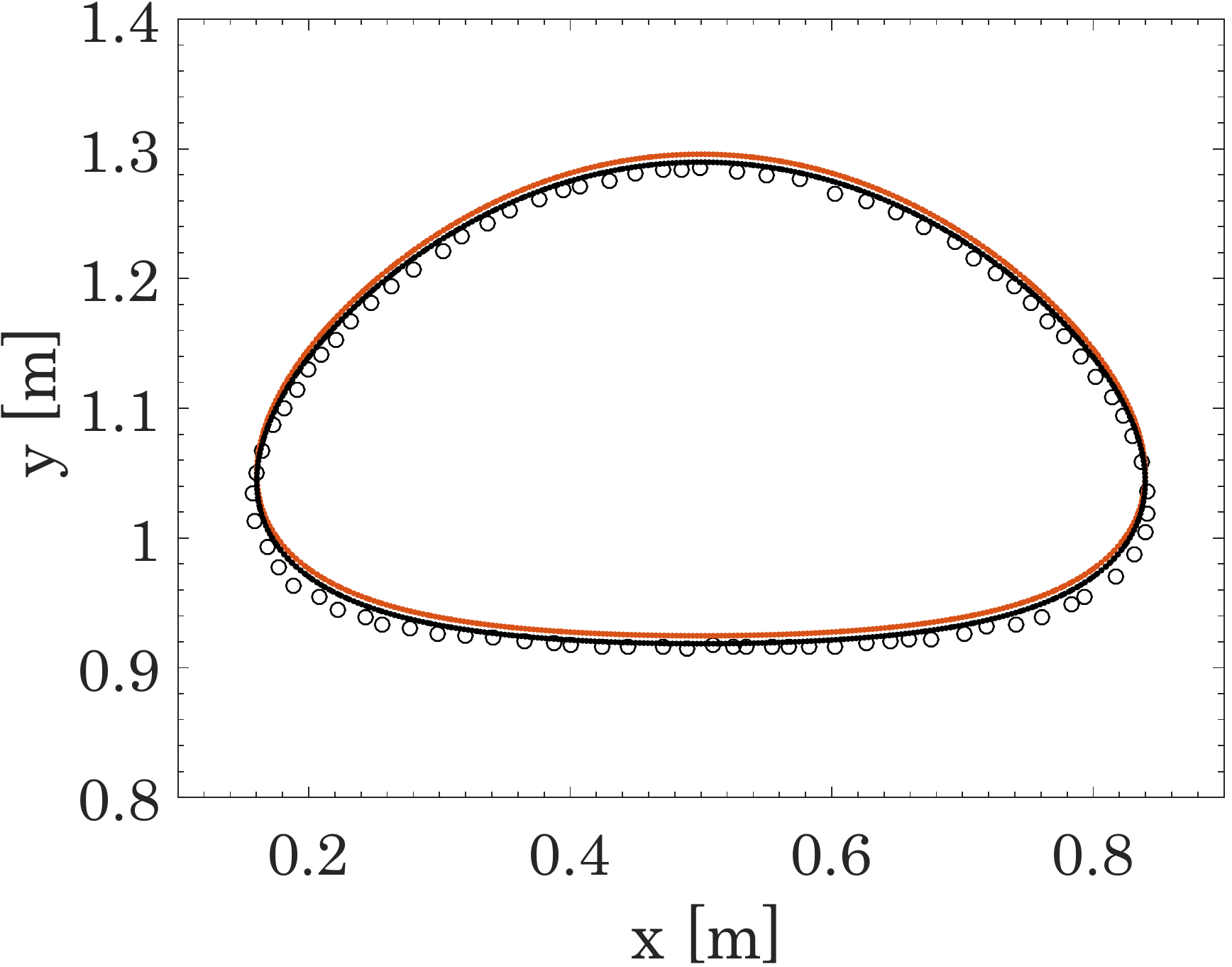}}
	\caption{Numerical results for the bubble center of mass $y_c$ (a), rising velocity $v_c$ (b), circularity $c$ (c) and final bubble shape (d) (at $t=3$ s). Coarsest mesh (orange line) and finest mesh (black line) compared with the reference results from Hysing et al. \cite{hysing2009quantitative} (circles). }
	\label{risingbubblecomparison}
\end{figure}

\begin{equation}
	c = \frac{\pi D}{p}
\end{equation}

The bubble perimeter $p$ is calculated as:

\begin{equation}
	p=\int_{S}^{}|\nabla\alpha|dS
	\label{perimeterFormula}
\end{equation}

where $S$ is the planar surface occupied by the bubble. These parameters  allow to track the motion of the bubble and to quantitatively evaluate the quality of the numerical results and the comparison with reference numerical works. A mesh convergence analysis is also performed, running the simulations at four different resolutions $D/\Delta x = 20, 40, 80, 160$. The results are presented in Figure \ref{risingbubblecomparison} and Table \ref{bubbleRisingTableValues}.\\
In the work of Hysing et al. \cite{hysing2009quantitative} three different numerical codes are used to assess the convergence of the results: TP2D (Transport Phenomena in 2D) \cite{turek1999efficient}, FreeLIFE (free-Surface Library of Finite Element) \cite{parolini2004computational} and MoonNMD (Mathematics and object-oriented Numerics in MagDeburg) \cite{john2004moonmd}. All of the three codes give almost identical results, therefore only the TP2D code will be used as a comparison in this work. It is based on a finite-element discretization method, adopting a level-set approach for the interface advection. The interface re-initialization (always needed for level-set approaches) is based on the fast marching method, while surface tension is included by a direct integration over the interface contour. The comparison in Figure \ref{risingbubblecomparison} shows a very good agreement between the results, especially regarding the center of mass of the bubble $y_c$ (a) (almost coincident). The rising velocity $v_c$ (b) in our work shows a slight delay in the maximum value, whereas the terminal velocity (asymptotic value) is well recovered. Most deviations are found in the circularity profile (c), probably due to the difficulty of using Equation \ref{perimeterFormula} to calculate the bubble perimeter: in fact, the bubble shape and position (d) is in excellent agreement with the reference work. It is worth noticing that quite good results are already obtained with the coarsest solution and they further improve with the refinement level.

\begin{table}
	\centering
	\begin{tabular}{lccccll}
		\toprule
		\multirow{2}*{Measure} \quad\quad\quad & \multicolumn{4}{c}{$D/\Delta x$}  & \quad\quad \multirow{2}*{O}   &\quad\quad \multirow{2}*{Ref. \cite{hysing2009quantitative}} \\
		\cmidrule(lr){2-5}
		& 20 \quad\quad & 40  \quad\quad& 80  \quad\quad & 160 \\
		\midrule
		$c_{min}$ & 0.9313 \quad\quad & 0.9213 \quad\quad & 0.9197\quad\quad & 0.9190 & \quad\quad 2.1 & \quad\quad 0.9012 $\pm$ 0.0001 \\
		$t|_{c_{min}}$ & 2.201 \quad\quad & 2.103\quad\quad & 2.042\quad\quad & 2.033 & \quad\quad 1.8 & \quad\quad 1.9 \\
		$v_{c,max}$ & 0.2421 \quad\quad & 0.2424 \quad\quad & 0.2430 \quad\quad & 0.2433 & \quad\quad 1.1  & \quad\quad 0.2419 $\pm$ 0.0002 \\
		$t|_{v_{max}}$ & 1.079 \quad\quad & 1.029 \quad\quad & 1.019 \quad\quad & 1.018 & \quad\quad 2.9  & \quad\quad 0.921 $\leqslant t \leqslant$ 0.932 \\
		$y_c|_{t=3}$ & 1.094 \quad\quad & 1.088 \quad\quad & 1.087 \quad\quad & 1.0879 & \quad\quad 1.7  & \quad\quad 1.081 $\pm$ 0.001 \\
		\bottomrule
	\end{tabular}
	\caption{Numerical results at different resolutions for the rising bubble test case. Comparison with the reference results from Hysing et al. \cite{hysing2009quantitative}.}
	\label{bubbleRisingTableValues}
	
\end{table}

Table \ref{bubbleRisingTableValues} shows the sensitivity of the results to the refinement level and the approximate order of convergence of the main measures. The numerical results converge towards a solution (with orders from $\sim$1 to $\sim$3) which is slightly different from the one obtained by Hysing et al. \cite{hysing2009quantitative}: while the final bubble position $y_c|_{t=3}$ and the maximum velocity $v_{c,max}$ are very similar, small deviations are present in the other quantities. Similar discrepancies from the reference work have been noticed by  Coquerelle et al. \cite{coquerelle2016fourth}, especially regarding the minimum in the circularity profile. Nevertheless, we believe the results are satisfactory and in line with the reference work, especially considering the multiple differences in the numerical codes (advection method, discretization approach, surface tension treatment, flow solvers etc.) which makes a complete adherence of the results very difficult.

\section{Contact angles}
\begin{figure}
	\centering
	{\includegraphics[width=.45\textwidth,height=0.3\textheight]{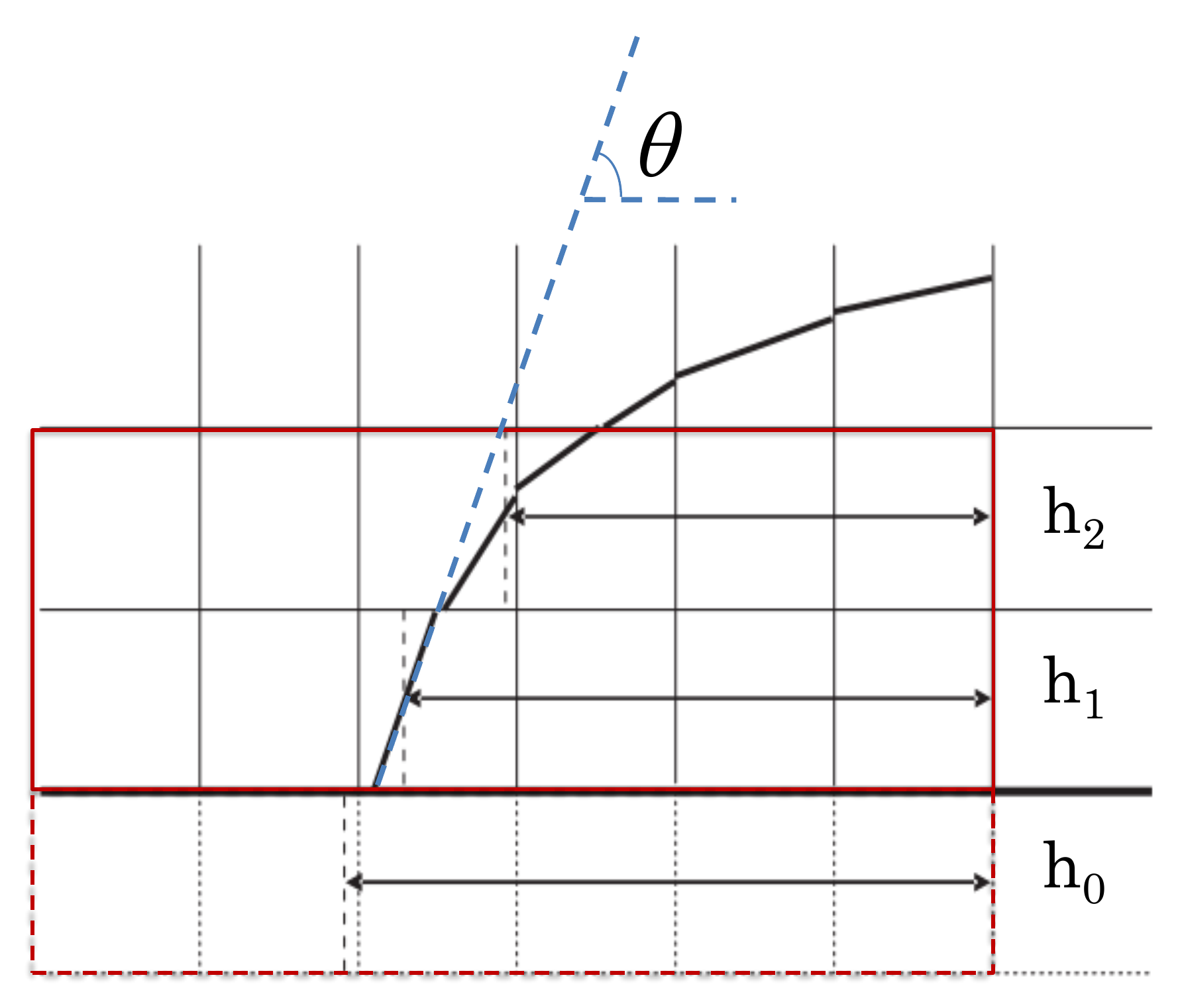}}
	\caption{Height functions implementation of the static contact angle.}
	\label{contactAngle}
\end{figure}

Referring to Figure \ref{contactAngle}, the contact angle is defined as the angle given by the intersection of the gas-liquid interface with the solid line. This defines a boundary condition for the VOF function, in which at the boundary cell the interface normal must satisfy the contact angle condition. As reported in the introduction, working with interface normals can be dangerous due to the large numerical errors arising from differentiating a discontinuous function. In this work the contact angle boundary condition is implemented as an extension of the Height Functions method for curvature calculation, following the work of Afkhami et al. \cite{afkhami2008height}. 
The Height Functions method requires three heights $h_0, h_1$ and $h_2$ for the curvature calculation. At the boundary cell only two heights are available $h_1, h_2$, requiring the construction of a "ghost" height $h_0$ which includes the contact angle condition:

\begin{equation}
	h_0=h_1+ \Delta x tan\theta 
	\label{contactAngleCondition} 
\end{equation}

It is very easy to implement this in $\texttt{OpenFOAM}^{\textregistered}$:

\begin{enumerate}
	\item In pre-processing, if a cell belongs to the boundary a complete stencil cannot be built. Fill the missing column (ghost) with a representative label (e.g. -1);
	\item During runtime, if a stencil contains the value -1 it means that the relative cell is at the boundary. Calculate the two available heights as usual, while the remaining one as:
		\begin{verbatim}
		heights[0] = heights[1]+Deltax*Foam::tan(theta);
		\end{verbatim}
	where \texttt{theta} is the assigned contact angle;	
	\item Once the three heights are available, the curvature at the boundary cell can be easily calculated (Equation \ref{curvaturecalculation}).

\end{enumerate}

\subsection{2D Sessile droplets}
A semicircular 2D liquid droplet ($D=1$ mm, $D/\Delta x =100$) is placed on a horizontal surface (sessile drop). The contact angle condition  (Equation \ref{contactAngleCondition}) is enforced at the boundary cell curvature calculation. The total simulation time is set to $t=0.1$ s. The curvature gradient at the boundary induces a flow, deforming the interface until a steady-state condition is reached (in which the curvature is uniform). The resulting circular cap is analyzed by means of  apparent radius $R$, height $e$, length $L$ and residual spurious currents. The analytical solution is available \cite{dupont2010numerical}:

\begin{equation}
	R=R_0\sqrt{\frac{\pi}{2\left(\theta-sin\theta cos\theta\right)}}
\end{equation}

\begin{equation}
	e=R\left(1-cos\theta\right)
\end{equation}

\begin{equation}
L=2Rsin\theta
\end{equation}

\begin{figure}
	\centering
	{\includegraphics[width=.4\textwidth,height=0.25\textheight]{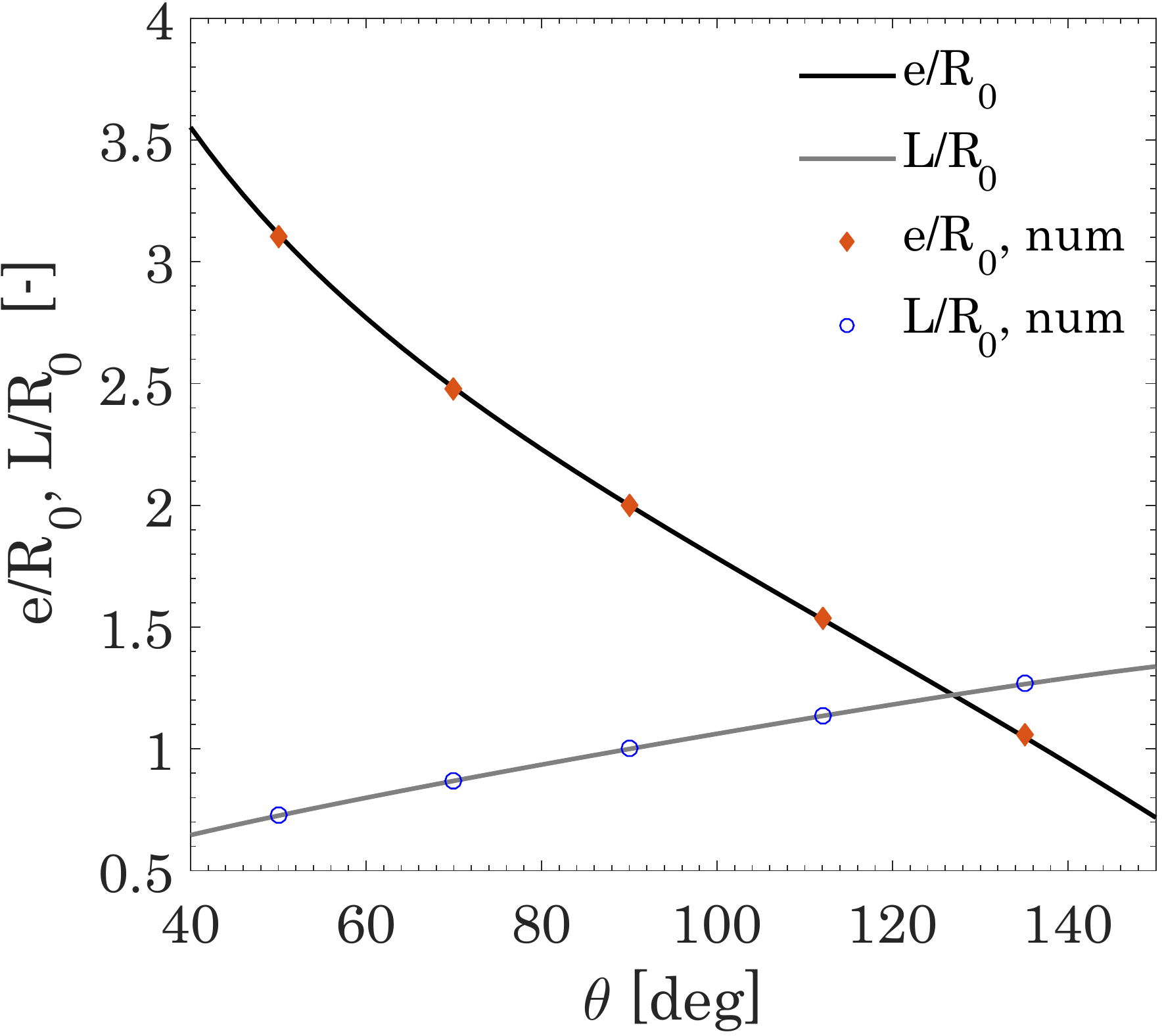}}\quad\quad
	\caption{Comparison between analytical and numerical solutions for static droplets contact angles in terms of $e/R_0$ (black line) and $L/R_0$ (gray line).}
	\label{contactAngleComparison}
\end{figure}

\begin{figure}
	\centering
	{\includegraphics[width=.9\textwidth,height=0.35\textheight]{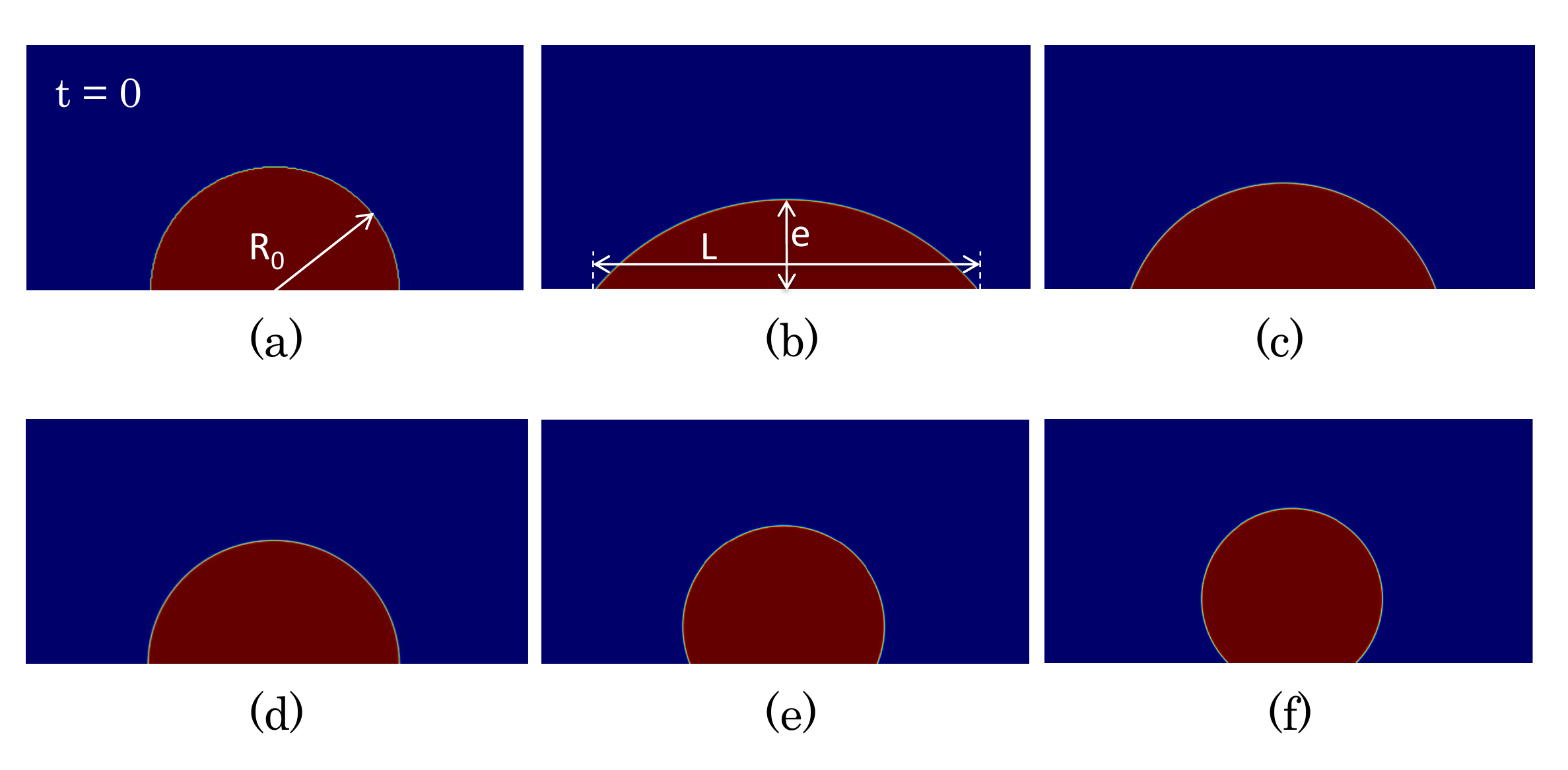}}
	\caption{$\alpha$ initial condition for static contact angles analysis (a). Steady-state droplet shapes for $\theta=50^\circ$ (b), $\theta=70^\circ$ (c), $\theta=90^\circ$ (d), $\theta=112^\circ$ (e) and $\theta=135^\circ$ (f).} 
	\label{staticAngles}
\end{figure}

The qualitative results (VOF function) are reported in Figure \ref{staticAngles} for different static contact angles $\theta$. The comparison with the analytical solution is very good for both for the $e, L$ parameters (Figure \ref{contactAngleComparison}). The residual spurious currents and the relative capillary number are reported in Table \ref{tableSpuriousCurrentsContactAngles}. Compared to the results for the equilibrium of a circular droplet (Table \ref{spuriousCurrentsTable}) we notice an increase in the capillary number, which however remains under $10^{-6}$ for all the cases. This has also been reported by Dupont et al. \cite{dupont2010numerical}.

\begin{table}
	\centering
	
	\begin{tabular}{lll}
		\toprule
		
		$\theta~ [deg]$ \quad\quad & $|\textbf{v}|_{max}$ [m/s] \quad\quad\quad & Ca \\
		\midrule
		50 & $2.2\cdot10^{-4}$  &   $3.14\cdot10^{-7}$    \\
		70 & $2.4\cdot10^{-4}$  &   $3.42\cdot10^{-7}$     \\
		90 & $2.9\cdot10^{-4}$  &   $4.14\cdot10^{-7}$     \\
		112 & $3.2\cdot10^{-4}$ &   $4.57\cdot10^{-7}$      \\
		135 & $3.8\cdot10^{-4}$ &   $5.42\cdot10^{-7}$      \\
		
		\bottomrule
		
	\end{tabular}	
	
	\caption{Residual spurious currents for 2D static sessile droplets (Figure \ref{staticAngles}).}
	\label{tableSpuriousCurrentsContactAngles}
\end{table}

\subsection{Droplet suspension on a vertical fiber}
\begin{figure}
  	\centering
  	{\includegraphics[width=.6\textwidth,height=0.35\textheight]{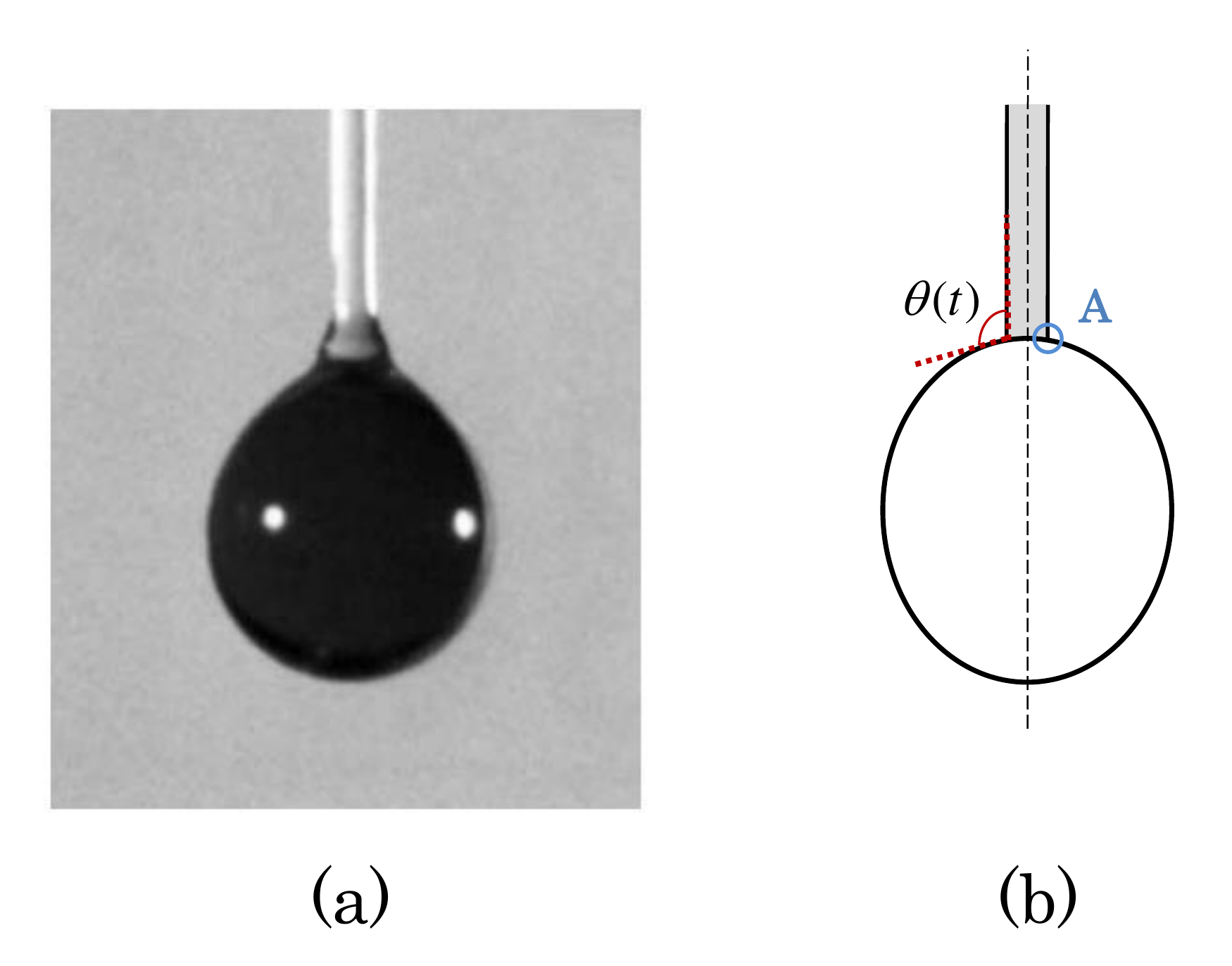}}
  	\caption{ Water droplet suspended on a vertical fiber \cite{walton2004evaporation} in Figure (a). Sketch of the same droplet highlighting the dynamic contact angle $\theta$ and the contact point $A$ in Figure (b).}
  	\label{suspendedDroplet}
  \end{figure}

A very stringent test is the suspension of small droplets against the gravity field on a thin vertical fiber, as usually done in many experimental works (Figure \ref{suspendedDroplet} a). In this case the contact angle with the solid fiber cannot be fixed, since it depends on multiple factors such as the droplet weight, the fiber geometry, the surface tension etc. For a 2D droplet having a radius $R$ and length $L$ it is possible to derive the equilibrium contact angle with a balance of forces:

\begin{equation}
	\rho_L\pi R^2L=2L\sigma cos\theta
\end{equation}

\begin{equation}
cos\theta = \frac{\rho_L\pi R^2}{2\sigma}
\label{static2Dcontactangle}
\end{equation}

  As described in the previous section (Figure \ref{contactAngle}), at the boundary cell we need to find the value of the "ghost height" $h_0$ that satisfies the contact angle condition (Equation \ref{contactAngleCondition}). Imposing Equation \ref{static2Dcontactangle} as a boundary condition is not sufficient, since it is an equilibrium solution which does not account for transient situations (e.g. convective flow around the droplet).  In fact, the numerical model should predict the contact angle dynamics depending on the local conditions (droplet weight and shape, surface tension, velocity etc.) and automatically find a steady-state situation independently of Equation \ref{static2Dcontactangle}. 
  Referring to Figure \ref{suspendedDroplet} (b), the numerical strategy consists in finding the value of the "ghost height" $h_0^*$ at the boundary cell (point $A$) \textit{that cancels the local momentum balance} \cite{dupont2010numerical}. For each boundary cell cut by the interface:

  \begin{enumerate}
    \item   We derive the pressure jump $H_f^*$ which provides $\phi_f=0$. The reconstructed face flux is given by Equations \ref{reconstructedVelocityGFM_wet} and \ref{reconstructedVelocityGFM_dry}:
      \begin{itemize}
      	\item if the owner is wet:

      	\begin{equation}
      	H_f^* = p_N-p_O-\frac{\frac{\textbf{H}\left(\textbf{v}_N\right)_f}{a_{P,f}}\cdot\textbf{S}_f}{\left(\frac{1}{a_{P}}\right)_f\frac{|\textbf{S}_f|}{|\textbf{d}|}\frac{\beta^L\beta^G}{\beta_w}}
      	\label{jumpForZeroVelocity_wet}
      	\end{equation}

      	\item if the owner is dry:
      	\begin{equation}
      	H_f^* = p_N-p_O-\frac{\frac{\textbf{H}\left(\textbf{v}_N\right)_f}{a_{P,f}}\cdot\textbf{S}_f}{\left(\frac{1}{a_{P}}\right)_f\frac{|\textbf{S}_f|}{|\textbf{d}|}\frac{\beta^L\beta^G}{\beta_d}}
      	\label{jumpForZeroVelocity_dry}
      	\end{equation}	
      	
      \end{itemize} 
      
      In $\texttt{OpenFOAM}^{\textregistered}$:
      
      \begin{verbatim}
      Jump.ref()[facei]= p.ref()[neighbour[facei]]-p.ref()[owner[facei]]
      -phiHbyA.ref()[facei]/laplacian.upper()[facei];
      \end{verbatim}  
      
      where the surfaceScalarField \texttt{phiHbyA} $=\frac{\textbf{H}\left(\textbf{v}_N\right)_f}{a_{P,f}}\cdot\textbf{S}_f$;
      
  \item  We calculate the value of the face-centered curvature $\kappa^*$ which provides an interfacial jump equal to $H_f^*$ (Equation \ref{pressurejump}):
  
  \begin{equation}
  \kappa^* =- \frac{H_f^*+\left(\rho_L-\rho_G\right)\textbf{g}\cdot\textbf{x}_f}{\sigma}
  \label{curvatureForZeroVelocity}
  \end{equation}
  
  In $\texttt{OpenFOAM}^{\textregistered}$:
  
    \begin{verbatim}
    if (alpha1.ref()[owner[facei]] > 0.5)
    double curvature = (Jump.ref()[facei]-
    (rhoL.ref()[owner[facei]]-rhoG.ref()[neighbour[facei]])
    *(g.value() & interfacePosition.ref()[facei]))/ 
    surfaceTension.ref()[facei];
    
    else
    double curvature = -(Jump.ref()[facei]+
    (rhoL.ref()[owner[facei]]-rhoG.ref()[neighbour[facei]])
    *(g.value() & interfacePosition.ref()[facei]))/ 
    surfaceTension.ref()[facei];
    \end{verbatim}  

The face-centered value \texttt{curvature} is then assigned to both the owner and the neighbour of the interfacial face \texttt{facei};

  \item  The value of the "ghost" height $h_0^*$ is obtained by solving the following non-linear algebraic equation:
  
  \begin{equation}
  \kappa^*-\frac{\frac{h_2-2h_1+h^*_0}{\Delta x^2}}{\left(1+\left(\frac{h_2-h^*_0}{2\Delta x}\right)^2\right)^{3/2}}=0
  \label{heightForZeroVelocity}
  \end{equation}
  
  \item   Finally, the contact angle can be easily derived as:
  
  \begin{equation}
  \theta^* = atan\left(\frac{h^*_0-h_1}{\Delta x}\right)
  \label{thetaForZeroVelocity}
  \end{equation}
  
  It is important to specify that only Equations \ref{jumpForZeroVelocity_wet}, \ref{jumpForZeroVelocity_dry}  are necessary for the method to work, since the jump $H_f^*$ the only information actually used by the GFM method. Equations \ref{curvatureForZeroVelocity}, \ref{heightForZeroVelocity} and \ref{thetaForZeroVelocity} are however useful to know the value of $\theta^*$ that cancels the momentum balance, which can be compared with the theoretical prediction.
  
  \end{enumerate}

 This is an iterative procedure, since the interfacial jump $H_f^*$  is derived explicitly, adopting the pressure field of the previous time step (Equations \ref{jumpForZeroVelocity_wet} and \ref{jumpForZeroVelocity_dry}). Multiple iterations of the Poisson equation are needed to find the correct pressure field that cancels the local momentum balance. However, we noticed that a single iteration is actually enough for the procedure to work. The actual face flux $\phi_f$ is slightly different from zero, but it tends to cancel over time. This allows to save significant computational time, avoiding the iterations.

      \begin{figure}
      	\centering
      	{\includegraphics[width=.66\textwidth,height=0.45\textheight]{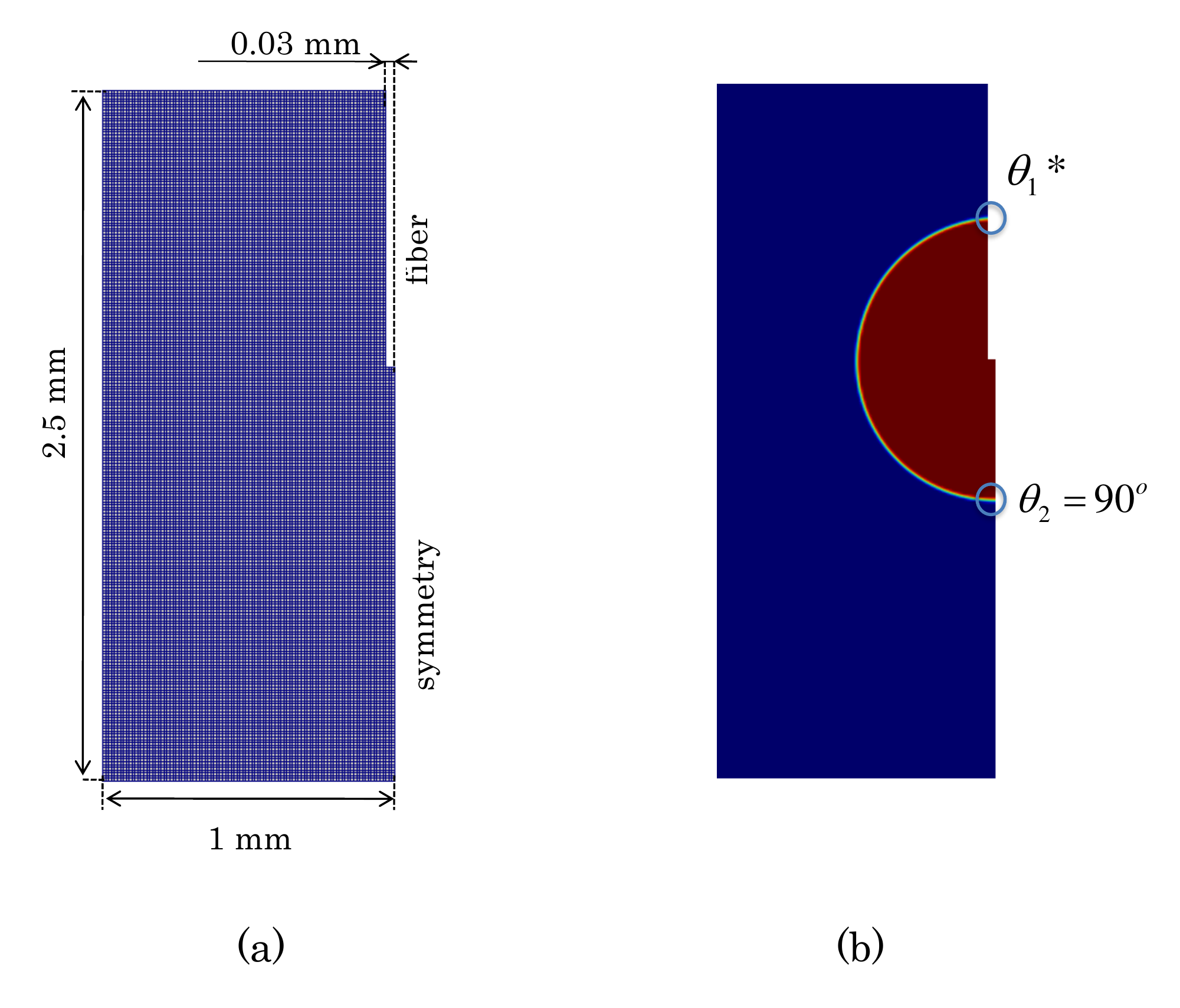}}
      	\caption{ 2D computational mesh used for suspended droplets simulations (a). VOF initial condition (b) for the droplet, with boundary conditions for contact angles.}
      	\label{meshSuspendedDroplet}
      \end{figure}

\subsubsection{2D droplet suspension}

\begin{figure}
	\centering
	\subfloat[]
	{\includegraphics[width=.4\textwidth,height=0.27\textheight]{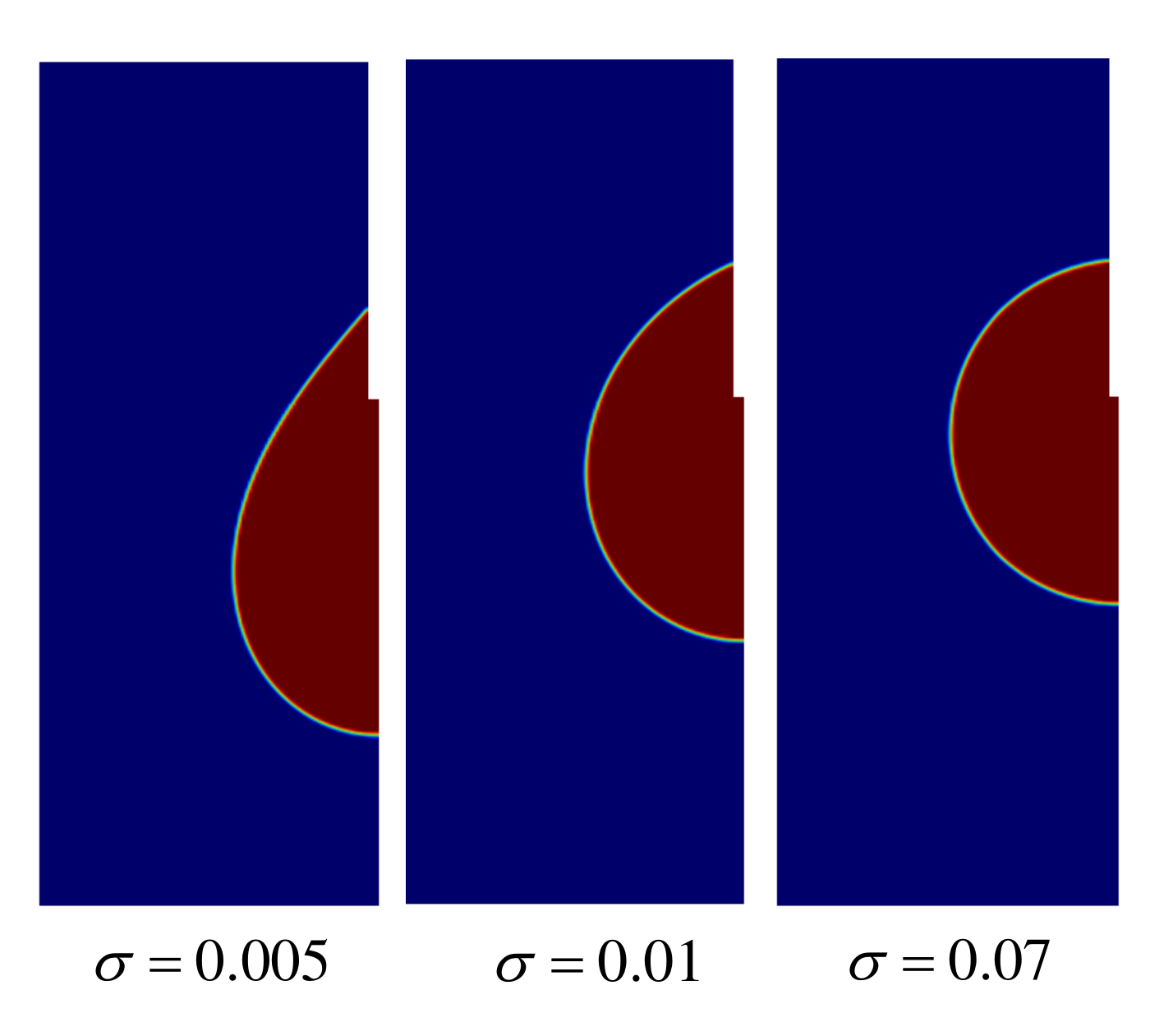}}\quad\quad
	\subfloat[]
	{\includegraphics[width=.43\textwidth,height=0.25\textheight]{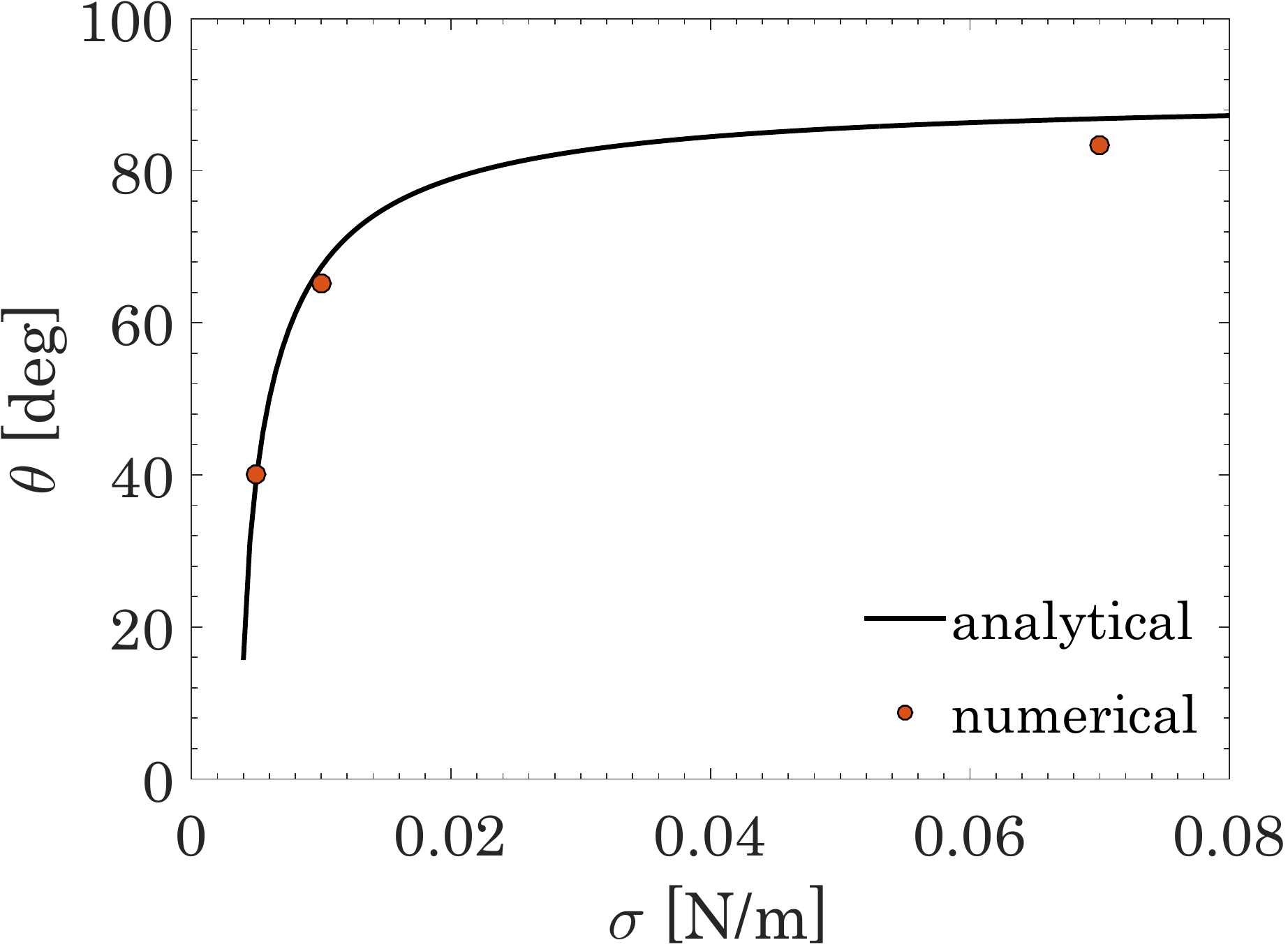}}
	\caption{Comparison between analytical and numerical solutions for 2D suspended droplets contact angles at different surface tension values. Steady-state condition.}
	\label{2DsuspendedDroplet}
\end{figure}

To validate the method, a 2D liquid droplet ($D= 1$ mm, $\rho_L=1000$ $kg/m^3$) is placed on a thin vertical fiber as shown in Figure \ref{meshSuspendedDroplet}. Only half of the droplet is modeled due to symmetry conditions. At the symmetry boundary the contact angle is imposed equal to $\theta_2=90^\circ$ (using the method reported for the sessile droplets). The fiber boundary is a wall, where a  \texttt{noSlip} condition apply for velocity. On this boundary, the contact angle $\theta_1^*$ is  calculated as jjust described in this section. The other boundaries are open. Three different surface tension values have been adopted ($\sigma=0.005, 0.01, 0.07$ N/m) and the numerical results are compared with the analytical solution given by Equation \ref{static2Dcontactangle}. The results are reported in Figure \ref{2DsuspendedDroplet}.

As expected, lower surface tensions provide a more deformed droplet shape. The contact point is not perfectly fixed due to the choice of avoid multiple iteration within the same time step. This is especially true for the lowest surface tension case ($\sigma=0.005$ N/m), because of the longer dynamics needed to reach the steady-state condition. The agreement with the theoretical prediction is good, with a slight deviation for   $\sigma=0.07$ N/m. In Table \ref{tableSpuriousCurrentsSuspendedDroplets} the residual spurious currents are reported, showing results comparable to the sessile droplets cases.

\begin{table}
	\centering
	
	\begin{tabular}{lll}
		\toprule
		
		$\sigma$ [N/m] \quad\quad\quad & $|\textbf{v}|_{max}$ [m/s] \quad\quad\quad & Ca \\
		\midrule
		$0.005$ & $6.1\cdot10^{-4}$  &   $1.2\cdot10^{-5}$    \\
		$0.01$ & $3.4\cdot10^{-4}$  &   $3.4\cdot10^{-6}$     \\
		$0.07$ & $4.3\cdot10^{-5}$  &   $6.1\cdot10^{-7}$     \\

		\bottomrule
		
	\end{tabular}	
	
	\caption{Residual spurious currents for 2D suspended droplets (Figure \ref{2DsuspendedDroplet}).}
	\label{tableSpuriousCurrentsSuspendedDroplets}
\end{table}

\subsubsection{2D axisymmetric droplet suspension}

The axisymmetric geometry is often used to save computational time with respect to full 3D simulations. In particular, suspended droplets usually exhibit  axisymmetric geometry if a vertical support fiber is adopted. From the theoretical and numerical points of view, the only difference is that a second curvature must be accounted for (perpendicular to the interface and to the plane containing the droplet). This is easily calculated from the 2D heights already available \cite{pozrikidis1997introduction}:

\begin{equation}
	\kappa_2=\frac{1}{|\textbf{x}|\sqrt{1+h'^2}}
\end{equation}

where $\textbf{x}$ is the distance of the interface perpendicularly to the symmetry axis. The interface position is defined by the exact location of the height, which must be used to compute $|\textbf{x}|$. Therefore:

\begin{itemize} 
	\item if a vertical stencil is adopted for the cell $(i,j)$:
\begin{equation}
|\textbf{x}| = x_{i,j}
\end{equation}

	\item if a horizontal stencil is adopted for the cell $(i,j)$:
\begin{equation}
|\textbf{x}| = x_{i,j}-3.5\Delta x + h_i
\end{equation}	

where $x_{i,j}$ is the $x$ coordinate of the interfacial cell center.

\end{itemize}

At the cell $(i,j)$ the total curvature $\kappa_{i,j}$ is then the sum of the two principal curvatures:

\begin{equation}
	\kappa_{i,j} = \frac{h''_{i,j}}{\left(1+h'^2_{i,j}\right)^{3/2}} + \frac{1}{|\textbf{x}|\sqrt{1+h'^2_{i,j}}}
\end{equation}

This second curvature represents an additional contribution to the pressure jump, which is then handled by the GFM. All the treatment of the contact angles remains the same. The only difference is that Equation \ref{heightForZeroVelocity} becomes:

\begin{equation}
  \kappa^*-\frac{\frac{h_2-2h_1+h^*_0}{\Delta x^2}}{\left(1+\left(\frac{h_2-h^*_0}{2\Delta x}\right)^2\right)^{3/2}}-\frac{1}{|\textbf{x}|\sqrt{1+\left(\frac{h_2-h^*_0}{2\Delta x}\right)^2}}=0
\end{equation}

in order to account for the second curvature. The theoretical value of the contact angle for 2D axisymmetric droplets (or 3D ones) can be easily derived by a balance of forces on the suspended droplet:

\begin{equation}
\frac{4}{3}\pi R^3 \rho_L=\sigma \pi d_F cos\theta
\end{equation}

where $d_F$ is the fiber diameter. We obtain:

\begin{equation}
	cos\theta=\frac{4\rho_LgR^3}{3\sigma d_F}
\end{equation}

\begin{figure}
	\centering
	\subfloat[]
	{\includegraphics[width=.4\textwidth,height=0.27\textheight]{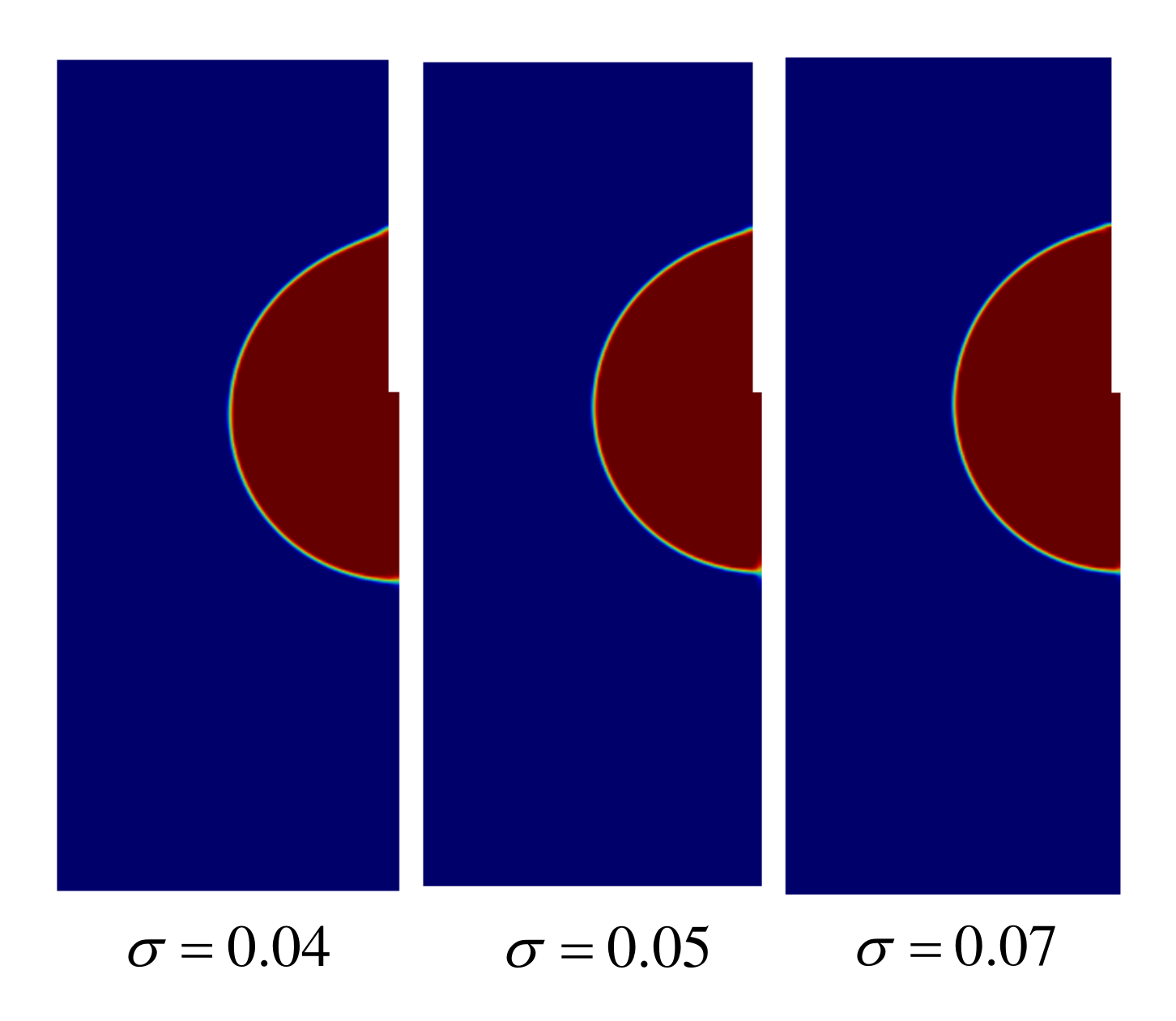}}\quad\quad
	\subfloat[]
	{\includegraphics[width=.43\textwidth,height=0.25\textheight]{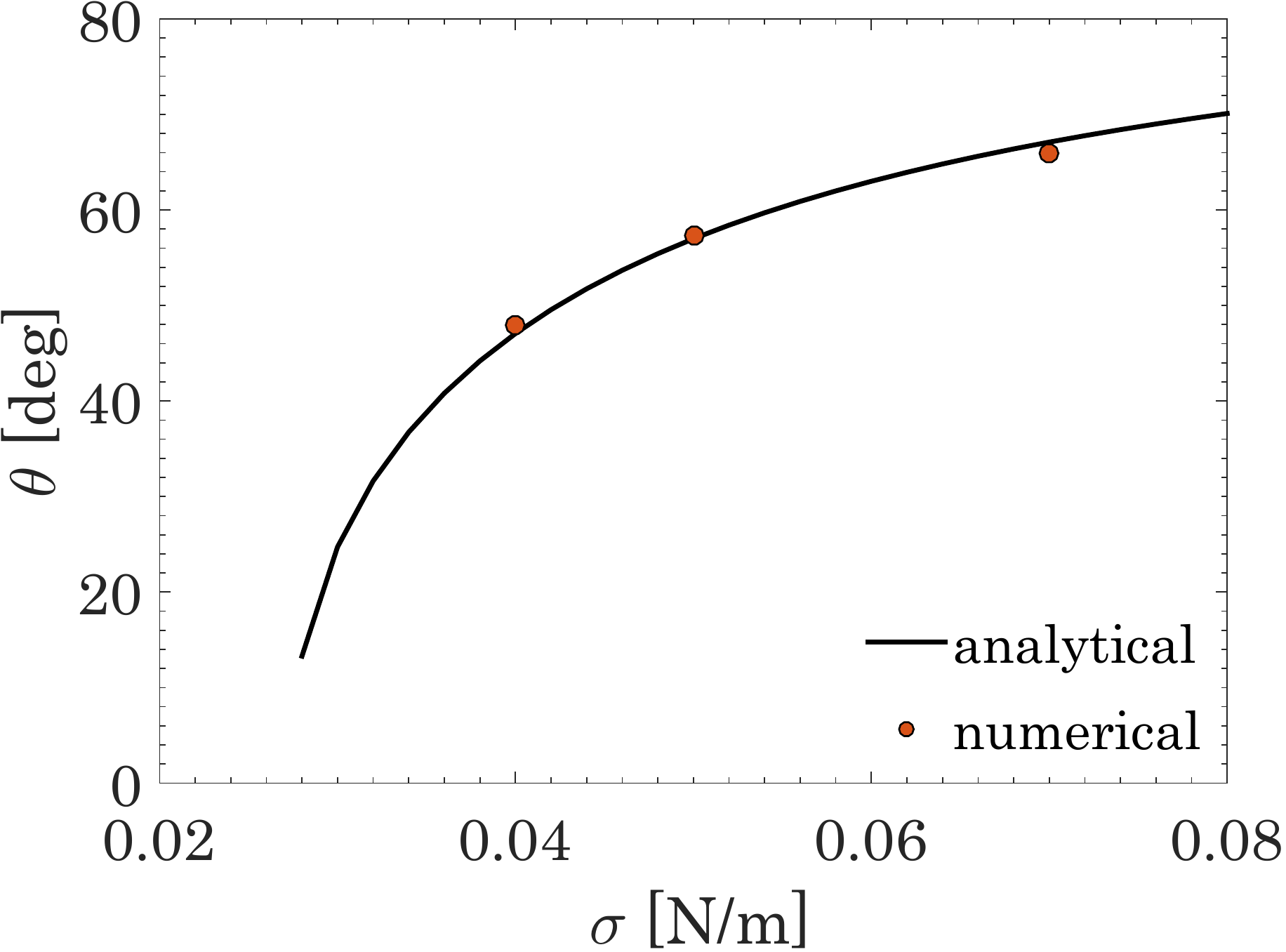}}
	\caption{Comparison between analytical and numerical solutions for 2D axisymmetric suspended droplets contact angles at different surface tension values. Steady state condition.}
	\label{3DsuspendedDroplet}
\end{figure}

We repeat the same test done for 2D droplets, suspending 2D axisymmetric droplets on a vertical fiber for different surface tension values ($\sigma=0.04, 0.05, 0.07$ N/m).  The computational mesh in Figure \ref{meshSuspendedDroplet} (a) remains the same, but the symmetry plane is collapsed into a symmetry axis. In this configuration the new domain is a slice of a cylinder with a thin vertical cylindrical fiber on the axis on which the droplet is suspended. The results are reported in Figure \ref{3DsuspendedDroplet} and Table \ref{tableSpuriousCurrentsAxiDroplets}.

The droplet deformation is now much less evident and only localized close to the contact point. The agreement with the theoretical value is satisfactory for all the three cases. The residual spurious currents are comparable to what obtained for the 2D droplet suspension. \\

\begin{table}
	\centering
	
	\begin{tabular}{lll}
		\toprule
		
		$\sigma$ [N/m] \quad\quad\quad & $|\textbf{v}|_{max}$ [m/s] \quad\quad\quad & Ca \\
		\midrule
		$0.04$ & $1.1\cdot10^{-4}$  &   $2.7\cdot10^{-7}$    \\
		$0.05$ & $6.5\cdot10^{-5}$  &   $1.3\cdot10^{-7}$     \\
		$0.07$ & $6.6\cdot10^{-4}$  &   $9.4\cdot10^{-7}$     \\		
		
		\bottomrule
		
	\end{tabular}	
	
	\caption{Residual spurious currents for 2D axisymmetric droplets (Figure \ref{3DsuspendedDroplet}).}
	\label{tableSpuriousCurrentsAxiDroplets}
\end{table}

\section{Conclusions}
In this work we presented an accurate methodology for the modeling of surface tension driven flows in the $\texttt{OpenFOAM}^{\textregistered}$ framework. The Ghost Fluid Method (GFM) is implemented for the numerical discretization of the pressure equation, including the jump due to surface tension and gravity, while the cell-centered values of the curvature are evaluated  adopting   Height Functions. The methodology shows a significant reduction of spurious currents for a 2D stationary droplet, close to machine accuracy. The curvature and the interface shape show second order convergence towards the exact solution, with a weak dependence on the Laplace number. The capillary number does not always converge with mesh refinement, especially at high Laplace numbers, probably due to the sensitivity of Height Functions to errors in the interface advection. This is confirmed by the translating droplet test case, which clearly indicates the interface advection as the main spurious currents magnifier rather than the curvature calculation errors, in line to what reported by other authors. \\
The methodology is also tested to analyze (i) capillary oscillations in a perturbed liquid droplet, exhibiting first order convergence on the frequency value and (ii) a bubble rising in a dense fluid, showing excellent agreement with the reference numerical solution in terms of bubble position, rising velocity and interface shape.   Finally, the Height Functions method is extended for the modeling of contact angles, with applications to sessile droplets and suspended droplets, both in 2D and 2D axisymmetric configurations. The agreement with the theoretical solution is satisfactory.\\ 

Future works will focus on the extension of the methodology to variable fluid properties and the implementation of phase-change.

\section*{Acknowledgments}
We acknowledge the CINECA award under the ISCRA initiative, for the availability of high performance computing resources and support.

	\printnomenclature	 	 
		 	 
\section*{References}
\bibliographystyle{elsarticle-num} 
\bibliography{bibliografia}

\begin{thebibliography}{10}
\expandafter\ifx\csname url\endcsname\relax
  \def\url#1{\texttt{#1}}\fi
\expandafter\ifx\csname urlprefix\endcsname\relax\def\urlprefix{URL }\fi
\expandafter\ifx\csname href\endcsname\relax
  \def\href#1#2{#2} \def\path#1{#1}\fi

\bibitem{abramzon1989droplet}
B.~Abramzon, W.~Sirignano, Droplet vaporization model for spray combustion
  calculations, International Journal of Heat and Mass Transfer 32~(9) (1989)
  1605--1618.

\bibitem{sirignano1999fluid}
W.~A. Sirignano, Fluid dynamics and transport of droplets and sprays, Cambridge
  university press, 1999.

\bibitem{juric1998computations}
D.~Juric, G.~Tryggvason, Computations of boiling flows, International journal
  of multiphase flow 24~(3) (1998) 387--410.

\bibitem{delale2005direct}
C.~Delale, S.~Nas, G.~Tryggvason, Direct numerical simulations of shock
  propagation in bubbly liquids, Physics of Fluids 17~(12) (2005) 121705.

\bibitem{hirt1981volume}
C.~W. Hirt, B.~D. Nichols, Volume of fluid ({VOF}) method for the dynamics of
  free boundaries, Journal of computational physics 39~(1) (1981) 201--225.

\bibitem{sussman1998improved}
M.~Sussman, E.~Fatemi, P.~Smereka, S.~Osher, An improved level set method for
  incompressible two-phase flows, Computers \& Fluids 27~(5-6) (1998) 663--680.

\bibitem{tryggvason2001front}
G.~Tryggvason, B.~Bunner, A.~Esmaeeli, D.~Juric, N.~Al-Rawahi, W.~Tauber,
  J.~Han, S.~Nas, Y.-J. Jan, A front-tracking method for the computations of
  multiphase flow, Journal of Computational Physics 169~(2) (2001) 708--759.

\bibitem{francois2006balanced}
M.~M. Francois, S.~J. Cummins, E.~D. Dendy, D.~B. Kothe, J.~M. Sicilian, M.~W.
  Williams, A balanced-force algorithm for continuous and sharp interfacial
  surface tension models within a volume tracking framework, Journal of
  Computational Physics 213~(1) (2006) 141--173.

\bibitem{popinet2009accurate}
S.~Popinet, An accurate adaptive solver for surface-tension-driven interfacial
  flows, Journal of Computational Physics 228~(16) (2009) 5838--5866.

\bibitem{popinet2018numerical}
S.~Popinet, Numerical models of surface tension, Annual Review of Fluid
  Mechanics 50 (2018) 49--75.

\bibitem{brackbill1992continuum}
J.~Brackbill, D.~B. Kothe, C.~Zemach, A continuum method for modeling surface
  tension, Journal of computational physics 100~(2) (1992) 335--354.

\bibitem{fedkiw1999non}
R.~P. Fedkiw, T.~Aslam, B.~Merriman, S.~Osher, A non-oscillatory eulerian
  approach to interfaces in multimaterial flows (the ghost fluid method),
  Journal of computational physics 152~(2) (1999) 457--492.

\bibitem{fedkiw1999ghost}
R.~P. Fedkiw, T.~Aslam, S.~Xu, The ghost fluid method for deflagration and
  detonation discontinuities, Journal of Computational Physics 154~(2) (1999)
  393--427.

\bibitem{liu2000boundary}
X.-D. Liu, R.~P. Fedkiw, M.~Kang, A boundary condition capturing method for
  poisson's equation on irregular domains, Journal of computational Physics
  160~(1) (2000) 151--178.

\bibitem{olsson2005conservative}
E.~Olsson, G.~Kreiss, A conservative level set method for two phase flow,
  Journal of computational physics 210~(1) (2005) 225--246.

\bibitem{desjardins2008accurate}
O.~Desjardins, V.~Moureau, H.~Pitsch, An accurate conservative level set/ghost
  fluid method for simulating turbulent atomization, Journal of Computational
  Physics 227~(18) (2008) 8395--8416.

\bibitem{bo2014volume}
W.~Bo, J.~W. Grove, A volume of fluid method based ghost fluid method for
  compressible multi-fluid flows, Computers \& Fluids 90 (2014) 113--122.

\bibitem{vukvcevic2017implementation}
V.~Vuk{\v{c}}evi{\'c}, H.~Jasak, I.~Gatin, Implementation of the ghost fluid
  method for free surface flows in polyhedral finite volume framework,
  Computers \& Fluids 153 (2017) 1--19.

\bibitem{lalanne2015computation}
B.~Lalanne, L.~R. Villegas, S.~Tanguy, F.~Risso, On the computation of viscous
  terms for incompressible two-phase flows with level set/ghost fluid method,
  Journal of Computational Physics 301 (2015) 289--307.

\bibitem{kang2000boundary}
M.~Kang, R.~P. Fedkiw, X.-D. Liu, A boundary condition capturing method for
  multiphase incompressible flow, Journal of Scientific Computing 15~(3) (2000)
  323--360.

\bibitem{sussman2007sharp}
M.~Sussman, K.~M. Smith, M.~Y. Hussaini, M.~Ohta, R.~Zhi-Wei, A sharp interface
  method for incompressible two-phase flows, Journal of computational physics
  221~(2) (2007) 469--505.

\bibitem{cummins2005estimating}
S.~J. Cummins, M.~M. Francois, D.~B. Kothe, Estimating curvature from volume
  fractions, Computers \& structures 83~(6-7) (2005) 425--434.

\bibitem{williams1998accuracy}
M.~Williams, D.~Kothe, E.~Puckett, Accuracy and convergence of continuum
  surface tension models, Fluid dynamics at interfaces (1998) 294--305.

\bibitem{tryggvason2011direct}
G.~Tryggvason, R.~Scardovelli, S.~Zaleski, Direct numerical simulations of
  gas--liquid multiphase flows, Cambridge University Press, 2011.

\bibitem{bornia2011properties}
G.~Bornia, A.~Cervone, S.~Manservisi, R.~Scardovelli, S.~Zaleski, On the
  properties and limitations of the height function method in two-dimensional
  cartesian geometry, Journal of Computational Physics 230~(4) (2011) 851--862.

\bibitem{chiodi2017reformulation}
R.~Chiodi, O.~Desjardins, A reformulation of the conservative level set
  reinitialization equation for accurate and robust simulation of complex
  multiphase flows, Journal of Computational Physics 343 (2017) 186--200.

\bibitem{marchandise2007stabilized}
E.~Marchandise, P.~Geuzaine, N.~Chevaugeon, J.-F. Remacle, A stabilized finite
  element method using a discontinuous level set approach for the computation
  of bubble dynamics, Journal of Computational Physics 225~(1) (2007) 949--974.

\bibitem{afkhami2008height}
S.~Afkhami, M.~Bussmann, Height functions for applying contact angles to 2d vof
  simulations, International journal for numerical methods in fluids 57~(4)
  (2008) 453--472.

\bibitem{huh1971hydrodynamic}
C.~Huh, L.~Scriven, Hydrodynamic model of steady movement of a
  solid/liquid/fluid contact line, Journal of colloid and interface science
  35~(1) (1971) 85--101.

\bibitem{afkhami2009mesh}
S.~Afkhami, S.~Zaleski, M.~Bussmann, A mesh-dependent model for applying
  dynamic contact angles to vof simulations, Journal of computational physics
  228~(15) (2009) 5370--5389.

\bibitem{wang20183d}
S.~Wang, O.~Desjardins, 3d numerical study of large-scale two-phase flows with
  contact lines and application to drop detachment from a horizontal fiber,
  International Journal of Multiphase Flow 101 (2018) 35--46.

\bibitem{renardy2001numerical}
M.~Renardy, Y.~Renardy, J.~Li, Numerical simulation of moving contact line
  problems using a volume-of-fluid method, Journal of Computational Physics
  171~(1) (2001) 243--263.

\bibitem{aboukhedr2018simulation}
M.~Aboukhedr, A.~Georgoulas, M.~Marengo, M.~Gavaises, K.~Vogiatzaki, Simulation
  of micro-flow dynamics at low capillary numbers using adaptive interface
  compression, Computers \& Fluids 165 (2018) 13--32.

\bibitem{raeini2012modelling}
A.~Q. Raeini, M.~J. Blunt, B.~Bijeljic, Modelling two-phase flow in porous
  media at the pore scale using the volume-of-fluid method, Journal of
  Computational Physics 231~(17) (2012) 5653--5668.

\bibitem{ferrari2017flexible}
A.~Ferrari, M.~Magnini, J.~R. Thome, A flexible coupled level set and volume of
  fluid (flexclv) method to simulate microscale two-phase flow in non-uniform
  and unstructured meshes, International Journal of Multiphase Flow 91 (2017)
  276--295.

\bibitem{albadawi2013influence}
A.~Albadawi, D.~Donoghue, A.~Robinson, D.~Murray, Y.~Delaur{\'e}, Influence of
  surface tension implementation in volume of fluid and coupled volume of fluid
  with level set methods for bubble growth and detachment, International
  Journal of Multiphase Flow 53 (2013) 11--28.

\bibitem{deshpande2012evaluating}
S.~S. Deshpande, L.~Anumolu, M.~F. Trujillo, Evaluating the performance of the
  two-phase flow solver interfoam, Computational science \& discovery 5~(1)
  (2012) 014016.

\bibitem{jamshidi2019suitability}
F.~Jamshidi, H.~Heimel, M.~Hasert, X.~Cai, O.~Deutschmann, H.~Marschall,
  M.~W{\"o}rner, On suitability of phase-field and algebraic volume-of-fluid
  openfoam{\textregistered} solvers for gas--liquid microfluidic applications,
  Computer Physics Communications 236 (2019) 72--85.

\bibitem{saufi2019dropletsmoke++}
A.~Saufi, A.~Frassoldati, T.~Faravelli, A.~Cuoci, Droplet{SMOKE}++: A
  comprehensive multiphase {CFD} framework for the evaporation of
  multidimensional fuel droplets, International Journal of Heat and Mass
  Transfer 131 (2019) 836--853.

\bibitem{dupont2010numerical}
J.-B. Dupont, D.~Legendre, Numerical simulation of static and sliding drop with
  contact angle hysteresis, Journal of Computational Physics 229~(7) (2010)
  2453--2478.

\bibitem{roenby2016computational}
J.~Roenby, H.~Bredmose, H.~Jasak, A computational method for sharp interface
  advection, Royal Society open science 3~(11) (2016) 160405.

\bibitem{damian2013extended}
S.~M. Dami{\'a}n, An extended mixture model for the simultaneous treatment of
  short and long scale interfaces, Ph.D. thesis, Universidad Nacional Del
  Litoral. Facultad de Ingenieria y Ciencias Hidricas (2013).

\bibitem{greenshields2015openfoam}
C.~J. Greenshields, Open{FOAM} user guide, Open{FOAM} Foundation Ltd, version
  3~(1).

\bibitem{jasak1996error}
H.~Jasak, Error analysis and estimation for the finite volume method with
  applications to fluid flows.

\bibitem{sussman2007improvements}
M.~Sussman, M.~Ohta, Improvements for calculating two-phase bubble and drop
  motion using an adaptive sharp interface method, Fluid Dyn. Mater. Process
  3~(1) (2007) 21--36.

\bibitem{owkes2015mesh}
M.~Owkes, O.~Desjardins, A mesh-decoupled height function method for computing
  interface curvature, Journal of Computational Physics 281 (2015) 285--300.

\bibitem{renardy2002prost}
Y.~Renardy, M.~Renardy, Prost: a parabolic reconstruction of surface tension
  for the volume-of-fluid method, Journal of computational physics 183~(2)
  (2002) 400--421.

\bibitem{ginzburg2001two}
I.~Ginzburg, G.~Wittum, Two-phase flows on interface refined grids modeled with
  vof, staggered finite volumes, and spline interpolants, Journal of
  Computational Physics 166~(2) (2001) 302--335.

\bibitem{coquerelle2016fourth}
M.~Coquerelle, S.~Glockner, A fourth-order accurate curvature computation in a
  level set framework for two-phase flows subjected to surface tension forces,
  Journal of Computational Physics 305 (2016) 838--876.

\bibitem{lopez2009improved}
J.~L{\'o}pez, C.~Zanzi, P.~G{\'o}mez, R.~Zamora, F.~Faura, J.~Hern{\'a}ndez, An
  improved height function technique for computing interface curvature from
  volume fractions, Computer methods in applied mechanics and engineering
  198~(33-36) (2009) 2555--2564.

\bibitem{bna2015numerical}
S.~Bn{\`a}, S.~Manservisi, R.~Scardovelli, P.~Yecko, S.~Zaleski, Numerical
  integration of implicit functions for the initialization of the vof function,
  Computers \& Fluids 113 (2015) 42--52.

\bibitem{bna2016vofi}
S.~Bn{\`a}, S.~Manservisi, R.~Scardovelli, P.~Yecko, S.~Zaleski, Vofi—a
  library to initialize the volume fraction scalar field, Computer Physics
  Communications 200 (2016) 291--299.

\bibitem{abadie2015combined}
T.~Abadie, J.~Aubin, D.~Legendre, On the combined effects of surface tension
  force calculation and interface advection on spurious currents within volume
  of fluid and level set frameworks, Journal of Computational Physics 297
  (2015) 611--636.

\bibitem{herrmann2008balanced}
M.~Herrmann, A balanced force refined level set grid method for two-phase flows
  on unstructured flow solver grids, Journal of computational physics 227~(4)
  (2008) 2674--2706.

\bibitem{harvie2006analysis}
D.~J. Harvie, M.~Davidson, M.~Rudman, An analysis of parasitic current
  generation in volume of fluid simulations, Applied mathematical modelling
  30~(10) (2006) 1056--1066.

\bibitem{popinet1999front}
S.~Popinet, S.~Zaleski, A front-tracking algorithm for accurate representation
  of surface tension, International Journal for Numerical Methods in Fluids
  30~(6) (1999) 775--793.

\bibitem{ubbink1999method}
O.~Ubbink, R.~Issa, A method for capturing sharp fluid interfaces on arbitrary
  meshes, Journal of Computational Physics 153~(1) (1999) 26--50.

\bibitem{lamb1945hydrodynamics}
H.~Lamb, Hydrodynamics, Dover Publications, New York, 1945.

\bibitem{hysing2009quantitative}
S.-R. Hysing, S.~Turek, D.~Kuzmin, N.~Parolini, E.~Burman, S.~Ganesan,
  L.~Tobiska, Quantitative benchmark computations of two-dimensional bubble
  dynamics, International Journal for Numerical Methods in Fluids 60~(11)
  (2009) 1259--1288.

\bibitem{turek1999efficient}
S.~Turek, Efficient Solvers for Incompressible Flow Problems: An Algorithmic
  and Computational Approache, Vol.~6, Springer Science \& Business Media,
  1999.

\bibitem{parolini2004computational}
N.~Parolini, Computational fluid dynamics for naval engineering problems, Tech.
  rep., EPFL (2004).

\bibitem{john2004moonmd}
V.~John, G.~Matthies, Moonmd--a program package based on mapped finite element
  methods, Computing and Visualization in Science 6~(2-3) (2004) 163--170.

\bibitem{walton2004evaporation}
D.~Walton, The evaporation of water droplets. a single droplet drying
  experiment, Drying technology 22~(3) (2004) 431--456.

\bibitem{pozrikidis1997introduction}
C.~Pozrikidis, J.~H. Ferziger, Introduction to theoretical and computational
  fluid dynamics, Physics Today 50 (1997) 72.

\end{thebibliography}


\begin{thebibliography}{1}
\expandafter\ifx\csname url\endcsname\relax
  \def\url#1{\texttt{#1}}\fi
\expandafter\ifx\csname urlprefix\endcsname\relax\def\urlprefix{URL }\fi
\expandafter\ifx\csname href\endcsname\relax
  \def\href#1#2{#2} \def\path#1{#1}\fi

\bibitem{ref_simpo_ben:15}
A.~Attili, F.~Bisetti, M.~E. Mueller, H.~Pitsch, Formation, growth and
  transport of soot in a three-dimensional turbulent non-premixed jet flame
  \textbf{161}  1841--1865.

\end{thebibliography}





\end{document}